\documentclass[reprint,aps,prc,amsmath,amssymb,nofootinbib,superscriptaddress]{revtex4-2}

\usepackage{mathtools}
\usepackage{braket} 
\usepackage[dvipsnames]{xcolor}  
\usepackage{float}
\usepackage[T1]{fontenc}
\usepackage[utf8]{inputenc}
\usepackage{graphicx,color,rotating,pifont}
\usepackage{amsmath,amssymb}
\usepackage{mathtools}
\usepackage{subfigure}
\usepackage{siunitx}
\usepackage{bm}
\usepackage{ae}
\usepackage{dcolumn}
\usepackage{txfonts}
\usepackage{tensor}
\usepackage{braket}
\usepackage{booktabs}
\usepackage{xspace}
\usepackage{soul}
\setstcolor{red}
\setul{}{.2ex} 

\usepackage[normalem]{ulem}

\usepackage{tikz}
\usetikzlibrary{positioning,calc,shapes,decorations.markings,decorations.pathmorphing}
\tikzset{asg/.cd,
  omega-vertex/.style={circle,solid,draw=black,fill=white,minimum size=5pt, inner sep=0pt},
  dbd-vertex/.style={coordinate},
  pline/.style={thick, postaction={decorate}, decoration={markings, mark=at position .5 with {\arrow[xshift=2pt]{stealth}}}},
  hline/.style={thick, postaction={decorate}, decoration={markings, mark=at position .5 with {\arrowreversed[xshift=-2pt]{stealth}}}},
  shift arrow/.style={/pgf/decoration/transform={xshift=#1}},
  shift arrow/.default=-2pt,
  dbd-2b/.style={decorate, decoration=snake},
  omega-2b/.style={densely dashed},
  neutron/.style={draw=blue},
  proton/.style={draw=red},
}

\usepackage[colorlinks,linkcolor=blue,anchorcolor=blue,citecolor=blue]{hyperref} 
\hypersetup{
   colorlinks=true, % false: boxed links; true: colored links
   linkcolor=blue,  % color of internal links
   citecolor=blue,  % color of links to bibliography
   filecolor=blue,  % color of file links
   urlcolor=blue    % color of external links
}

\DeclareMathSymbol{\NS}{\mathord}{AMSb}{"4E}

\DeclareSIUnit{\fm}{\femto\meter}

\newcommand{\beq}{\begin{equation}}
\newcommand{\eeq}{\end{equation}}
\newcommand{\beqn}{\begin{eqnarray}}
\newcommand{\eeqn}{\end{eqnarray}}
\newcommand{\bsub}{\begin{subequations}}
\newcommand{\esub}{\end{subequations}}
\newcommand{\bpm}{\begin{pmatrix}}
\newcommand{\epm}{\end{pmatrix}}

\newcommand\identity{1\kern-0.25em\text{l}}

\begin{document}

\title{Optimization of generator coordinate method with machine-learning techniques for nuclear spectra and  neutrinoless double-beta decay: ridge regression for nuclei with axial deformation}
 
 \author{X. Zhang}  
 \affiliation{School of Physics and Astronomy, Sun Yat-sen University, Zhuhai 519082, P.R. China} 

   \author{W. Lin}  
  \affiliation{School of Physics and Astronomy, Sun Yat-sen University, Zhuhai 519082, P.R. China}

  \author{J. M. Yao}   
  \email{Corresponding author: yaojm8@sysu.edu.cn}
  \affiliation{School of Physics and Astronomy, Sun Yat-sen University, Zhuhai 519082, P.R. China}  
  
  \author{C. F. Jiao}    
  \affiliation{School of Physics and Astronomy, Sun Yat-sen University, Zhuhai 519082, P.R. China}

  \author{A. M. Romero}  
  \affiliation{Department of Physics and Astronomy, University of North Carolina,  Chapel Hill, North Carolina 27516-3255, USA}
  \affiliation{Departament de Física Quàntica i Astrofísica (FQA), Universitat de Barcelona (UB), c. Martí i Franqués, 1, 08028 Barcelona, Spain}
  \affiliation{Institut de Ciències del Cosmos (ICCUB), Universitat de Barcelona (UB), c. Martí i Franqués, 1, 08028 Barcelona, Spain}
  
    \author{T. R. Rodr\'iguez} 
  \affiliation{Departamento de Estructura de la Materia, F\'isica T\'ermica y Electr\'onica, Universidad Complutense de Madrid, E-28040 Madrid, Spain}
\affiliation{Departamento de F\'isica Te\'orica, Universidad Aut\'onoma de Madrid, E-28049 Madrid, Spain}
\affiliation{Centro de Investigaci\'on Avanzada en F\'isica Fundamental-CIAFF-UAM, E-28049 Madrid, Spain}

  \author{H. Hergert}  
  \affiliation{Facility for Rare Isotope Beams, Michigan State University, East Lansing, Michigan 48824-1321, USA}
 \affiliation{Department of Physics \& Astronomy, Michigan State University, East Lansing, Michigan 48824-1321, USA}

\date{\today}

\begin{abstract}
\begin{description}

\item[Background] 
The generator coordinate method (GCM) is an important tool of choice for modeling large-amplitude collective motion in atomic nuclei. Recently, it has attracted increasing interest as it can be exploited to extend {\em ab initio} methods to the collective excitations of medium-mass and heavy deformed nuclei, as well as the nuclear matrix elements (NME) of candidates for neutrinoless double-beta ($0\nu\beta\beta$) decay.

\item[Purpose] The computational complexity of the GCM increases rapidly with the number of collective coordinates. It imposes a strong restriction on the applicability of the method.  We aim to exploit  statistical machine-learning (ML) algorithms  to speed up GCM calculations and ultimately provide a more efficient description of nuclear energy spectra and other observables such as the NME of $0\nu\beta\beta$ decay without loss of accuracy.

\item[Method]  In this work, we propose a subspace-reduction algorithm that employs optimal statistical ML models as surrogates for exact quantum-number projection calculations for norm and Hamiltonian kernels.
The model space of the original GCM is reduced to a subspace relevant for nuclear low energy spectra and the NME of ground state to ground state $0\nu\beta\beta$ decay based on the orthogonality condition (OC) and the energy-transition-orthogonality procedure (ENTROP), respectively. Nuclear energy spectra are determined by the GCM through the configuration mixing within this subspace.  For simplicity, the polynomial ridge regression (RR) algorithm is used to learn the norm and Hamiltonian kernels of axially deformed configurations. The efficiency and accuracy of this algorithm are illustrated for \nuclide[76]{Ge} and \nuclide[76]{Se} by comparing results obtained using the optimal  RR models to direct GCM calculations. The non-relativistic Gogny force D1S and relativistic energy density functional PC-PK1, a valence-space shell-model Hamiltonian, and a modern nuclear interaction derived from chiral effective field theory are employed.  

\item[Results] The low-lying energy spectra of $^{76}$Ge and $^{76}$Se, as well as the $0\nu\beta\beta$-decay NME between their ground states, are computed.  The results show that the performance of the GCM+OC/ENTROP+RR is more robust than that of the GCM+RR alone, and the former can reproduce the results of the original GCM calculation accurately with a significantly reduced computational cost. 

\item[Conclusions] Statistical ML algorithms, when implemented properly, can accelerate GCM calculations without loss of accuracy. In applications with axially deformed states, the computation time can be reduced by a factor of three to nine for energy spectra and NMEs, respectively. This factor is expected to increase significantly with the number of employed generator coordinates. 
 
\end{description}
\end{abstract}

\maketitle

\section{Introduction}
The core idea of the generator coordinate method (GCM) is that the wave functions of nuclear states can be represented as a superposition of a set of nonorthogonal basis functions, such as Slater determinants, that are generated by some continuously changing parameters called generator coordinates {\em}~\cite{Hill:1953,Griffin:1957}. In practical applications,  one often chooses macroscopic quantities that define global nuclear properties such as deformation, and discretizes the associated coordinates on meshes. The dynamics are then described by the Ritz variational principle, where the variation is usually performed with respect to the expansion coefficients in the chosen basis, and sometimes the basis configurations themselves~\cite{Ring:1980,Reinhard:1987RPP},  resulting in a generalized eigenvalue equation (GEE). 
In this way, the GCM provides a general approach for solving many-body problems in both nuclear physics~\cite{Ring:1980} and quantum chemistry~\cite{Capelle:2003,Alon:2005,Orestes:2007} due to the great flexibility of choosing basis functions or generator coordinates. In nuclear physics, the GCM combined with quantum number projections has been extensively employed in studies of the energies and transition rates of low-lying states (see, for instance, Refs. \cite{Bender:2003RMP,Niksic:2011PPNP,Egido:2016PS,Robledo:2018JPG,Sheikh:2021qv}). State-of-the-art GCM applications range from the structure of nuclei with triaxial deformation~\cite{Yao:2014,Bally:2014,Rodriguez:2014Kr,Egido:2016PRL}, to quadrupole-octupole deformed even-even nuclei~\cite{Yao:2015,Zhou:2016,Bernard:2016}, certain odd-mass nuclei~\cite{Bally:2014,Borrajo:2017PLB} to the the computation of nuclear matrix elements (NMEs) of $0\nu\beta\beta$ decay \cite{Rodriguez:2010,Vaquero:2013,Song:2014,Yao:2015,Yao:2016PRC,Jiao:2017,Yao:2018wq,Yao:2020PRL}. The latter are vital for interpreting and planning the current- and next-generation tonne-scale experiments for $0\nu\beta\beta$ decays (see the recent reviews~\cite{Engel:2017,Yao:2022PPNP,Agostini:2022RMP}). GCM calculations most frequently use modern energy density functionals (EDFs) and effective Hamiltonians as inputs, but there have been several works that employ nuclear forces from chiral Effective Field Theory (EFT) in recent years, as the GCM has attracted interest as an pathway for extending nuclear {\em ab initio} calculations to deformed nuclei~\cite{Yao:2018wq,Yao:2020PRL,Romero:2021PRC,Frosini:2022EPJA,Frosini:2022xg,Frosini:2022EPJA3,Duguet:2022}. 

 The exact wave functions of nuclear states can in principle be well represented with the GCM ansatz if one chooses a sufficient number of generator coordinates, but this comes at the price of increasing both complexity and computational time. It makes the problem hard to handle exactly because the kernels in the GEE usually require multidimensional integrals of overlap functions over the collective coordinates. For this reason, it is a challenge to extend multi-dimensional GCM to atomic nuclei throughout the nuclear chart. Practical applications are usually limited to only one or two generator coordinates~\cite{Bender:2004Global,Rodriguez:2014Global}. Therefore, a good choice of generator coordinates or a subset defined by the basis functions becomes important, and this choice is usually based on an {\em  educated guess}, unfortunately.

In applications, one often observes that many of the basis functions connected by the generator coordinates have little contribution to the wave functions of low-lying states and can therefore be safely omitted --- see, for instance, Refs.~\cite{Broeckhove:1979,Romero:2021PRC,Martinez-Larraz:2022}. In other words, a careful selection of the basis functions can reduce the dimensions of the GEE, and therefore the computational cost.  It is worth noting that similar considerations apply to the eigenvector continuation (EC) method~\cite{Frame:2018PRL,Sarkar:2021}, which finds the eigenvalues and eigenvectors of a Hamiltonian with one or more control parameters, usually the coupling constants of chiral Hamiltonians. In this context, the EC method has been extensively applied to emulate few- and many-body calculations for nuclear structure and scattering  \cite{Ekstrom:2019PRL,Konig:2020fn,Furnstahl:2020PLB,Drischler:2021PLB,Bai:2021PRC} in recent years. These parameters of EC are analogous to the generator coordinates in GCM, hence finding an efficient way for sampling the basis functions that define a subspace to represent the states of interest is important for both EC and GCM~\cite{Broeckhove:1979}. Several algorithms have been proposed, including the variation-after-projection algorithm~\cite{Kanada-Enyo:1998PRL,Ohta:2004PRC,Gao:2022PLB}, the stochastic sampling with  Monte-Carlo techniques~\cite{Otsuka:2001PPNP,Shinohara:2006PRC,Ichikawa:2021},  the choice of low-lying quasiparticle Tamm-Dancoff modes~\cite{Jiao:2019PRC},  the  energy-transition-orthogonality procedure (ENTROP)~\cite{Romero:2021PRC}, and the discrete nonorthogonal shell model (DNO-SM)~\cite{Dao:2022}. We note that many of these algorithms still require a substantial computational effort for the subspace determination.

  In the past decade, machine learning (ML) techniques combined with statistical methods have been applied to a variety of nuclear physics problems, ranging from the smallest constituents of matter to the physics of dense astronomical objects --- see, for instance, the recent review~\cite{Boehnlein:2022RMP} and the references therein. In some of these applications, a specific statistical ML model is trained to predict nuclear observables directly, including nuclear masses~\cite{Utama:2016PRC,Niu:2018PLB,Neufcourt:2018PLB,Niu:2022PRC,Wu:2022PLB}, charge radii~\cite{Utama:2016,Wu:2020,Dong:2022PRC}, $\beta$-decay half-lives~\cite{Costiris:2009PRC,Niu:2018PRC_BetaDecay}, fission yields~\cite{Wang:2019PRL} and many others. In other approaches, statistical ML can be used to enhance nuclear many-body calculations as a surrogate model for expensive computational steps. Examples are the use of a deep neural network committee to optimize collective Hamiltonians for low-lying nuclear states~\cite{Lasseri:2020PRL}, or back-propagation neural networks~\cite{Yang:2022} and kernel Ridge regression~\cite{Wu:2022PRC} to determine density profiles as inputs for nuclear radii and binding energies in the framework of density functional theory.

In the present work,  we present the first application of statistical ML techniques to optimize GCM calculations for the low-lying states and the NMEs of $0\nu\beta\beta$ decay of realistic candidate nuclei. The norm and Hamiltonian kernels will be learned by a statistical model.  Here, we are confronted with several challenges originating from the norm kernels, which are nonlocal in the collective coordinate space and vary by several orders of magnitude. The solutions of GCM are sensitive to any noise in the norm kernels, which may affect the linear dependence among basis functions and their nontrivial coherence with Hamiltonian kernels, and potentially spoil the GCM description entirely. To tackle these challenges, we propose a subspace reduction algorithm, that uses statistical ML  models together with an orthogonality condition (OC) method as an efficient tool to determine the subspace in which the wave functions of nuclear low-lying states can be well represented. Its efficiency and accuracy are illustrated for nuclear energy spectra and $0\nu\beta\beta$ decay NMEs starting from two different EDFs, a valence-space shell-model interaction, and a realistic two- plus three-nucleon Hamiltonian from chiral EFT. 

The article is organized as follows. In Sect.~\ref{sec:GCM}, the formalism for the GCM, the NME of $0\nu\beta\beta$ decay, and the statistical ML model are introduced. In Sect.~\ref{sec:results}, the performance of the subspace reduction algorithm is illustrated with two different EDFs, and two different Hamiltonians. A summary of our findings and an outlook are given in Sect.~\ref{sec:summary}. 

 \section{Formalism}
\subsection{The generator coordinate method} 
\label{sec:GCM}

In a GCM calculation with quantum-number projections, the nuclear wave function is constructed as follows:
\begin{equation}
\label{eq:wfs}
    \left|\Psi_{\alpha}^{JM N Z}\right\rangle
    = \sum^{N_q}_{\boldsymbol{q}=1} \sum_{K=-J}^{J} 
    f^{J \alpha}_{K}(\boldsymbol{q})\ket{J(MK)NZ, \boldsymbol{q}}
\end{equation}
where $\alpha=1, 2, \ldots$ distinguishes the states with the same angular momentum $J$ and the numbers of nucleons. The basis functions $\ket{J(MK)NZ, \boldsymbol{q}}$ are quantum-number projected Hartree-Fock-Bogoliubov (HFB) states labeled by the generator coordinate $\boldsymbol{q}$
\begin{equation}
    \left|J(MK)NZ, \boldsymbol{q}\right\rangle
    =\hat{P}_{M K}^{J} \hat{P}^{N} \hat{P}^{Z}\ket{\Phi\left(\boldsymbol{q}\right)}.
\end{equation}
The operator $\hat P^J_{MK}$  extracts from the HFB wave function  $\ket{\Phi\left(\boldsymbol{q}\right)}$ the component whose angular momentum along the intrinsic $z$ axis is given by $K$. The $\hat{P}^{N,Z}$ are the particle number projection (PNP) operators that extract the component with the appropriate neutron number $N$ and proton number $Z$,  respectively.  The weight function $f_{K}^{J \alpha}(\boldsymbol{q})$ in the GCM states given by Eq. (\ref{eq:wfs}) is determined by the variational principle, which leads to the discretized Hill-Wheeler-Griffin (HWG) equation~\cite{Hill:1953,Griffin:1957},
\begin{equation}
    \label{eq:HWG}
    \sum_{\boldsymbol{q}^{\prime}K^{\prime}}
    \left[\mathcal{H}_{K K^{\prime}}^{J}\left(\boldsymbol{q}, \boldsymbol{q}^{\prime}\right)
    -E_{\alpha}^{J} \mathcal{N}_{K K^{\prime}}^{J}\left(\boldsymbol{q}, \boldsymbol{q}^{\prime}\right)\right] f^{J \alpha}_{K^{\prime}}(\boldsymbol{q}^{\prime})=0.
\end{equation}
The Hamiltonian and norm kernels $\mathcal{H}$  and  $\mathcal{N}$  are defined as
\bsub
\label{eq:kernels}
 \begin{align}
 \label{eq:H_kernel}
 \mathcal{H}_{K K^{\prime}}^{J}\left(\boldsymbol{q}, \boldsymbol{q}^{\prime}\right)
&\equiv\bra{J(MK)NZ, \boldsymbol{q}} \hat{\mathbf{H}} \ket{J(MK')NZ, \boldsymbol{q}^{\prime}},\\
\label{eq:Norm_kernel}
 \mathcal{N}_{K K^{\prime}}^{J}\left(\boldsymbol{q}, \boldsymbol{q}^{\prime}\right)
&\equiv\bra{J(MK)NZ, \boldsymbol{q}} \hat\identity \ket{J(MK')NZ, \boldsymbol{q}^{\prime}},
\end{align}
\esub
where $\hat{\mathbf{H}}$ and $\hat\identity$ are the Hamiltonian and identity operators, respectively. We note that in the subsequent EDF-based calculations, the Hamiltonian kernels are evaluated based on the {\em mixed density prescription}: the Hamiltonian overlaps between two different configurations are replaced with the energy which is a functional of mixed densities and currents defined by the two configurations~\cite{Robledo:2018JPG}. 

The HWG equation (\ref{eq:HWG}) is solved as follows. We first diagonalize the norm kernel matrix $\mathcal{N}_{K K^{\prime}}^{J}\left(\boldsymbol{q}, \boldsymbol{q}^{\prime}\right)$ and use its eigenvalues and eigenvectors to construct a set of orthonormal bases $\{\ket{k}\}$, called 'natural states'. To remove the overcompleteness of the original basis functions $\ket{J(MK)NZ, \boldsymbol{q}}$ that stems from the use of continuous quantum numbers, only eigenvectors whose corresponding eigenvalue is larger than a chosen cutoff are included. Then we evaluate the elements of the Hamiltonian matrix in this new subspace $H_{kk'}$, whose eigenvalues then define the energies $E_{\alpha}^{JNZ}$ of the GCM states. The corresponding eigenvectors are used to determine the weight function $f_{K}^{J \alpha}(\boldsymbol{q})$. More details can be found in Refs.~\cite{Ring:1980,Yao:2022PPNP}, for instance.

\subsection{The nuclear matrix element of $0\nu\beta\beta$ decay}

Here, we only consider the NME of  $0\nu\beta\beta$ decay corresponding to the transition from the ground state of an even-even nucleus to that of a neighboring even-even nucleus. The spin-parity of both ground states is $0^+$. Their wave functions are given by Eq.~(\ref{eq:wfs}) with $J=K=0$. For convenience, we simply use the symbol $f(\mathbf{q})$ to replace $f^{J=0, \alpha=1}_{K=0}(\boldsymbol{q})$ and ${\cal N}$ for ${\cal N}^{J=0}_{K=0,K'=0}$. Therefore, one finds the following expression for the NME:
\begin{eqnarray}
    M^{0\nu}
    &=&\sum_{\mathbf{q}_F,\mathbf{q}_I}  f^\ast(\mathbf{q}_F)f(\mathbf{q}_I) \langle \Phi(\mathbf{q}_F)|\hat O^{0\nu}\hat P^{N_I}\hat P^{Z_I}\hat P^{J  = 0}|\Phi(\boldsymbol{q}_I)\rangle \nonumber\\
    &=& \sum_{\mathbf{q}_F,\mathbf{q}_I} f^\ast(\mathbf{q}_F)f(\mathbf{q}_I)
 \mathcal{N}^{1/2}\left(\boldsymbol{q}_I, \boldsymbol{q}_I\right)\mathcal{N}^{1/2}\left(\boldsymbol{q}_F, \boldsymbol{q}_F\right) 
     {\cal M}^{0\nu} (\bm{q}_F, \bm{q}_I),\nonumber\\
\end{eqnarray}
where the kernel ${\cal M}^{0\nu} (\bm{q}_F, \bm{q}_I)$  for the NME can be computed with the help of two-body transition matrix elements $O^{0\nu}_{pp'nn'}$ in a spherical harmonic oscillator basis, 
 \begin{eqnarray}
   {\cal M}^{0\nu} (\bm{q}_F, \bm{q}_I) 
 = \frac{1}{4}\sum_{pp'nn'}  O^{0\nu}_{pp'nn'} 
  \rho_{pp'nn'}(\bm{q}_F, \bm{q}_I).
\end{eqnarray}
The two-body transition density $\rho_{pp'nn'}(\bm{q}_F, \bm{q}_I)$ is determined by
\begin{widetext}
\begin{eqnarray}
    \label{eq:config_dependent_TD2B}
     \rho_{pp'nn'}(\bm{q}_F, \bm{q}_I)  
      &=& \dfrac{1}{\mathcal{N}^{1/2}(\boldsymbol{q}_I, \boldsymbol{q}_I)\mathcal{N}^{1/2}(\boldsymbol{q}_F, \boldsymbol{q}_F)} 
      \int^{2\pi}_0 \dfrac{e^{-iN_I\varphi_N}}{2\pi}d\varphi_N
      \int^{2\pi}_0 \dfrac{e^{-iZ_I\varphi_Z}}{2\pi} d\varphi_Z\nonumber\\
      &&\times \int d\Omega  
         \bra{\Phi(\bm{q}_{F})}   c^\dagger_p c^\dagger_{p'}c_{n'}c_n e^{i\varphi_N\hat N}e^{i\varphi_Z \hat Z} \hat R(\Omega) \ket{\Phi(\bm{q}_{I})}.
\end{eqnarray}
\end{widetext}
where the rotation operator reads $\hat R(\Omega)=e^{i\varphi\hat J_z}e^{i\theta\hat J_y}e^{i\psi\hat J_z}$, $(\varphi,\theta,\psi)$ are the three Euler angles, and $\varphi_{N/Z}$ are the gauge angles that define the projection operators $\hat P^{N/Z}$. The symbols $N_I, Z_I$ are the neutron and proton numbers of the initial nucleus. The $\ket{\Phi(\bm{q}_{I/F})}$ are the HFB wave functions for the initial and final nucleus, respectively. In the present work, only the long-range transition operator $O^{0\nu}$ in the standard mechanism of light Majorana neutrino exchange is considered. See Ref.~\cite{Yao:2022PPNP} for details.

\subsection{Learning kernels with the polynomial ridge regression}
\label{framework:ML}
 
 For the present proof-of-concept study, we only consider the  quadrupole deformation parameter $\beta$ as a generator coordinate $\mathbf{q}$ in Eq.(\ref{eq:wfs}), and we will switch labels in the basis configurations accordingly in the following discussion. Because of this specialization, we have $K=0$ and the norm kernel simplifies into the following form,
 \beqn
 \mathcal{N}_{00}^{J}\left(\beta, \beta^{\prime}\right)
 &=&\bra{J(MK=0)NZ, \beta} \hat\identity \ket{J(MK'=0)NZ, \beta^{\prime}}\nonumber\\
 &=& \frac{2J+1}{2}\int^\pi_0 d\theta d^J_{00}(\theta) \bra{\Phi(\beta)} 
 e^{i\theta \hat J_y}\hat P^Z\hat P^N\ket{\Phi(\beta^\prime)},
 \eeqn
 where $d^J_{00}(\theta)=\langle J0\vert e^{i\theta\hat J_y}\vert J0\rangle$ is the Wigner (small) d-matrix. This norm kernel will be learned with a statistical ML model in the following.
 
 Previous GCM studies have established the following features of the norm kernel:
 
 \begin{itemize}
     \item For the diagonal element of the norm kernel with $\beta=\beta'$ but without PNP operators, the overlap function can be well approximated with a Gaussian function~\cite{Islam:1979NPA,Ring:1980,Yao:2009PRC},
 \beq
  \bra{\Phi(\beta)} 
 e^{i\theta \hat J_y}\ket{\Phi(\beta)}
 \simeq \exp\left(-\frac{\theta^2}{2} \bra{\Phi(\beta)}\hat J^2_y\ket{\Phi(\beta)} \right).
 \eeq
 The non-diagonal overlap with $\beta\neq\beta'$ can be parameterized with an extension of the Gaussian overlap approximation~\cite{Hagino:2003,Sabbey:2007}.
 \item For the non-diagonal element with $\beta\neq\beta'$ and without any projections, the norm kernel can also be approximated with a Gaussian function~\cite{Brink:1968NPA,Ring:1980,Reinhard:1987RPP},
  \beq
  \mathcal{N}\left(\beta, \beta^{\prime}\right)
 \simeq \exp\left[-\frac{\gamma(q)}{2}(\beta-\beta')^2 \right],
 \eeq
 where $\gamma(q)$ is a function of the deformation parameter $q=(\beta+\beta')/2$ that can be calculated using the corresponding HFB wave functions.  
 \end{itemize}
 Of course, the actual norm kernels with the projections of particle-number and angular momentum are expected to have a much more complicated expression that is to be learned by  statistical ML models.
 
In light of the Gaussian structure that was found in the aforementioned results, it is reasonable to attempt to train the logarithm of the norm kernels to avoid dealing with data that spans several orders of magnitude. We expect this logarithm to be well approximated with a polynomial function of the quadrupole deformation parameters $\beta$ and $\beta'$. Therefore, a basic machine-learning algorithm, i.e., {\em the polynomial ridge regression (RR)}~\cite{Shwartz:2014Book,Aurelien:2017Book} is adopted for our purpose. As a test, we have validated that this algorithm exactly reproduces the kernels of a simple model with quadratic approximation~\cite{Griffin:1957}. We have also employed other algorithms such as support vector regression, which yielded similar results, but at the cost of significantly increased training time.

 Here we present some details on our implementation of the polynomial RR model. In the data preparation,  we  compute all the kernels ${\cal N}^J(\beta, \beta')$ and ${\cal H}^J(\beta, \beta')$ exactly with (\ref{eq:kernels}) for a specific nucleus $(Z,A)$ and a given interval in $\beta$ large enough to include all the relevant configurations. There are $N^2_q$ norm kernels and  $N^2_q$ Hamiltonian kernels, where $N_q$ is the number of mesh points in the deformation parameter $\beta$. For simplicity, these mesh points are equally distributed with step size $\Delta\beta$. Among all the kernels, those equally distributed in the entire deformation space with the step size of $2\Delta\beta$ or $4\Delta\beta$ are selected as a training set, while the remaining data are used for testing and validation. The number of training data and the number of test data are denoted $M_{\rm train}$ and $M_{\rm test}$, respectively. Thus, we have the relation $N^2_q=M_{\rm train}+M_{\rm test}$. 
     
 The $i$-th predicted value $\hat y^{(i)}$ in the polynomial RR is given by the following hypothetical $N$-degree polynomial function   
 \begin{equation}
 \label{eq:ML_output}
     \hat{y}^{(i)}({\boldsymbol{\theta}};N) 
     = \boldsymbol{X}^{(i)}_N\boldsymbol{\theta}_{N},
 \end{equation}
 where the $i$-th input vector $\boldsymbol{X}^{(i)}_N$ is defined as
  \begin{equation}
  \centering
  \label{eq:PolynomialFeatures}
    \boldsymbol{X}^{(i)}_N 
   =\Bigg(1, \beta, \beta^\prime, \cdots, \beta^N, \beta^{N-1}\beta^\prime, \cdots, \beta\beta^{\prime N-1}, \beta^{\prime N}\Bigg)^{(i)}
  \end{equation}  
  with $(N+2)(N+1)/2$ features, and the corresponding weight parameters
   \begin{equation}
 \boldsymbol{\theta}^T_N=\left(\theta^{0}_0, \theta^{1}_{0}, \theta^{0}_{1}\cdots, \theta^{N}_0, \theta^{N-1}_1\cdots  \theta^{1}_{N-1}, \theta_{N}\right). 
 \end{equation}  
 
 The degree $N$ of the polynomial is a hyper-parameter controlling the complexity of the model. We note from (\ref{eq:PolynomialFeatures}) that the polynomial RR is more flexible than the ridge regression with the $N$-degree polynomial kernel~\cite{Shwartz:2014Book,Aurelien:2017Book}.
 Besides, one can see that the polynomial RR fits the single, high-degree polynomial function (\ref{eq:ML_output}) to all of the values in the training set. Compared to the spline interpolation which fits low-degree polynomials to small subsets of the training set, the polynomial RR works equally well when the degree of polynomials is chosen appropriately. However, the polynomial RR is easier to be extended to the case with multiple coordinates and the possible overfitting problem is avoided by adding the Tikhonov regularization term~\cite{Shwartz:2014Book} to the mean-square error (MSE) in the definition of loss function,
 \begin{eqnarray}
 \label{eq:loss}
     {\cal L}(\boldsymbol{\theta; N, \alpha})
     &=&  \sum_{i=1}^{M_{\rm train}}\left[\hat{y}^{(i)}({\boldsymbol{\theta}};N) - y^{(i)}\right]^{2}+  \alpha \boldsymbol{\theta}^T_N\boldsymbol{\theta}_N \nonumber\\
     &=&  (\hat{\boldsymbol{Y}}-\boldsymbol{Y})^T (\hat{\boldsymbol{Y}}-\boldsymbol{Y})+ \alpha \boldsymbol{\theta}^T_N\boldsymbol{\theta}_N,
 \end{eqnarray} 
   where the vector $\boldsymbol{Y}$ is defined by
   \begin{equation}
   \boldsymbol{Y}=\left( y^{(1)}, \cdots , y^{(M_{\rm train})}\right)^{T},
    \end{equation} 
    with  $y^{(i)}$ chosen as the logarithmic value of the $i$-th  norm kernel, and $\hat y^{(i)}$ is the output (\ref{eq:ML_output}) of the RR model. The summation in (\ref{eq:loss}) runs through all the kernels in the training set. 
    
    The minimization of the above loss function leads to a normal equation with the solution given by~\cite{Aurelien:2017Book}
    \begin{equation}
  \label{eq:normal_equation}
   {\boldsymbol{\theta}}_N = (\boldsymbol{X}^T\boldsymbol{X} + \alpha \mathbf{I})^{-1} \boldsymbol{X}^T \boldsymbol{Y},
   \end{equation} 
   where $\mathbf {I}$  is the identity matrix. The ridge parameter $\alpha\ge0$ controls how much one wants to regularize the model. When the $\alpha$ is zero, the RR is simplified into the linear regression~\cite{Barlow:1989}. With the increase of $\alpha$, the regularization term dominates the squared loss function and the weight coefficients ${\boldsymbol{\theta}}_N$ tend to be small, providing a way to prevent the overfitting problem. In practice it is necessary to tune $\alpha$ in such a way that a balance is maintained between both. Instead of determining the weight coefficients ${\boldsymbol{\theta}}_N$ analytically according to Eq. (\ref{eq:normal_equation}) by  computing the inverse of the normal matrix $(\boldsymbol{X}^T\boldsymbol{X} + \alpha \mathbf{I})$,   the vector parameters {\em} $\boldsymbol{\theta}_N$ can also be determined by minimizing the loss function ${\cal L}$ with the gradient descent (GD) method for given hyperparameters $(N, \alpha)$.   We note that in general both methods give the same solution to the model parameters. Compared to the GD method, the normal-equation method is usually employed in the regression with a small number of features. However, with the increase of the hyperparameter $N$ and the size of training set, the inverse of the normal matrix becomes difficult to compute. In contrast, the computation complexity of the GD method grows moderately with the number of model parameters.  In this work, we employ the GD method to determine the weight parameters in the polynomial RR. More detailed introduction to  ML models in physics can be found in, for instance, Ref.~\cite{Mehta2019PR}.

\section{Illustrations}
\label{sec:results}

\begin{figure}
    \centering
    \includegraphics[width=\columnwidth]{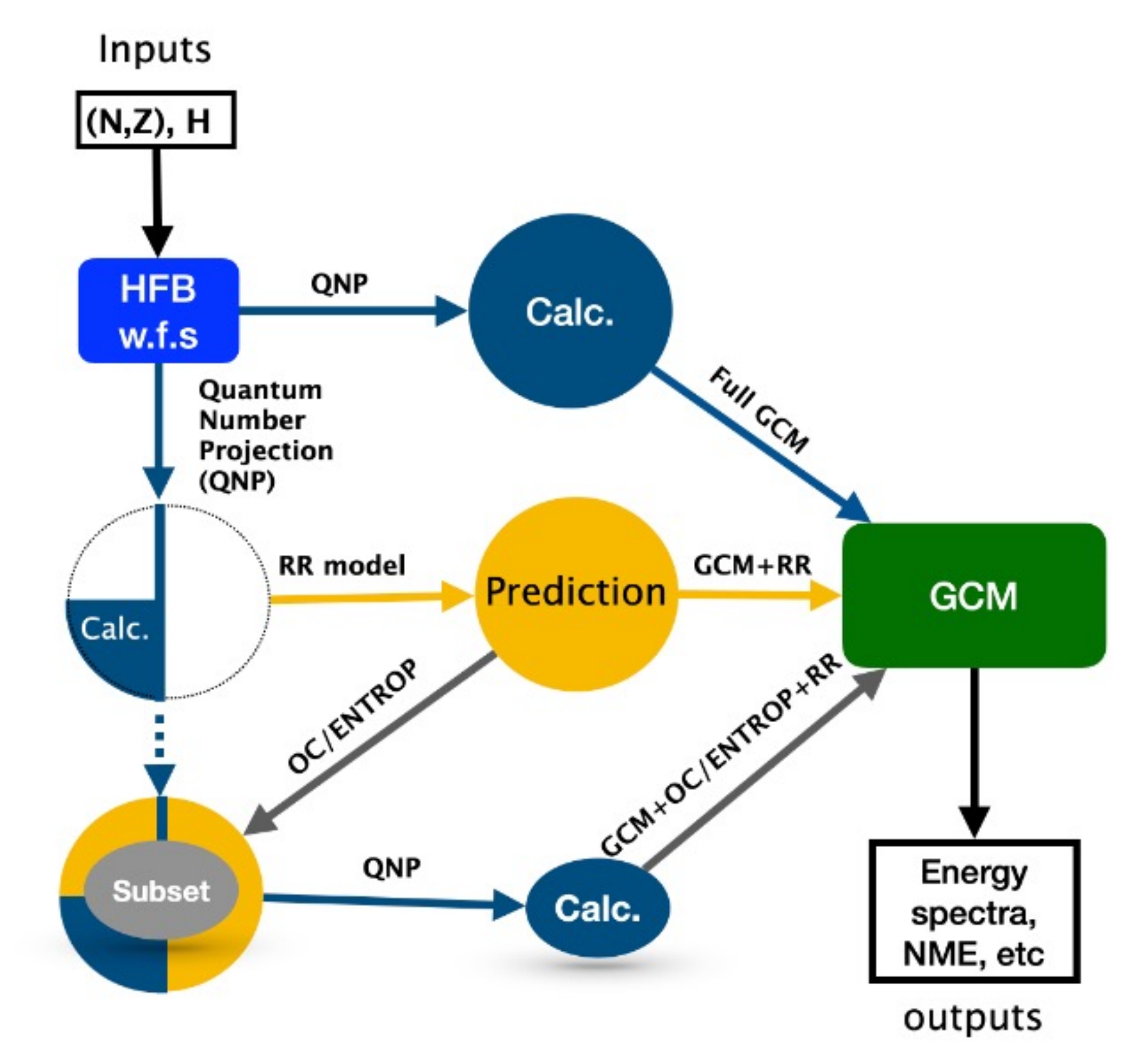}
    \caption{(Color online) Comparison of the flowcharts for the full GCM, GCM+RR and GCM+OC/ENTROP+RR approaches. The area of different shapes represents the domain of kernels obtained from different methods indicated with colored arrows. The vertical line indicates the diagonal elements of kernels. See text for details.}
    \label{fig:flowchart}
\end{figure}

\begin{figure*}[bt]
	\centering
	\hspace{-0.3cm}
	\subfigure[$0^{+}$ of $^{76}$Ge(D1S)]{
		\includegraphics[width=2.3 in]{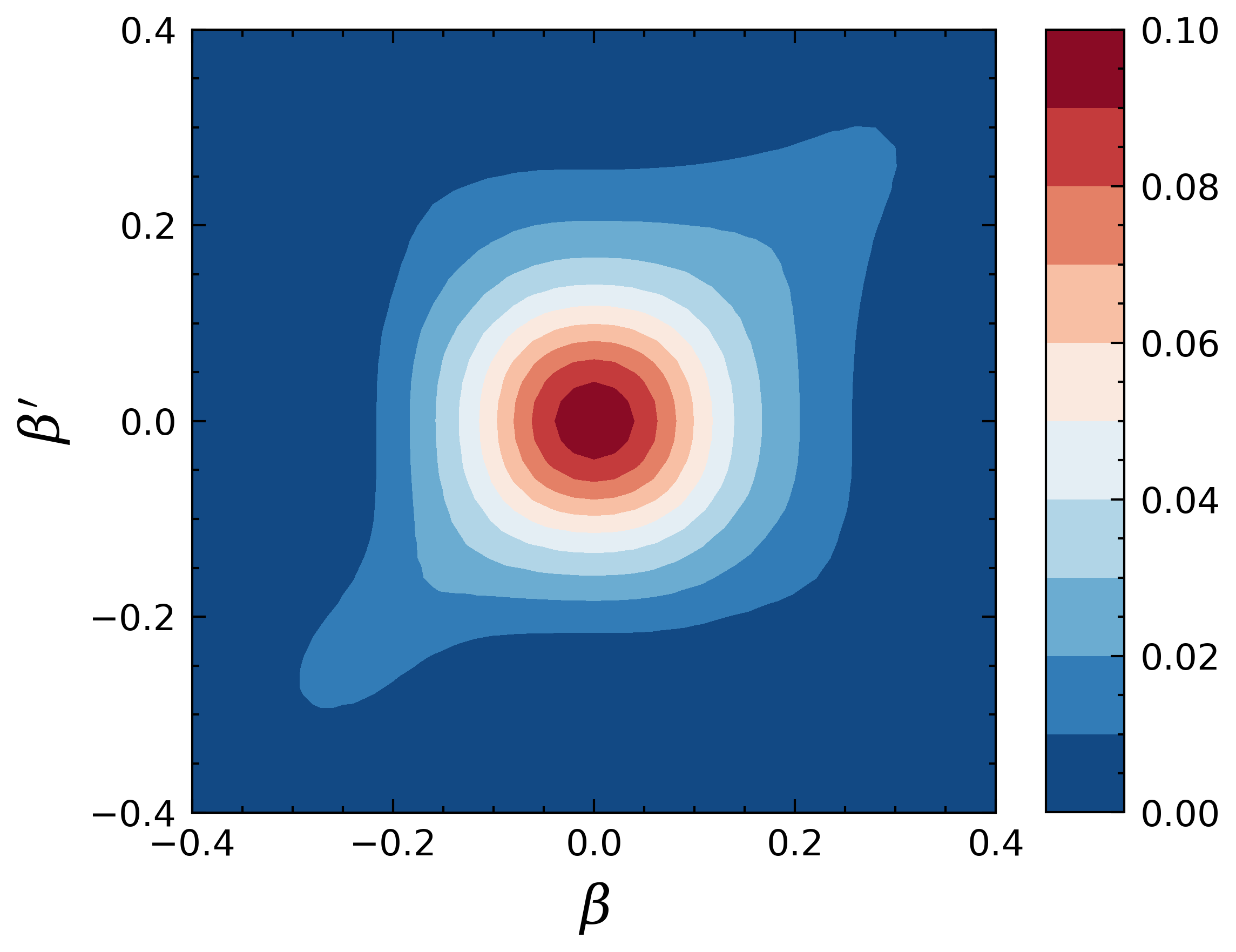}}
	\subfigure[$2^{+}$ of $^{76}$Ge(D1S)]{
		\includegraphics[width=2.3 in]{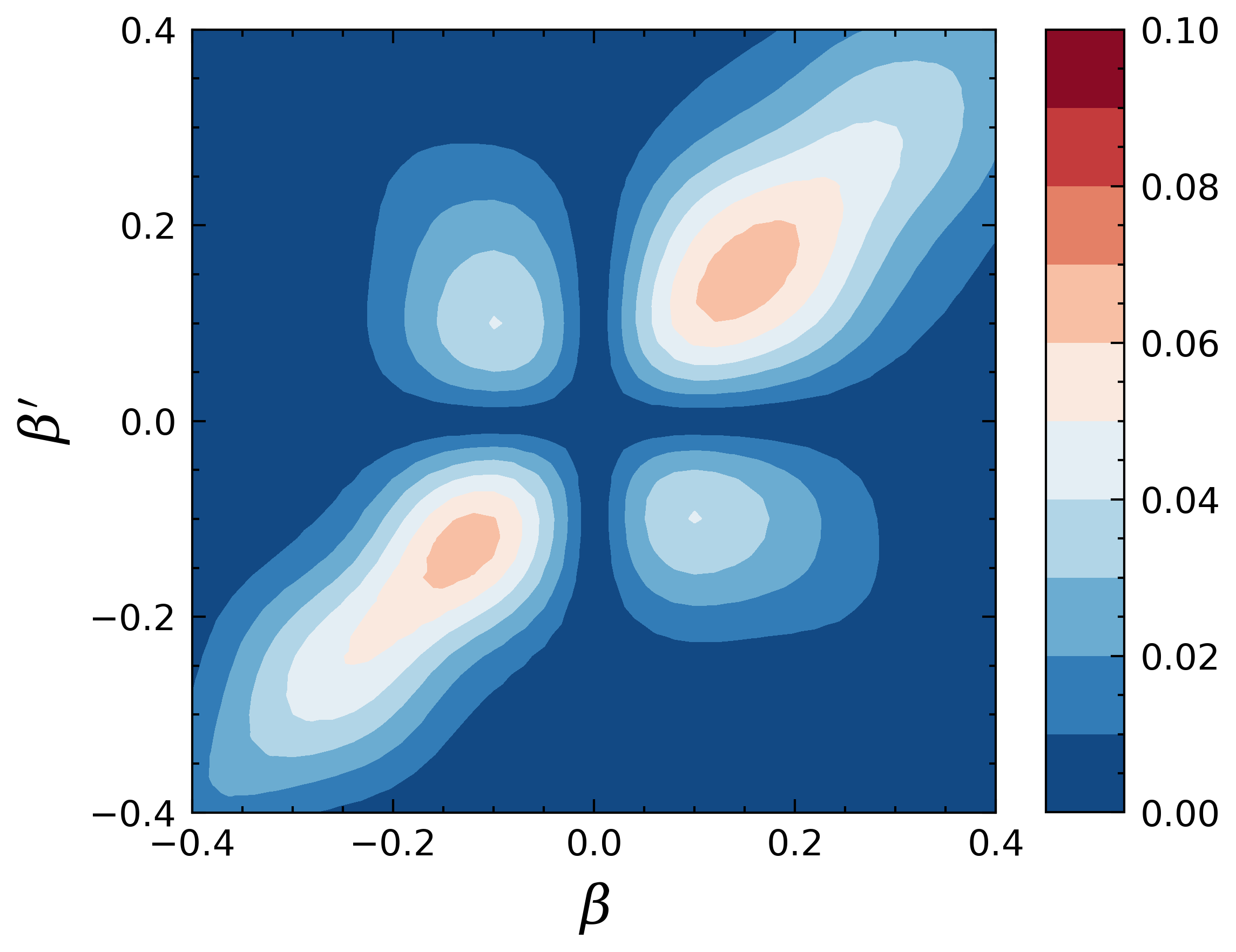}}
	\subfigure[$4^{+}$ of $^{76}$Ge(D1S)]{
		\includegraphics[width=2.3 in]{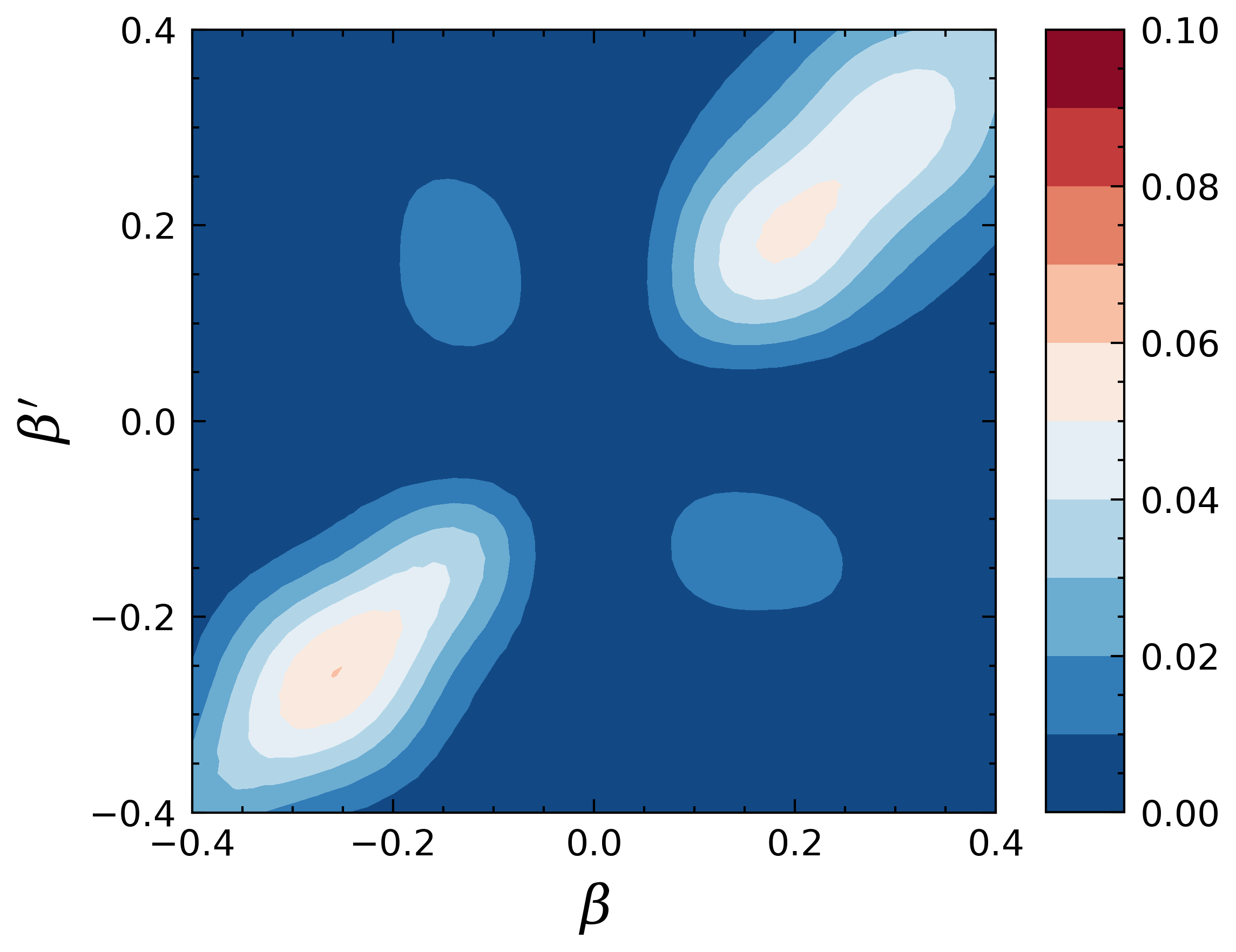}}
		
	\quad
	\hspace{-0.56cm}
	\subfigure[$0^{+}$ of $^{76}$Se(D1S)]{
		\includegraphics[width=2.3 in]{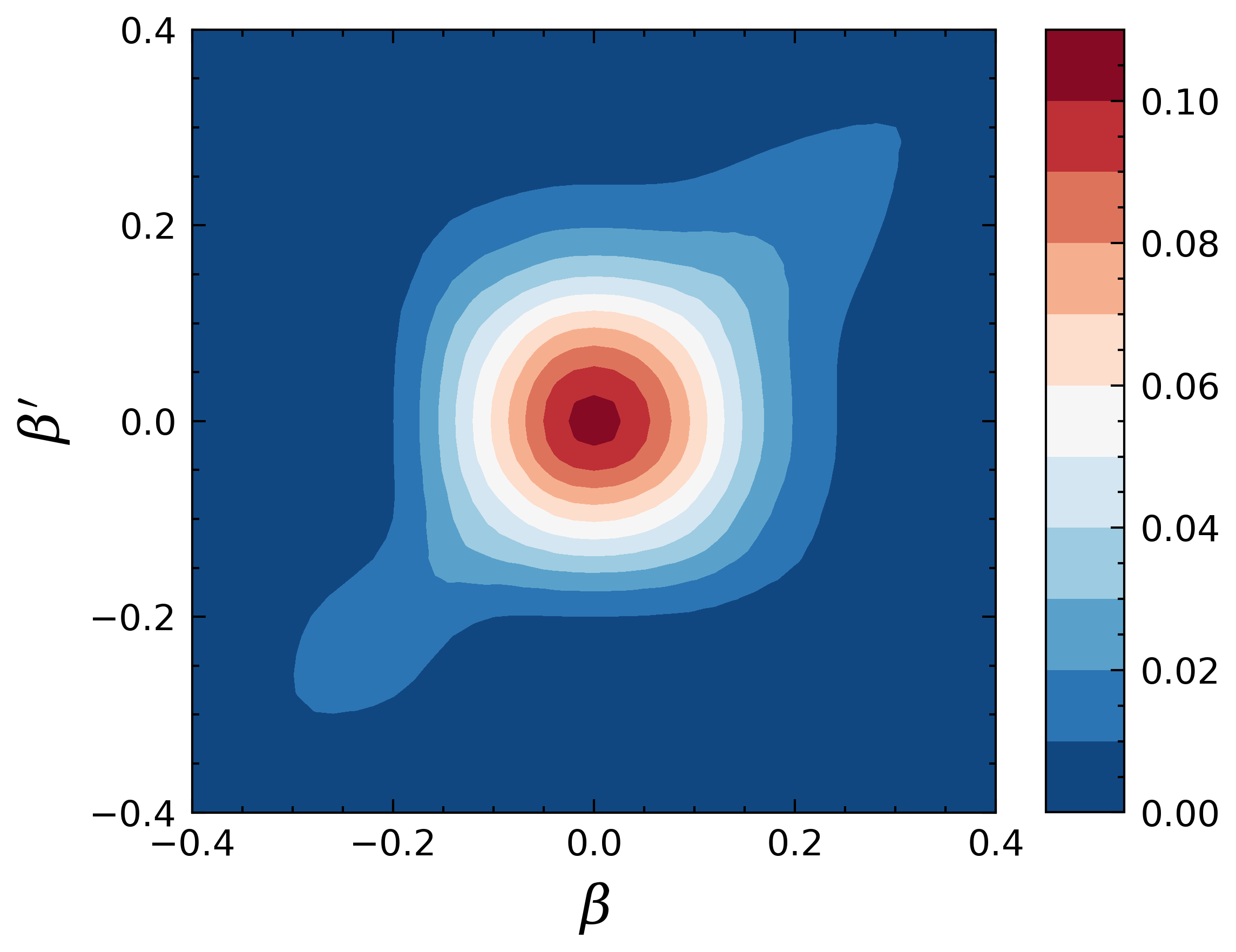}}
	\subfigure[$2^{+}$ of $^{76}$Se(D1S)]{
		\includegraphics[width=2.3 in]{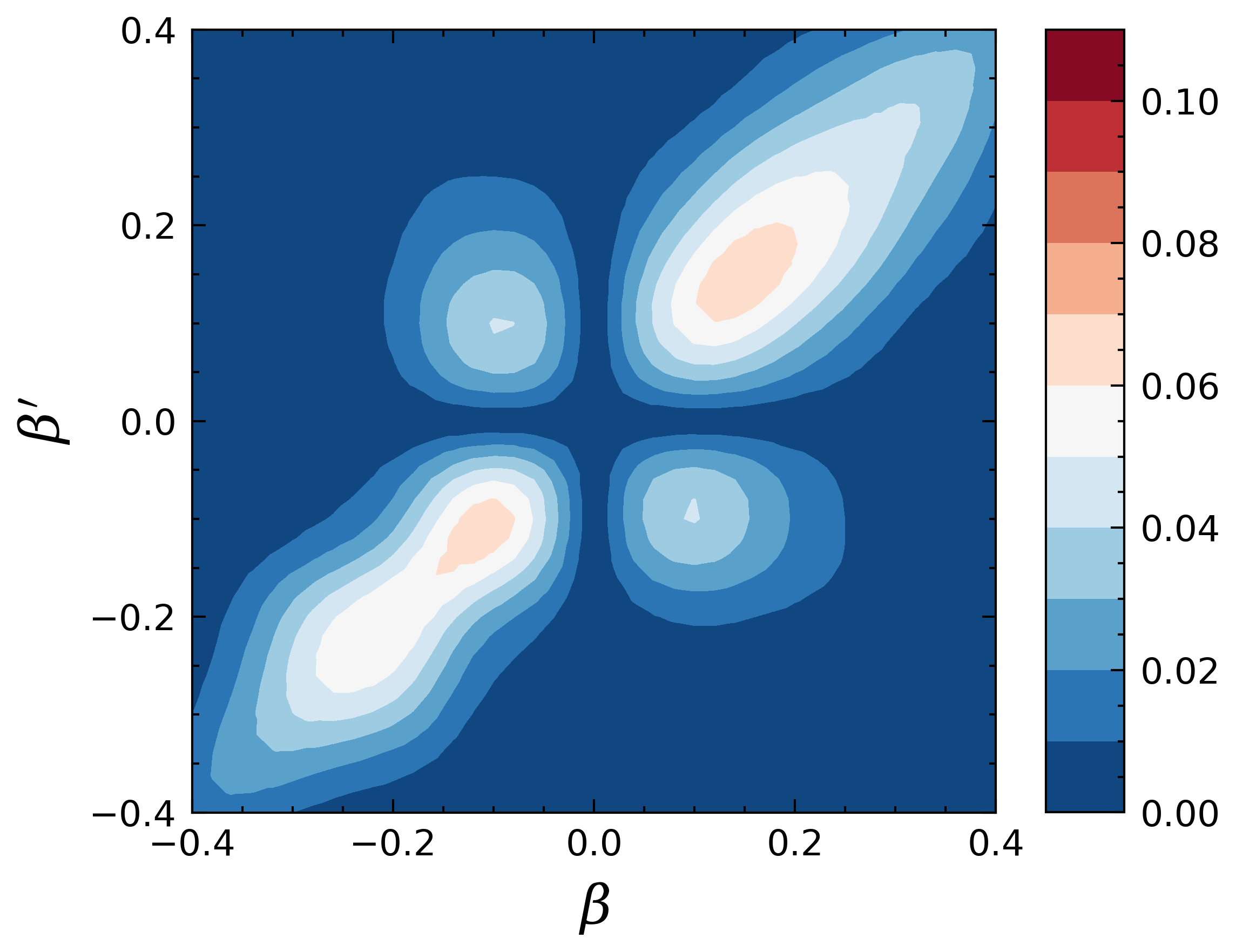}}
	\subfigure[$4^{+}$ of $^{76}$Se(D1S)]{
		\includegraphics[width=2.3 in]{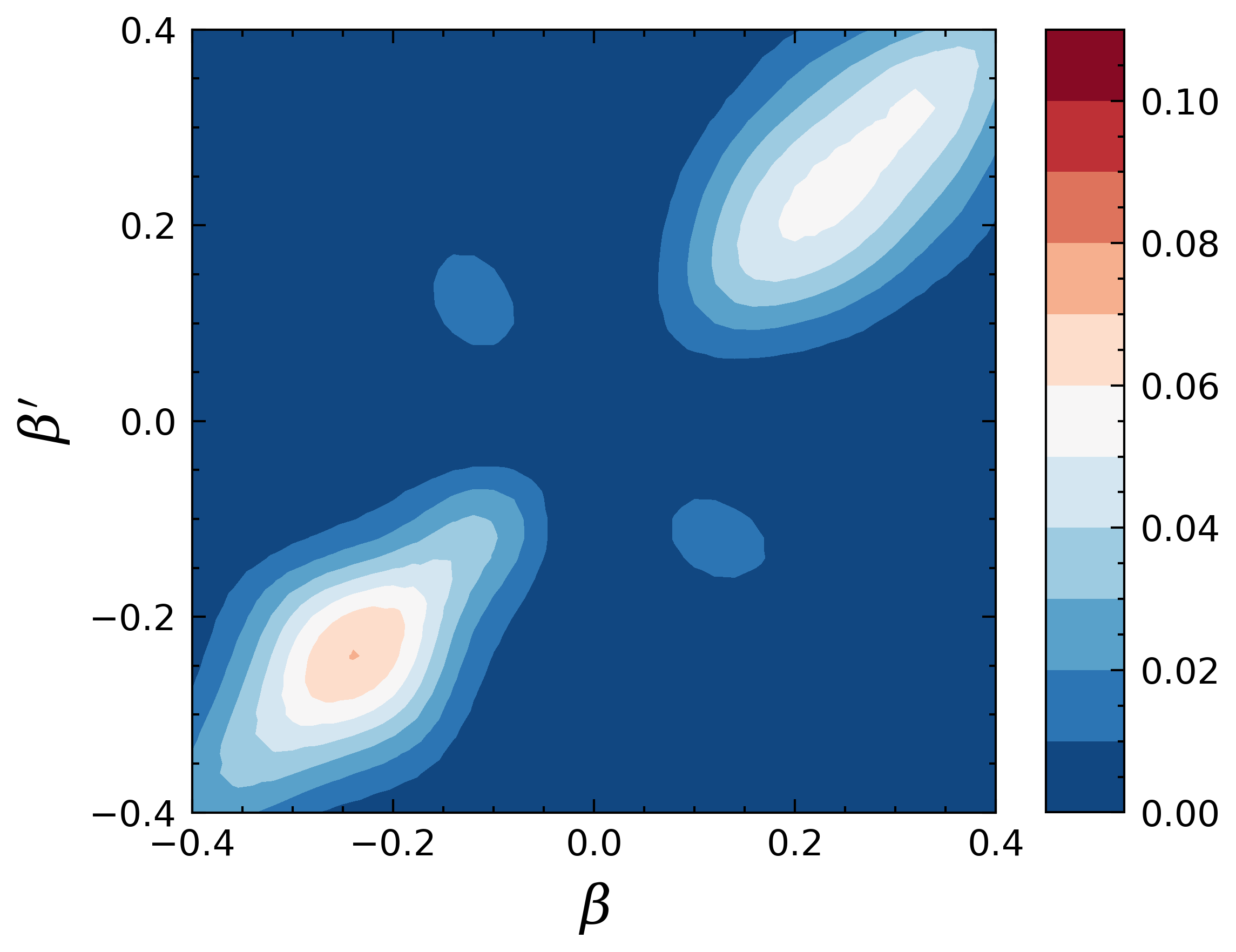}}
	\caption{(Color online) The distribution of norm kernels ${\cal N}^{J}_{00}(\beta, \beta^\prime)$ for $^{76}$Ge (a,b,c) and $^{76}$Se (d,e,f) from the HFB calculation using the Gogny D1S force as a function of the quadrupole deformation parameters $\beta, \beta'$, where the angular momentum is $J=0$ (a,d), $2$ (b,c), and $4$ (e,f), respectively. 
	}
	\label{fig:D1S_norm}
\end{figure*}
 
\begin{figure}[] 
		\includegraphics[width=0.48\columnwidth]{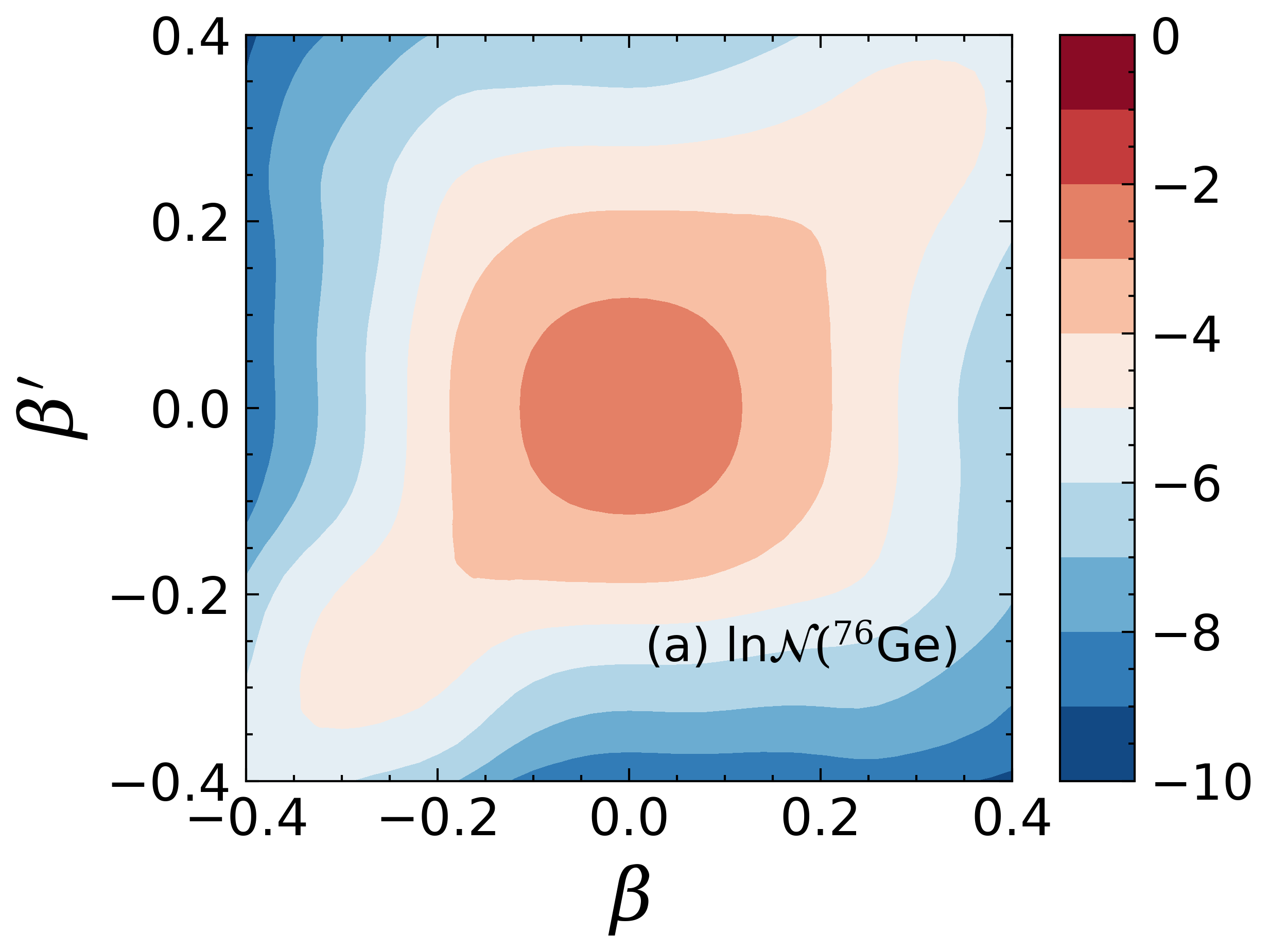}
		\includegraphics[width=0.48\columnwidth]{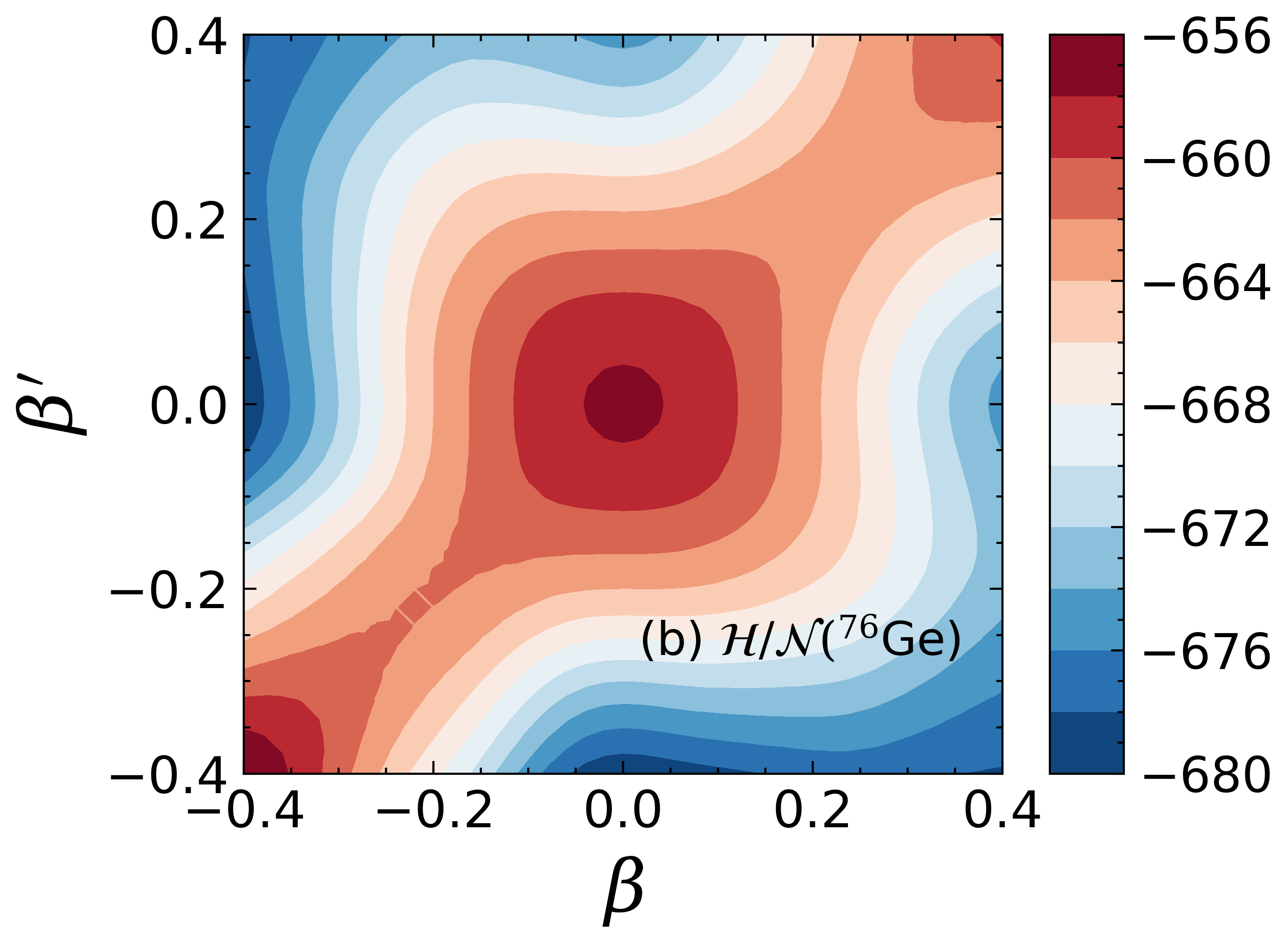}
		\includegraphics[width=0.48\columnwidth]{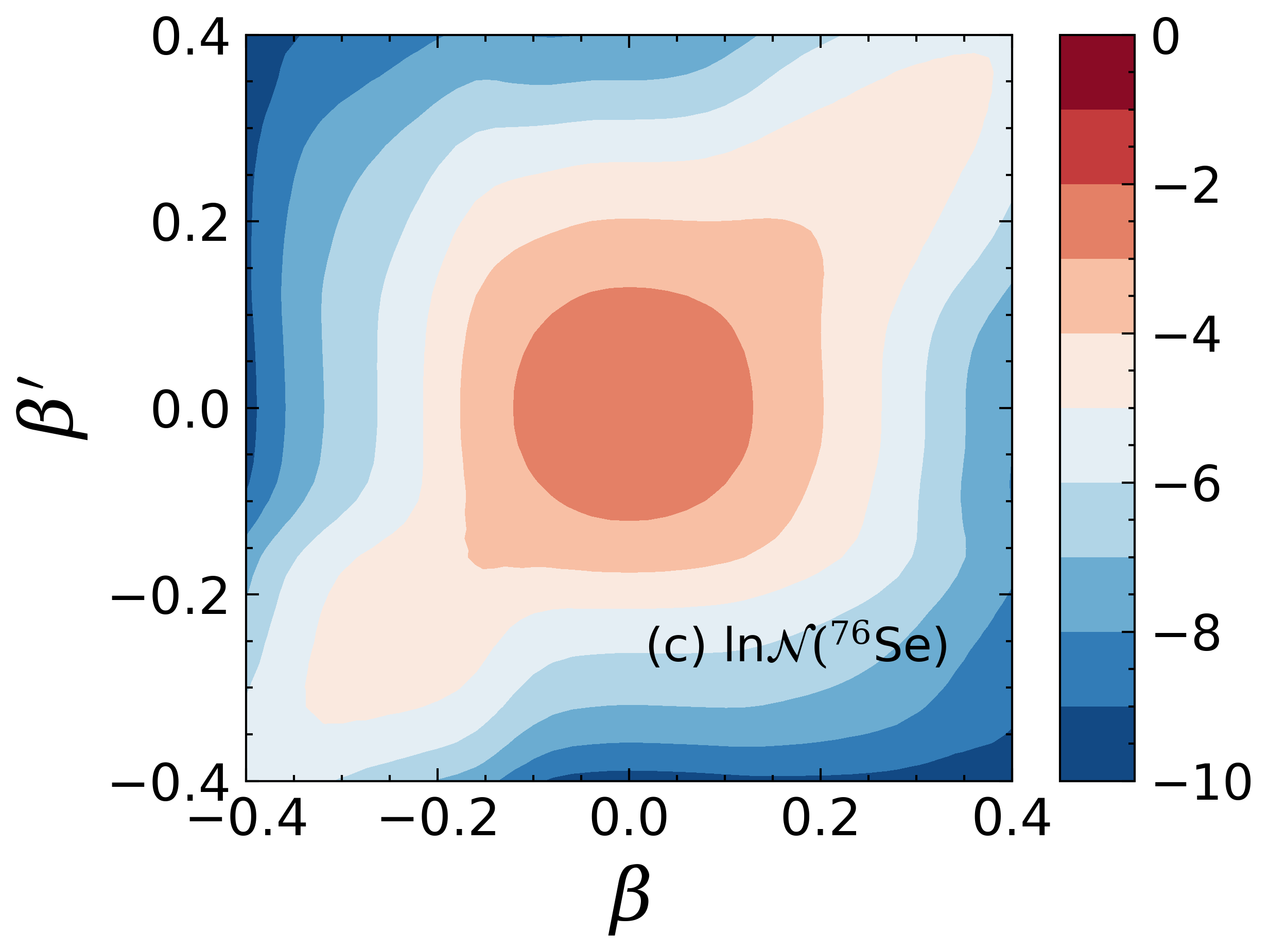} 
		\includegraphics[width=0.48\columnwidth]{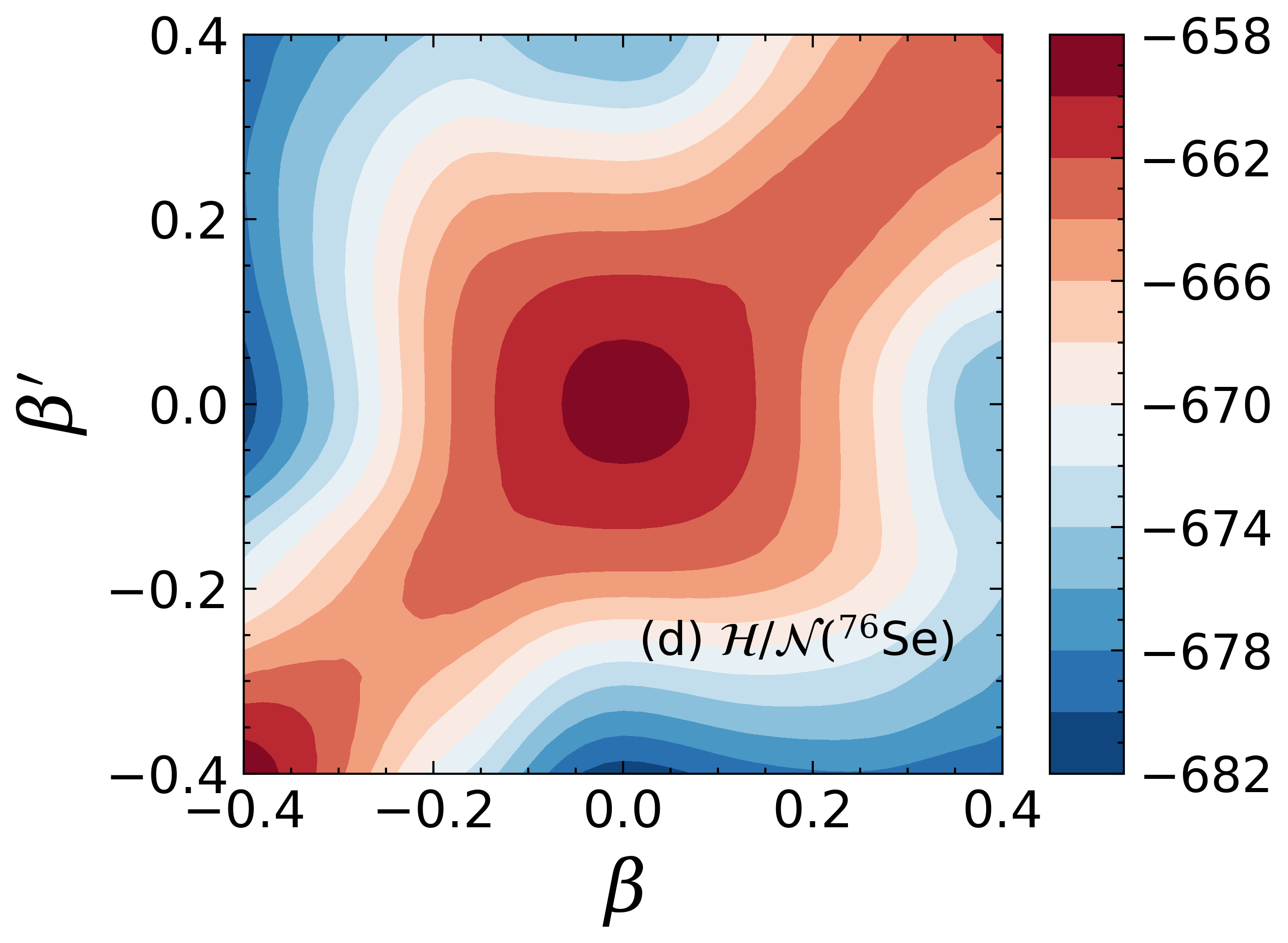} 
	\caption{(Color online) The distributions of norm kernels $\ln({\cal N})$ and the  ratio of kernels ${\cal H}(\beta, \beta^\prime)/{\cal N}(\beta, \beta^\prime)$ with $J=0$ for $^{76}$Ge (a,b) and $^{76}$Se (c,d)  as a function of the quadrupole deformation parameters $(\beta, \beta')$ from the HFB calculation using the Gogny D1S force. 
	}
	\label{fig:D1S_hamiltonian}
\end{figure}
 
In the following, we will study three different procedures for implementing the polynomial RR into GCM calculations, starting from two EDFs, a shell-model interaction, and a chiral NN+3N interaction.
They are illustrated with a flow chart shown in Fig.~\ref{fig:flowchart}. In the GCM+RR, a small portion (about $1/4$) of the norm and Hamiltonian kernels are calculated exactly with quantum number projection (QNP) method, providing inputs for training the RR model parameters. Once the model parameters are optimized, they are used to predict all the norm and Hamiltonian kernels which serve as inputs of the GCM calculation.  In the GCM+RR procedure, the noise introduced by the RR model is expected to generate errors in the final results. To mitigate these errors, we propose a combination of GCM, OC/ENTROP, and RR, in which a subset is selected based on either the OC method or ENTROP algorithm. An exact GCM calculation is carried out within this subset afterward. 

% In the GCM + OC + RR, we use norm and Hamiltonian kernels which are mixed with exact data used for training and predicted data.
 
 The procedure of the GCM+OC+RR method for nuclear 
 low-lying states is as follows: 
\begin{enumerate}
    \item [($\textrm{i}$)] All the configurations $\ket{\Phi(\beta)}$ are sorted by their projected energies, i.e., the ratios of the kernels ${\cal H}^J_{00}(\beta, \beta)/{\cal N}^J_{00}(\beta, \beta)$ for the nucleus of interest. To this end, one needs to calculate these ratios for all the configurations, which requires ${\cal O}(N_q)$ computational effort. We note that the ordering of the configurations is different for different angular momenta $J$.
    
    \item [($\textrm{ii}$)] Starting from the configuration with the lowest energy and stepping from low to high energy,  the $(n+1)$-th configuration $\ket{n+1}$ is  added into the subspace if its projection onto the subspace spanned by the already selected $n$ configurations, defined by
    \begin{align} 
    L(n,n+1) 
    &=\frac{\left\langle n+1\left|P^{(n)}\right| n+1\right\rangle}{\langle n+1 \mid n+1\rangle}
    \notag\\
    &=\frac{\boldsymbol{\gamma}^{(n) \dagger}\left(\boldsymbol{S}^{(n)}\right)^{-1} \boldsymbol{\gamma}^{(n)}}{\langle n+1 \mid n+1\rangle}\,,
    \label{eq:cutoff}
    \end{align}
    is smaller than a pre-selected cutoff parameter $L_{c}$ \cite{Romero:2021PRC}. This implies that the new configuration is approximately orthogonal to the previous configurations, hence the name of this stage. In Eq. \eqref{eq:cutoff}, $S_{i j}^{(n)}=\langle i \mid j\rangle$ and $\gamma_{i}^{(n)}=\langle i \mid n+1\rangle$ are nothing but the matrix elements of the norm kernel ${\cal N}^J_{00}(\beta, \beta')$.  
    Using the orthogonality criterion, a subspace ${\cal S}_{L_c}=\left\{\ket{1},\ket{2},\dots, \ket{n}\right\}_{L_c}$ is eventually determined for a given value of  $L_{c}$. In practice, we employ the pre-calculated norm kernels by the QNP method with the rest given by the RR model in the subspace-selection procedure.

    \item [($\textrm{iii}$)] The norm kernels and Hamiltonian kernels for the configurations within the subspace ${\cal S}_{L_c}$ are determined by the QNP method and they are used in the solution of the HWG equation (\ref{eq:HWG}). The convergence of each observable against the cutoff parameter $L_c$ is examined. 
\end{enumerate}

In the calculation of the NME of $0\nu\beta\beta$ decay, for comparison, we also employ the recently developed ENTROP algorithm to select the subspaces for the initial and the final nuclei simultaneously. Details about the ENTROP algorithm can be found in Ref.~\cite{Romero:2021PRC}.

\subsection{EDF-based GCM calculations}
\subsubsection{A non-relativistic EDF: Gogny D1S}

\begin{figure}[] 
		\includegraphics[width=0.48\columnwidth]{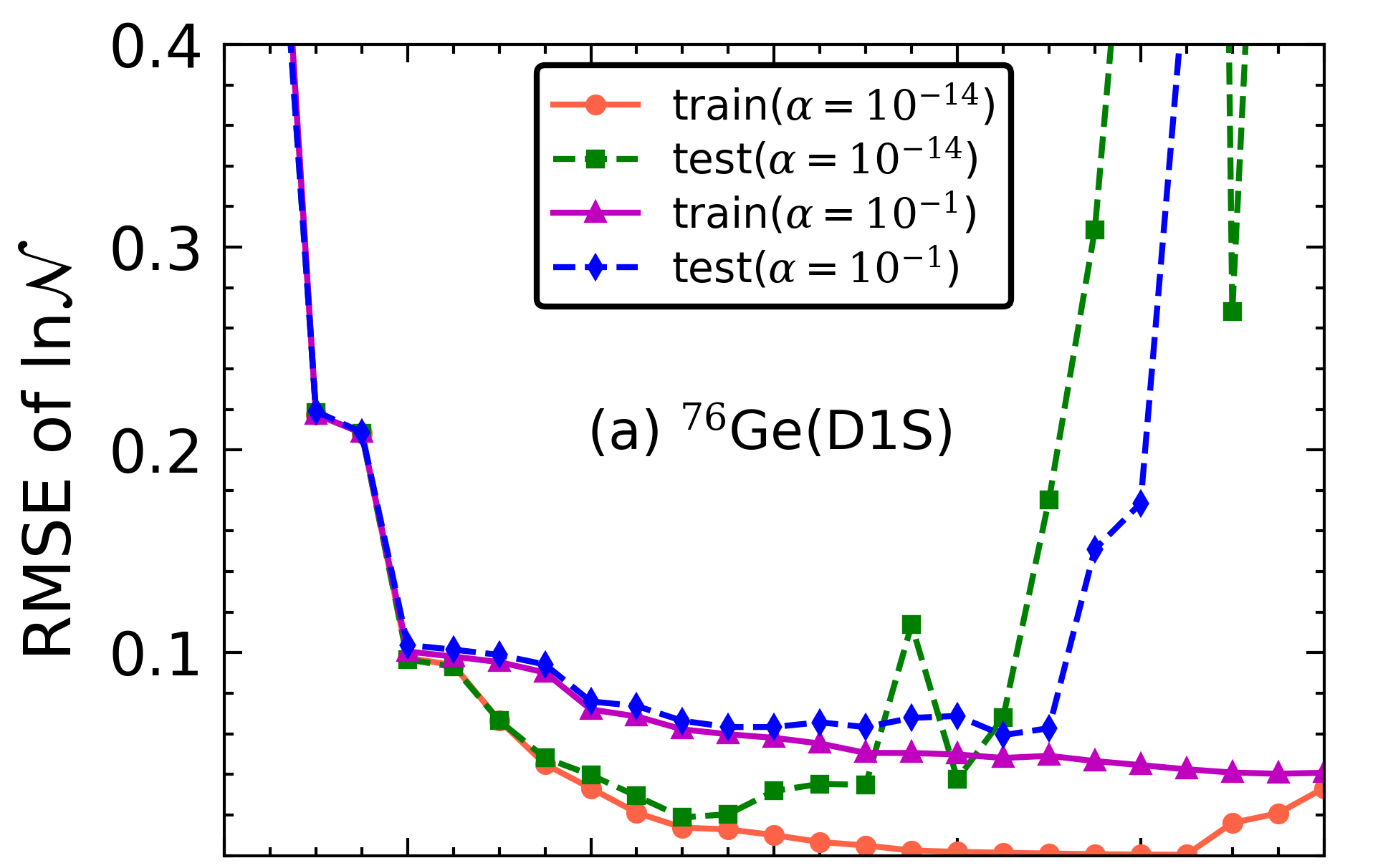}
		\includegraphics[width=0.48\columnwidth]{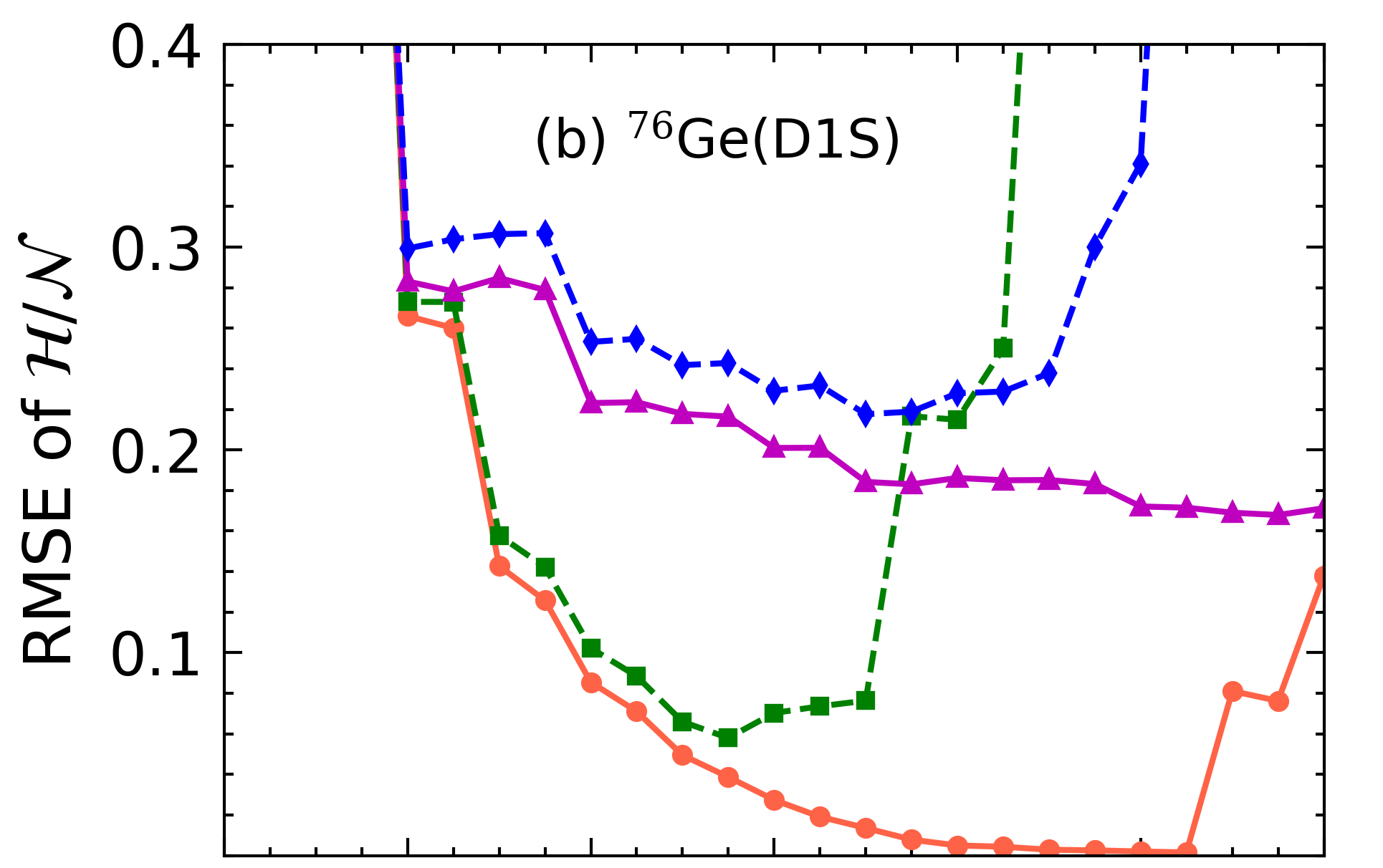}
		\includegraphics[width=0.48\columnwidth]{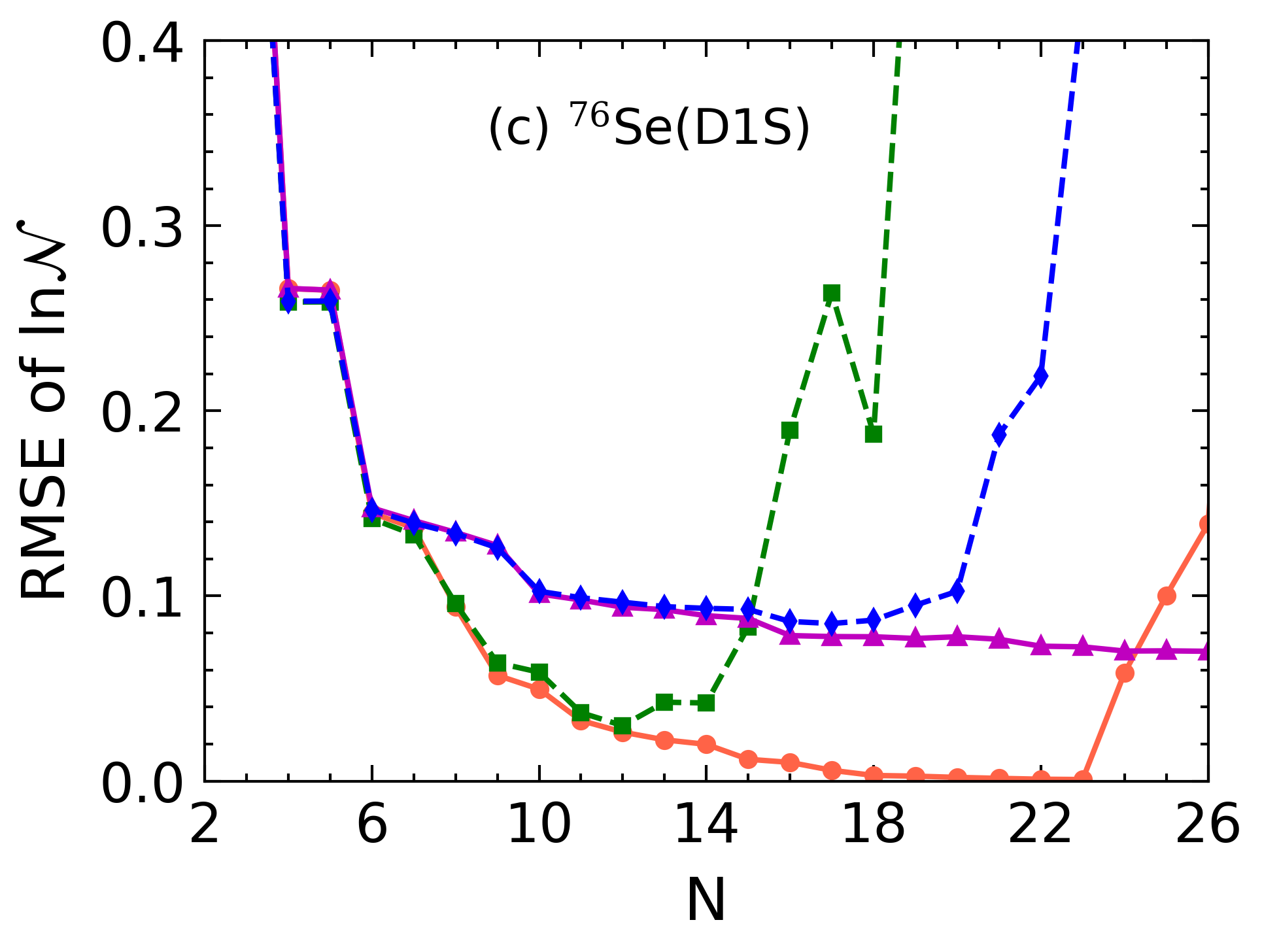} 
		\includegraphics[width=0.48\columnwidth]{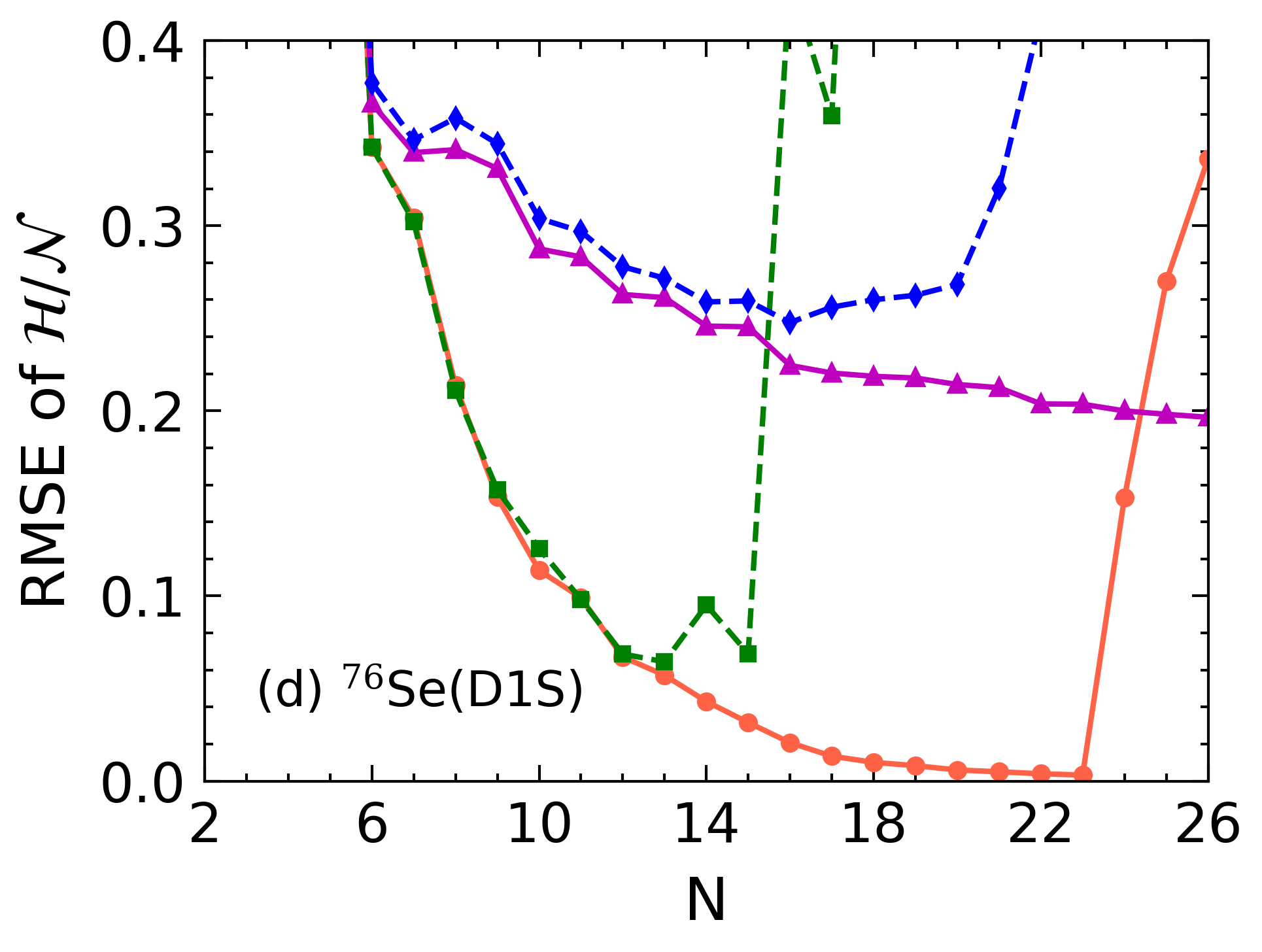}  
	\caption{(Color online) The root-mean-square error (RMSE) of the RR models for  the training set (red filled circles) and test set (green filled squares) of the norm kernels $\ln{\cal N}$ (a,c) and the  ratio of kernels ${\cal H}(\beta, \beta^\prime)/{\cal N}(\beta, \beta^\prime)$ (b,d) with $J=0$ for $\nuclide[76]{Ge}$ (a,b) and $\nuclide[76]{Se}$ (c,d) as a function of the degree parameter $N$ of the polynomials,  where the results with the ridge parameter $\alpha$ chosen as $10^{-14}$ and $10^{-1}$ respectively are given for comparison. }
	\label{fig:D1S_ML_RMS}
\end{figure}

      \begin{figure}[] 
     \includegraphics[width=\columnwidth]{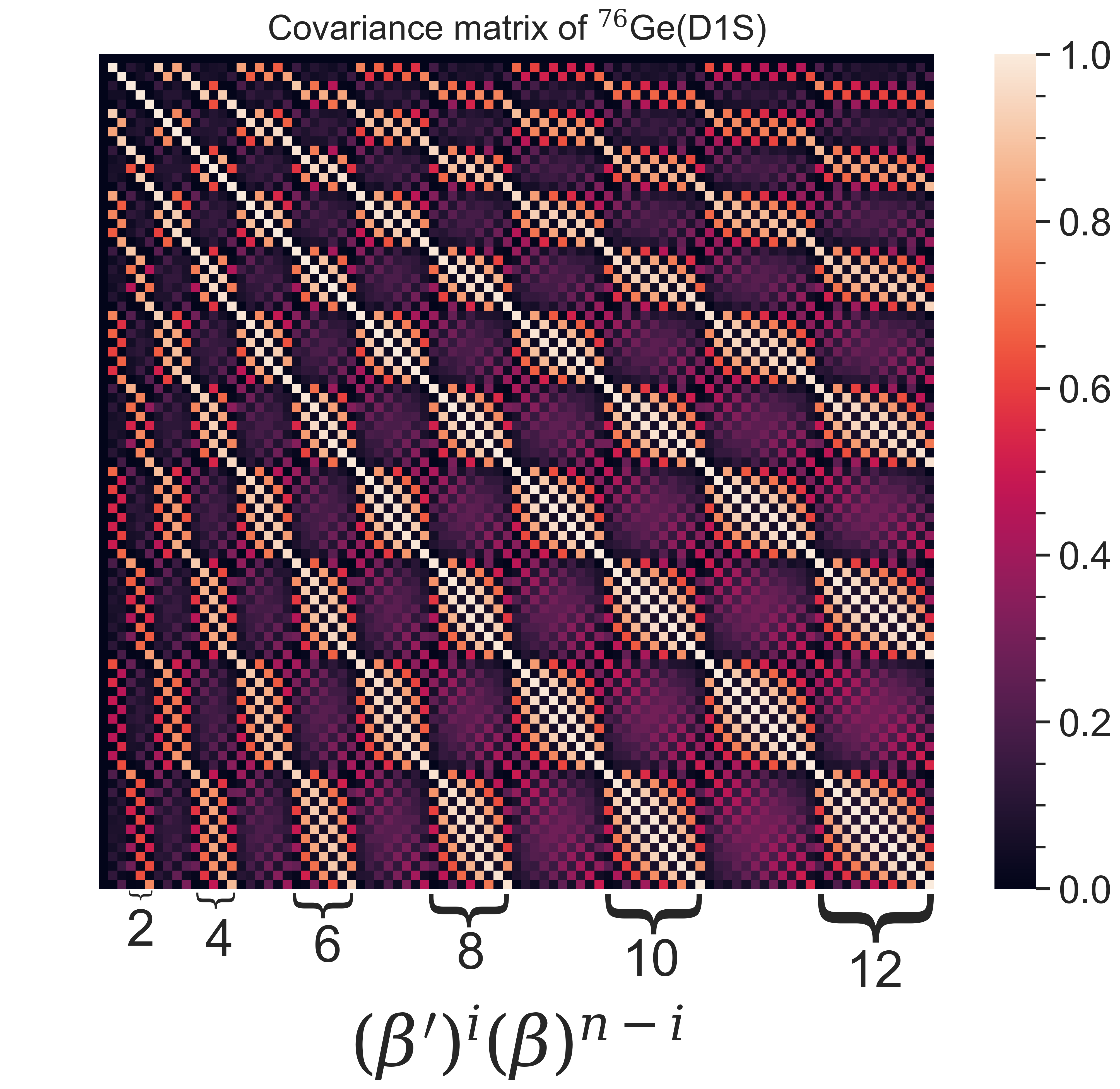}  
	\caption{(Color online)  The covariance matrix of  the RR model (\ref{eq:loss}) for  the norm kernels of $\nuclide[76]{Ge}$ with $J=0$ by the Gogny D1S force, where the degree parameter $N=12$ and ridge parameter $\alpha=10^{-14}$.  The number of features $(\beta^\prime)^i(\beta)^{n-i}$ in (\ref{eq:PolynomialFeatures}) is $(N+2)(N+1)/2$, where the integer number $n\in[0, N]$ is shown in the bottom of the figure with the integer number $i$ varying  from 0 to $n$. }
	\label{fig:D1S_Covariance_Matrix}
\end{figure}

\begin{figure}[]
                 \includegraphics[width=0.48\columnwidth]{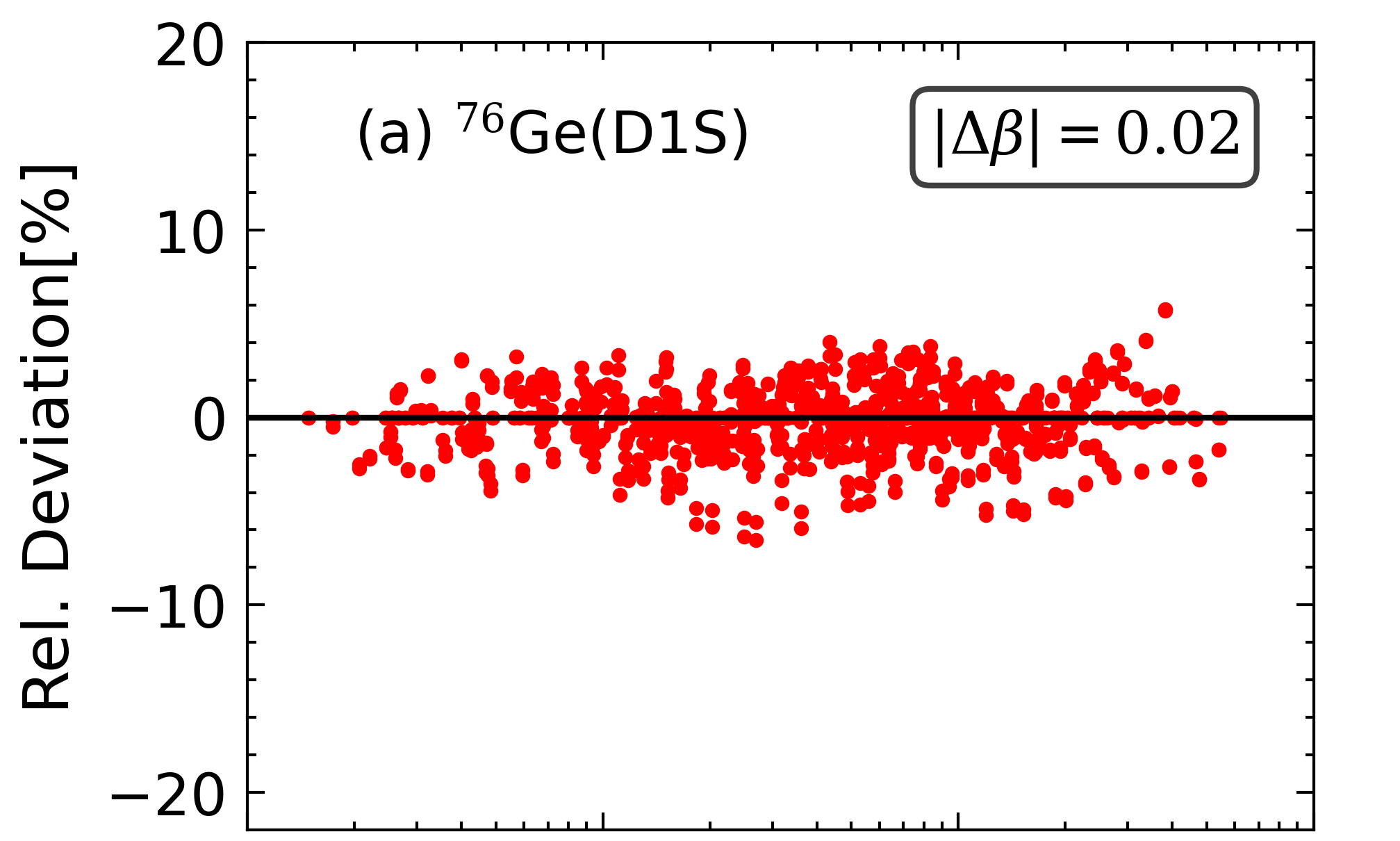}
		\includegraphics[width=0.48\columnwidth]{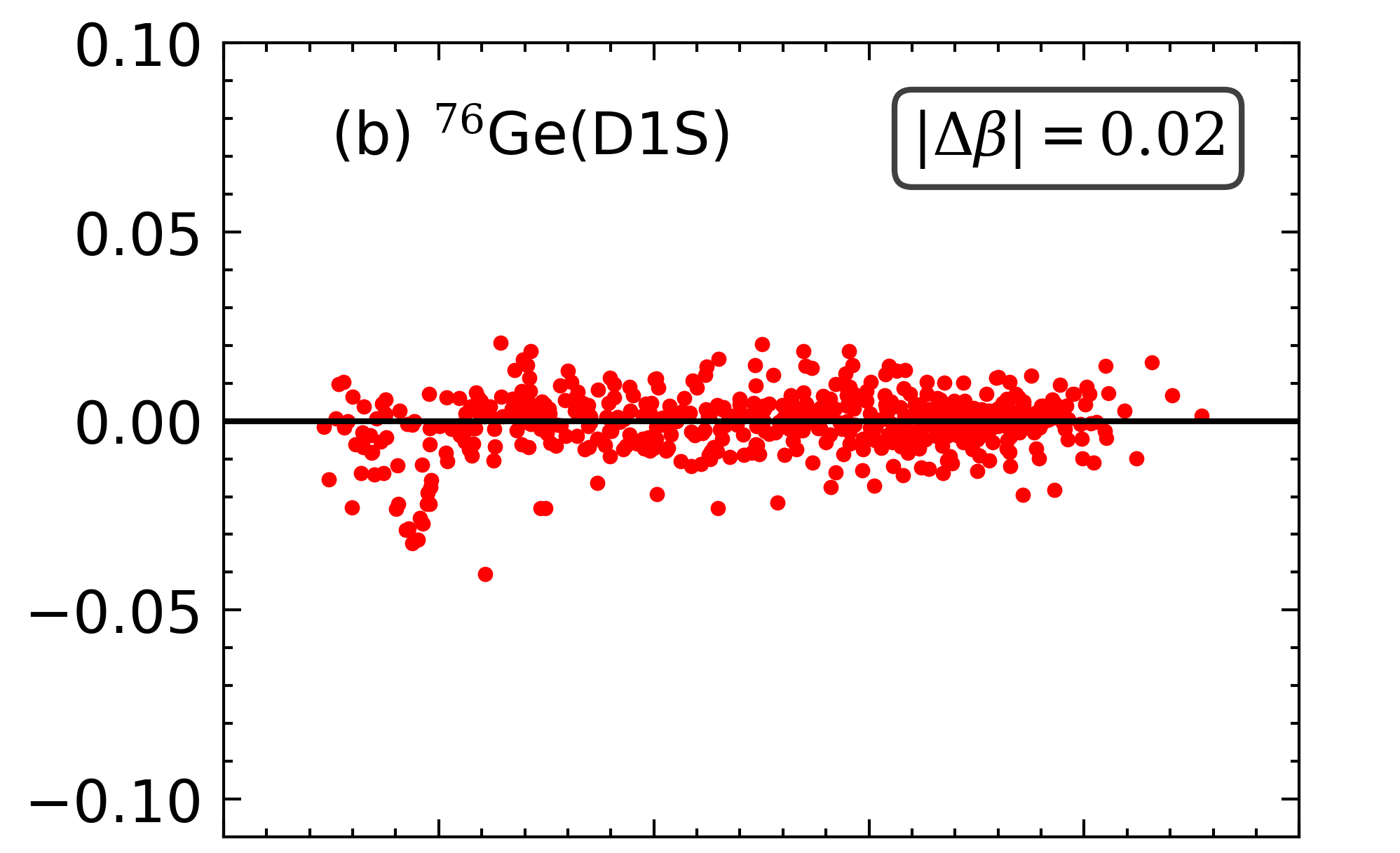}
                 \includegraphics[width=0.48\columnwidth]{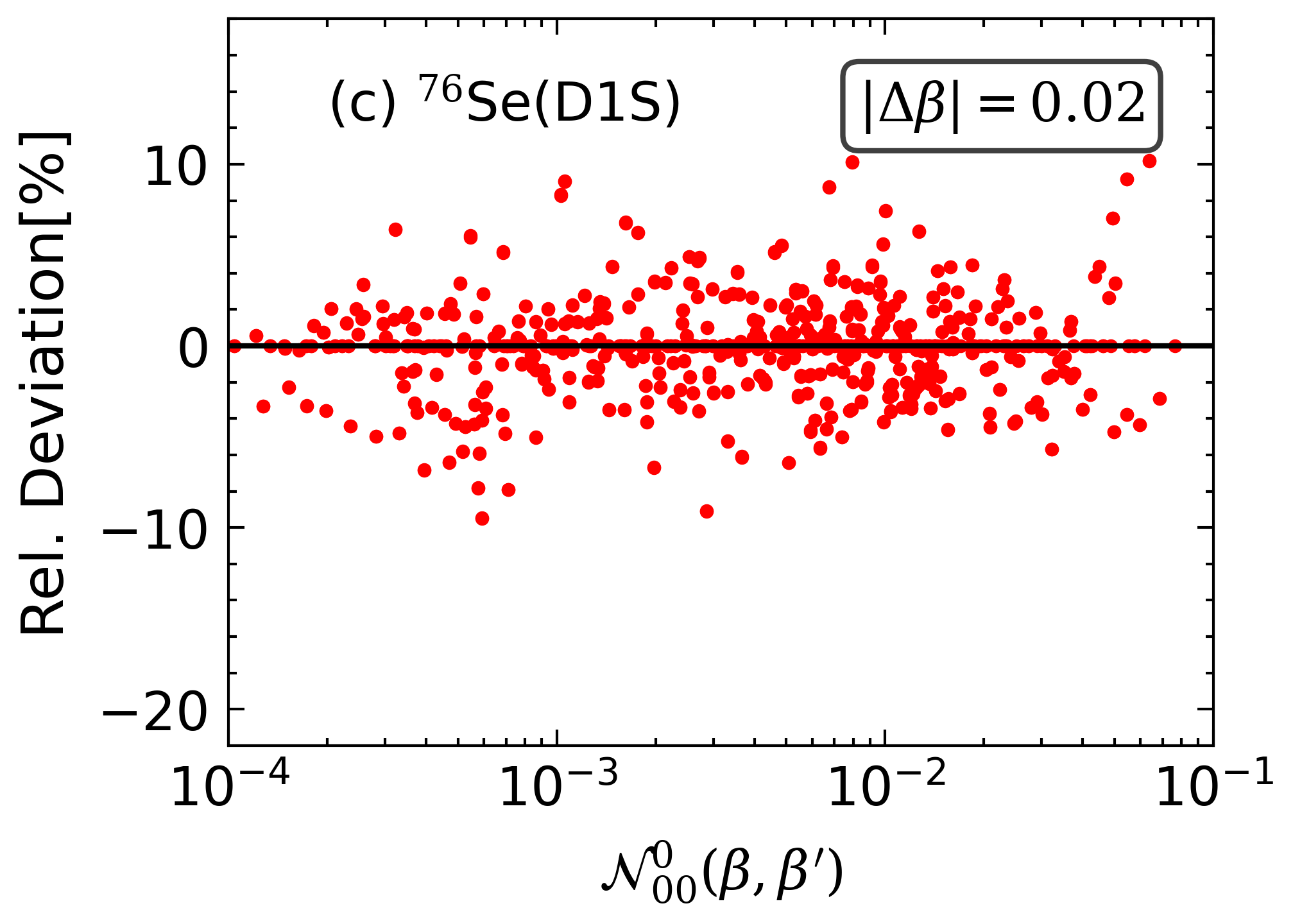}
		\includegraphics[width=0.48\columnwidth]{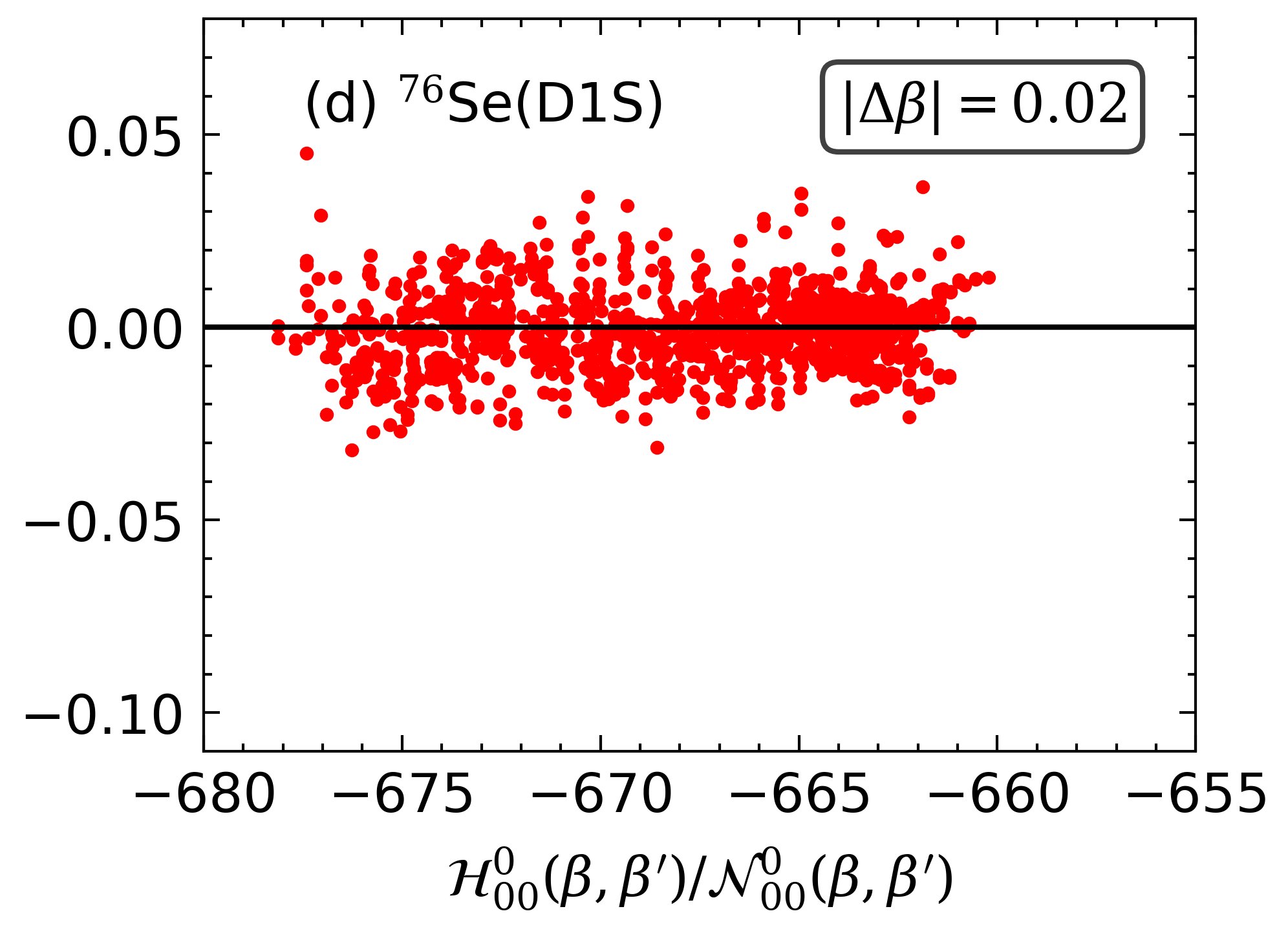}
		\caption{(Color online) The relative deviation $\delta$ of the kernels predicted by the optimal RR model for (a,b) $\nuclide[76]{Ge}$  and (c,d) $\nuclide[76]{Se}$, where the $\delta$ is defined as $\delta^{(i)}=(\hat y^{(i)}-y^{(i)})/y^{(i)}$ with $y$ being the  (a,c) norm kernel ${\cal N}$ or (b, d) the ratio of kernels ${\cal H}/{\cal N}$, respectively. See  main text for details.}
	\label{fig:D1S_deviations}
\end{figure}

Figures~\ref{fig:D1S_norm} and \ref{fig:D1S_hamiltonian} show the distributions of the norm kernels ${\cal N}^J(\beta, \beta^{\prime})$ and the ratio  of kernels ${\cal H}^J(\beta, \beta^{\prime})/{\cal N}^J(\beta, \beta^{\prime})$ with different angular momentum $J$ from quantum-number-projection calculations for the  HFB states based on the Gogny D1S force \cite{Decharge:1980zm,Berger:1991by}. The norm kernels with different angular momenta $J$ are distributed differently, but they share a common feature in that they are mainly concentrated along the diagonal line with $\beta=\beta'$.
For the $J=0$ case, the norm kernel is dominated by a product of two Gaussian functions centered at $\beta=\beta'=0$. It can be understood that the spherical state only contains a $J=0$ component. 
The distributions of the diagonal element ${\cal N}^{J}_{00}(\beta, \beta)$ with $J\neq0$ share the similar feature
that the peak locates at a deformed state, approximately symmetric with respect to $\beta=0$. This is a general feature of norm kernels~\cite{Yao:2022HBNP}.
 One may expect that this feature can be well captured by the polynomial regression on their logarithmic values. Nevertheless, as shown in Fig.~\ref{fig:D1S_hamiltonian}(a) and (c), the matrix elements of norm kernels vary by several orders of magnitude in the entire deformation space, which is a challenge for ML algorithms. A small error in the norm kernel may degrade the description of GCM, as discussed in detail later on. In contrast, the variation of the ratios ${\cal H}^{J=0}_{00}(\beta, \beta^\prime)/{\cal N}^{J=0}_{00}(\beta, \beta^\prime)$ with $\beta$ and $\beta'$ is moderate and thus expected to be more readily learned by ML algorithms.

\begin{figure}[]
\includegraphics[width=0.48\columnwidth]{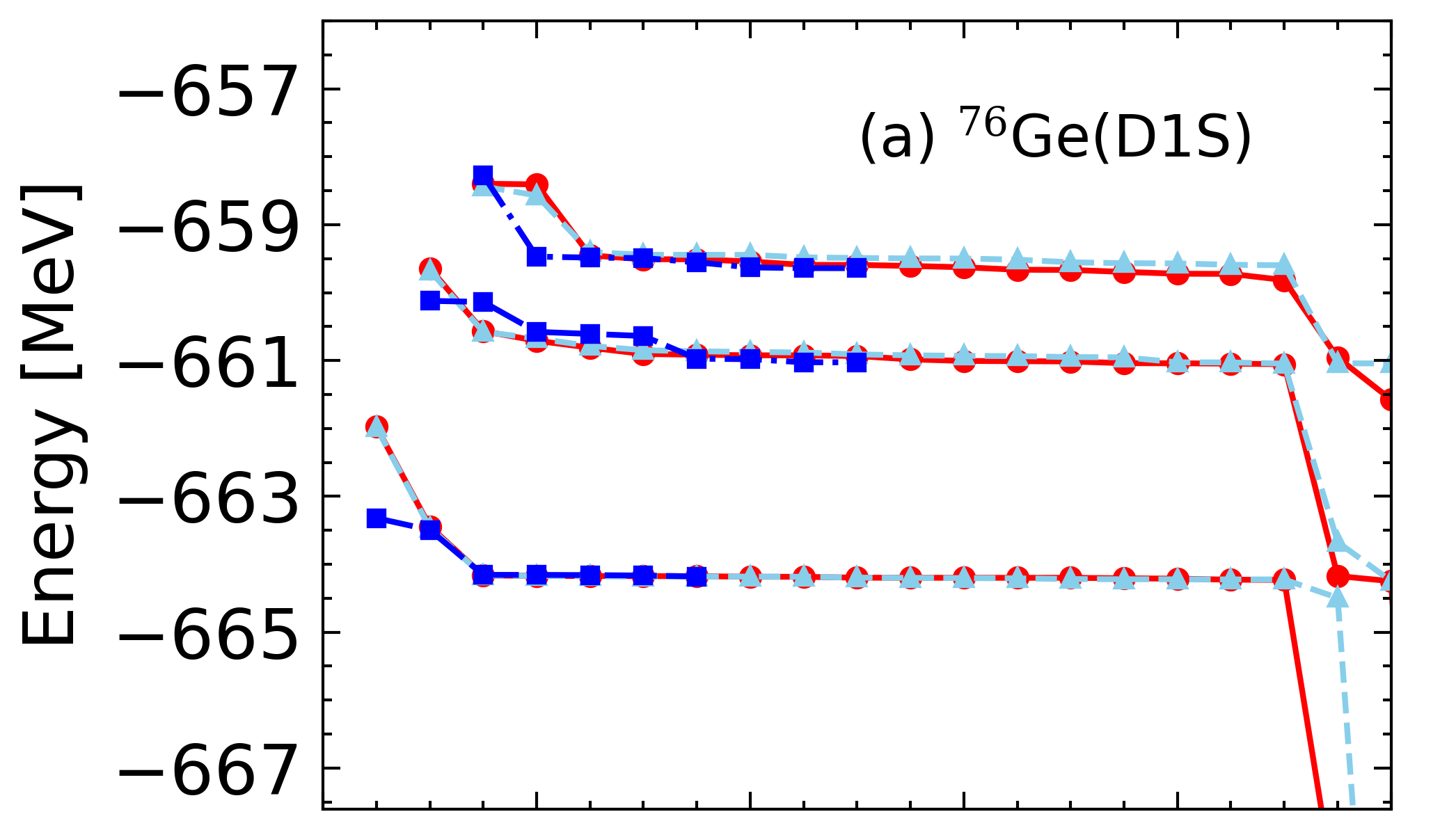}
\includegraphics[width=0.48\columnwidth]{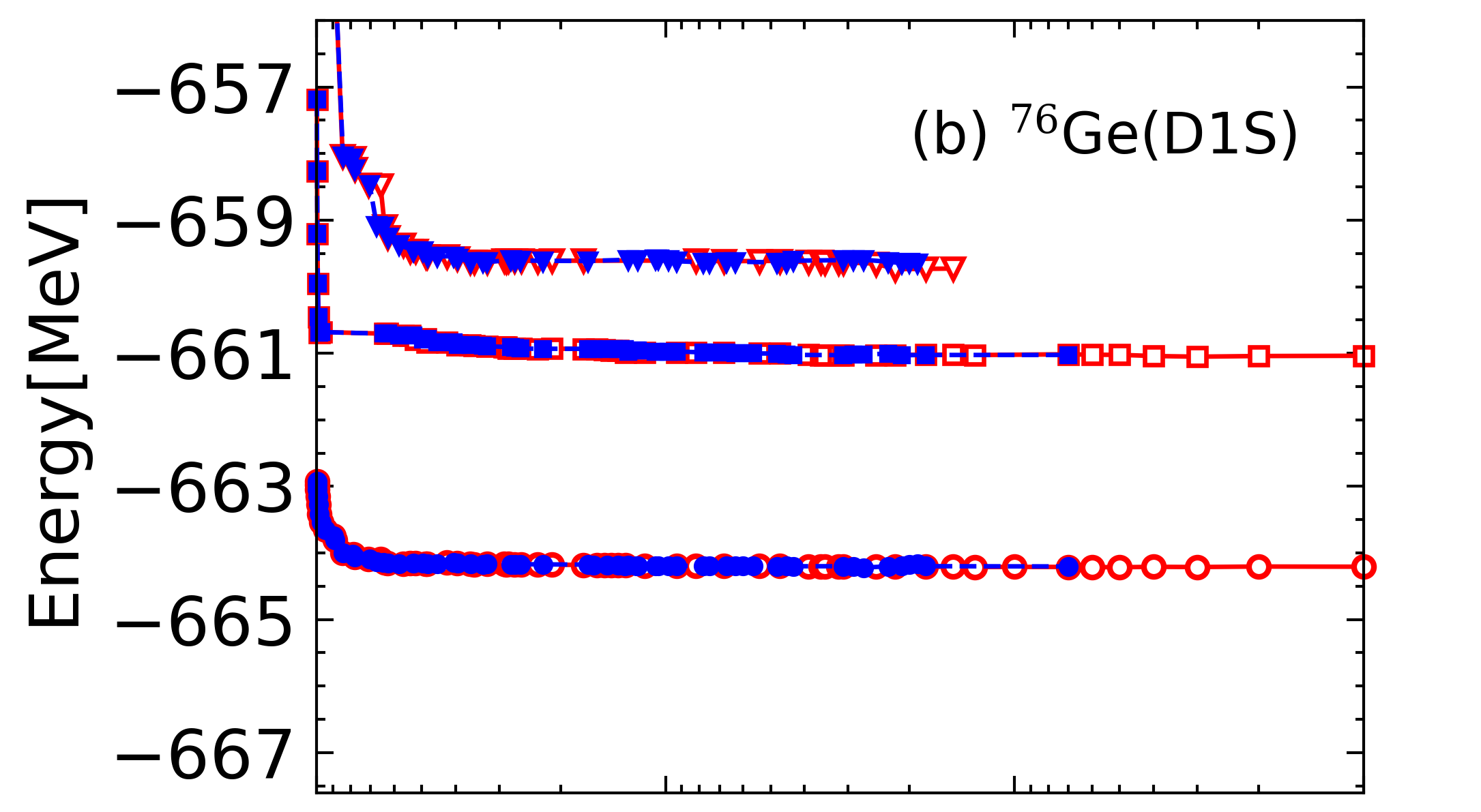}
\includegraphics[width=0.48\columnwidth]{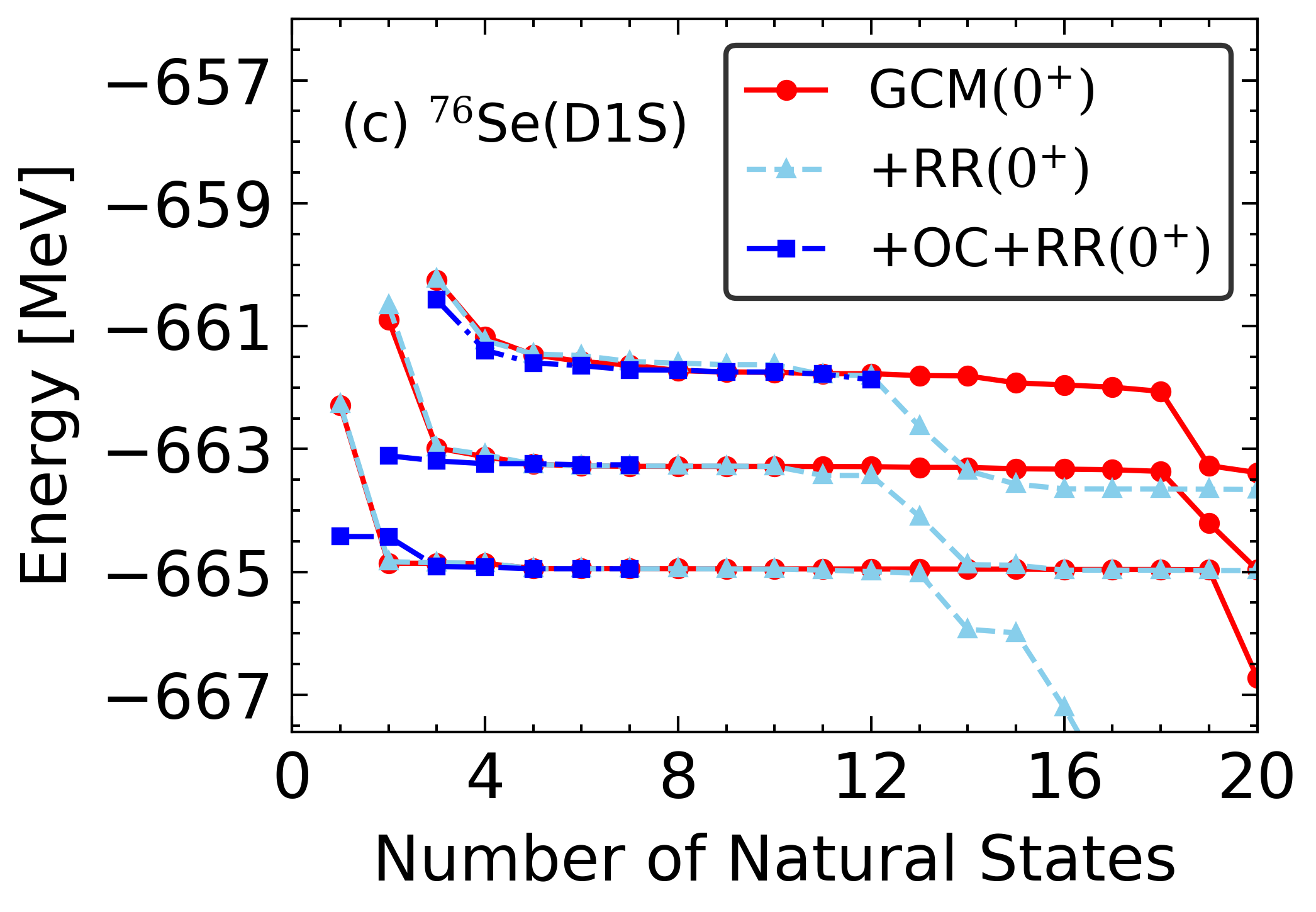}
\includegraphics[width=0.48\columnwidth]{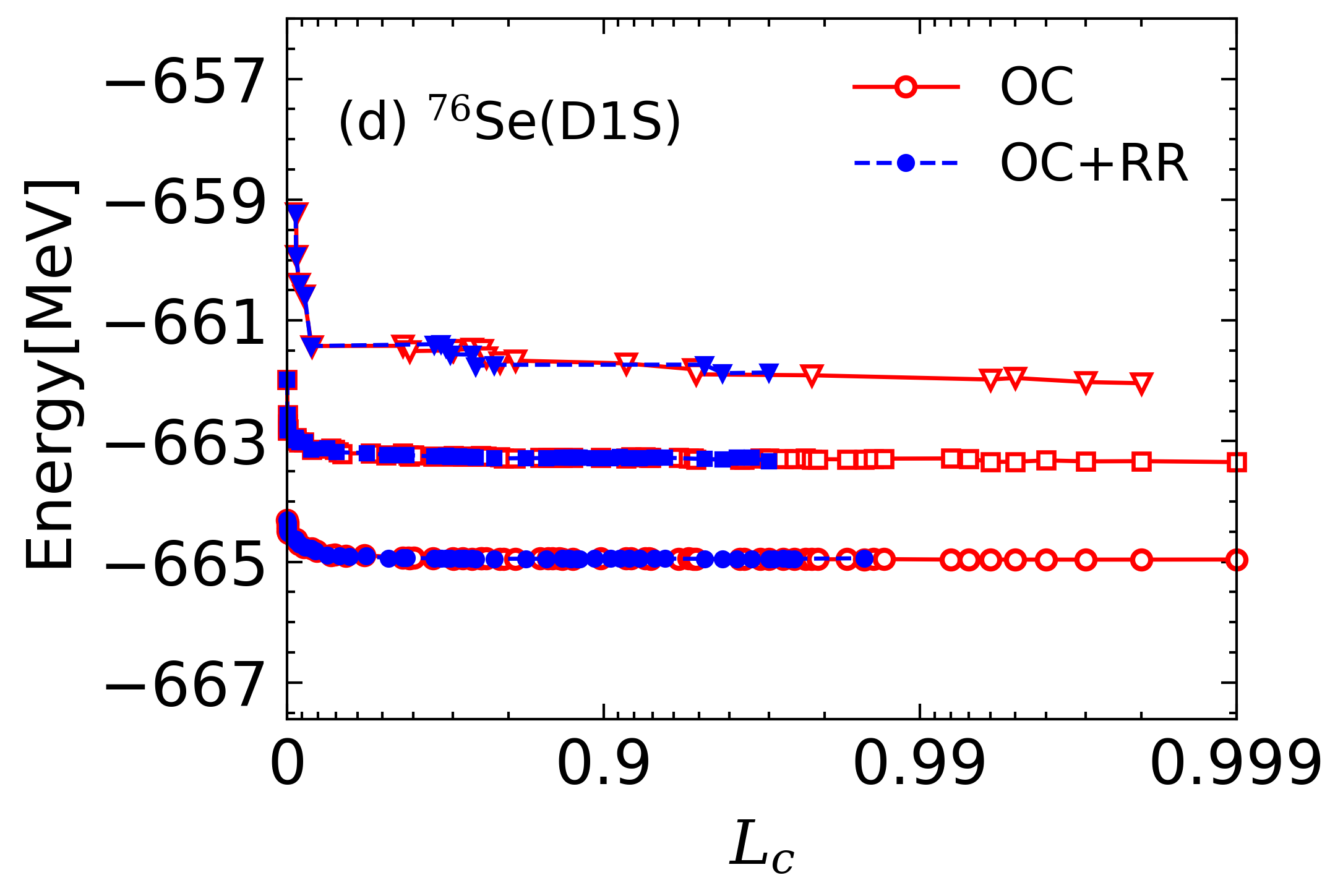}
 \caption{(Color online)  The energies of states with angular momentum $J=0$ as a function of (a,c) the number of natural states and (b,d) the cutoff parameter $L_{c}$ from different calculations for (a,b) $^{76}$Ge and (c,d) $^{76}$Se respectively.  See main text for details.}
	\label{fig:D1S_renormalized_energy_platform}
\end{figure}

\begin{figure}[]
 \includegraphics[width=\columnwidth]{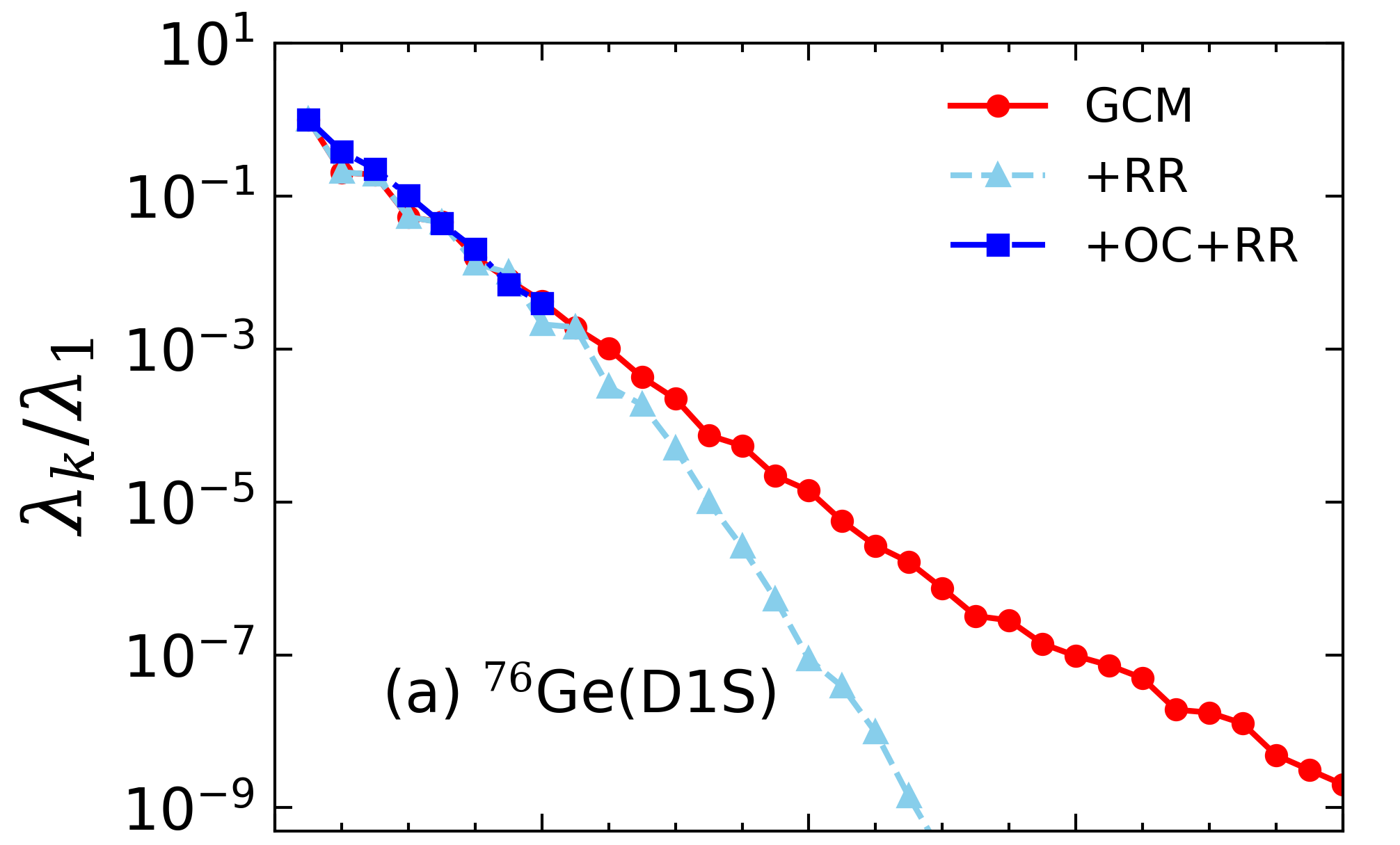}
 \includegraphics[width=\columnwidth]{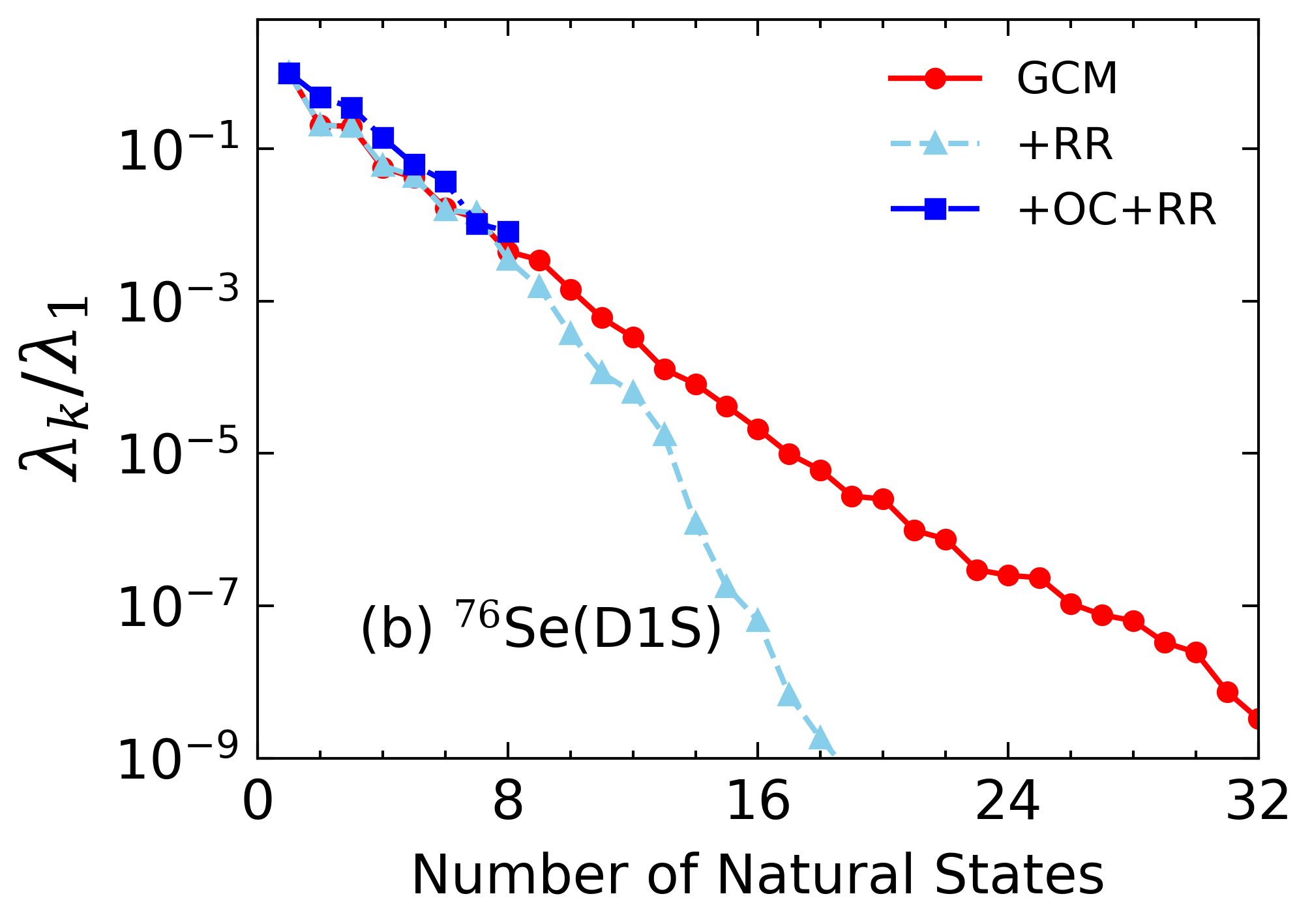}
 \caption{(Color online) The distribution of eigenvalues $\lambda_k$ (normalized to the maximal one) of the norm kernels ${\cal N}^{J=0}_{00}(\beta, \beta^\prime)$ in the full GCM, GCM+RR and GCM+OC+RR calculations for $^{76}$Ge (a) and  $^{76}$Se (b), where different subspaces are used in the three calculations. }
 \label{fig:norm_kernels_eigenvalues}
\end{figure}

\begin{figure}[]
 \includegraphics[width=0.48\columnwidth]{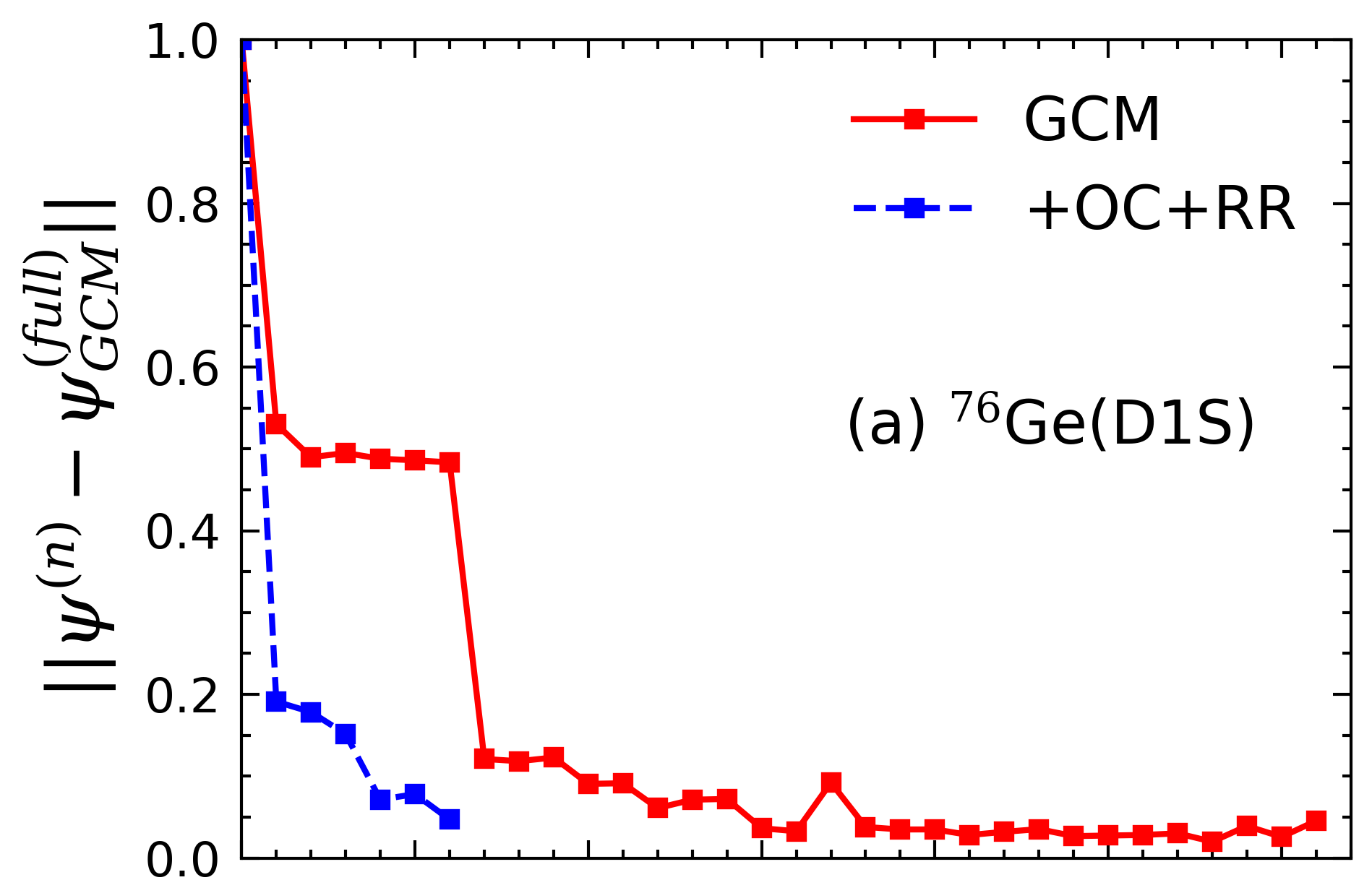}
 \includegraphics[width=0.48\columnwidth]{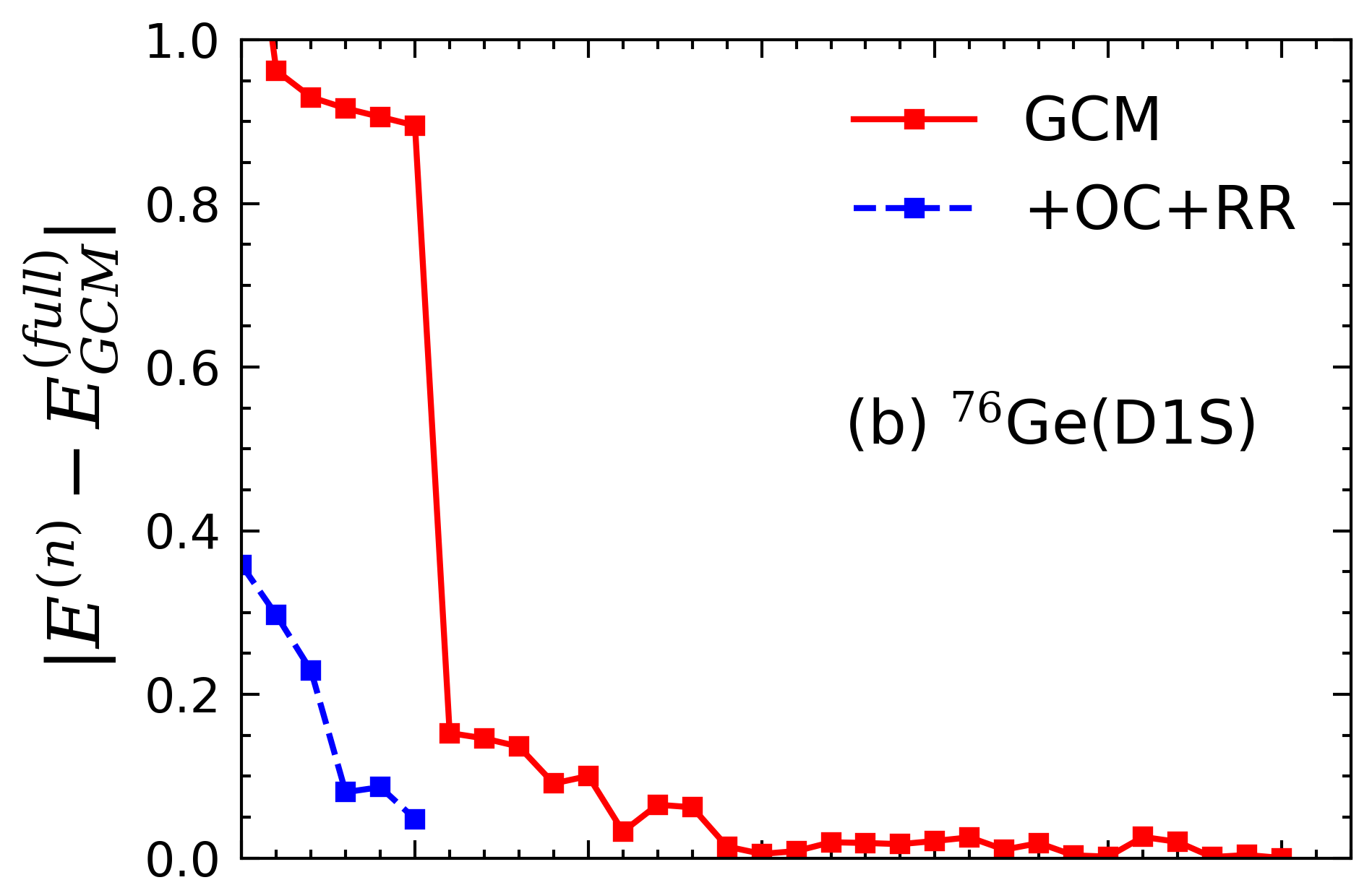}
 \includegraphics[width=0.48\columnwidth]{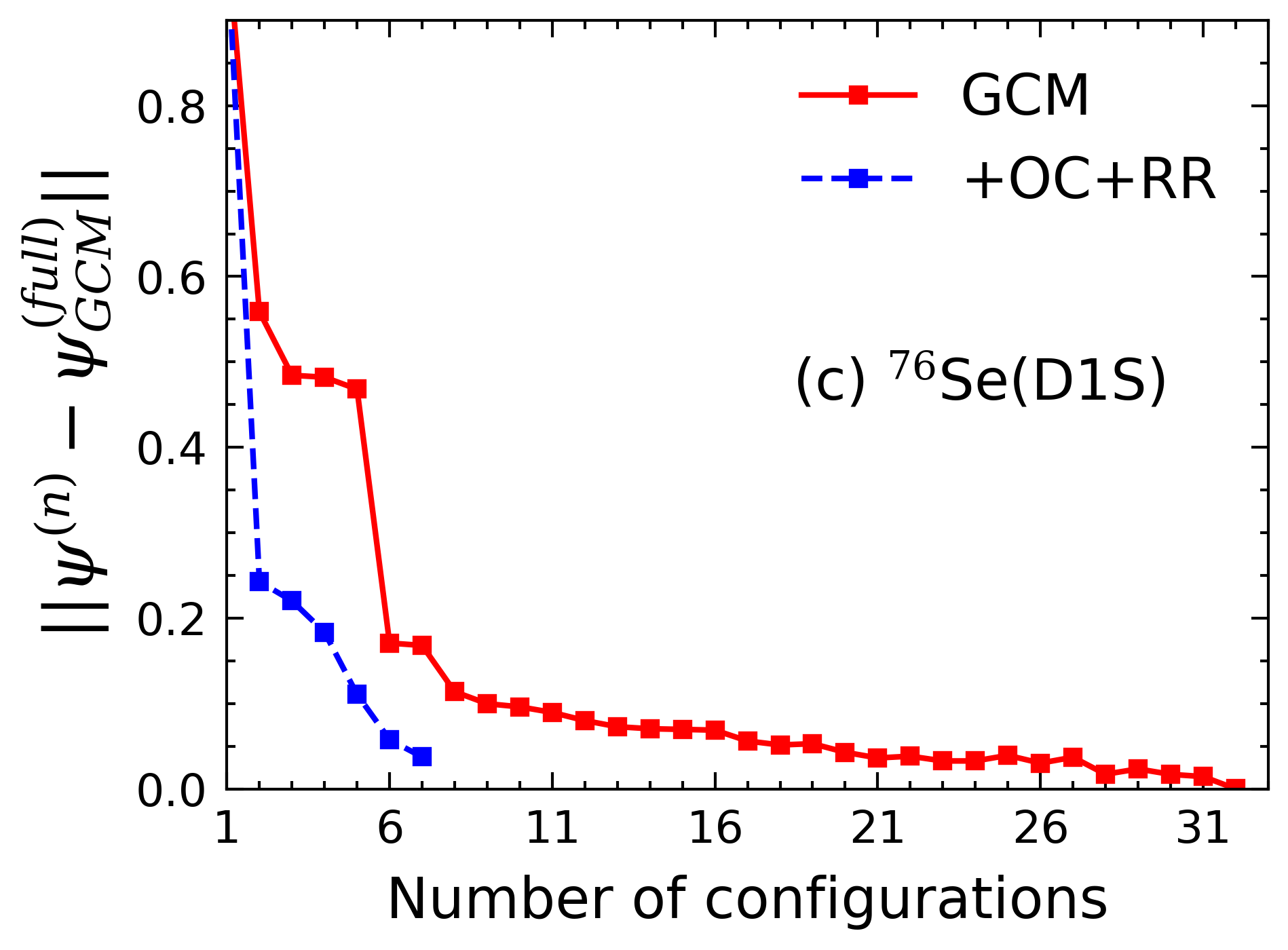}
 \includegraphics[width=0.48\columnwidth]{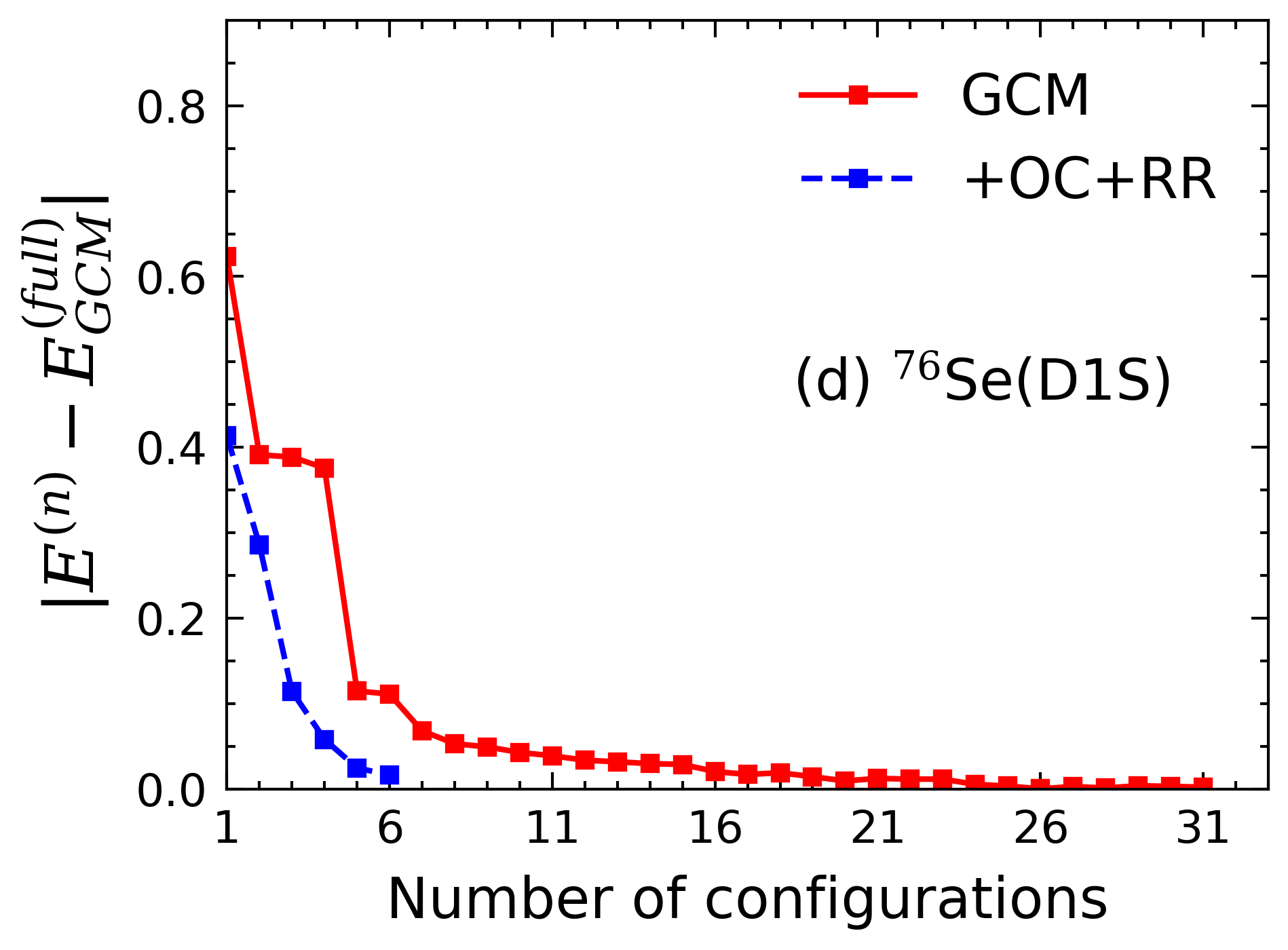}
 \caption{(Color online) Comparison of the convergence of the ground-state wave function (a, c) and energy (b, d) with respect to that by the full GCM calculation as a function of the number of HFB configurations in both GCM and GCM+OC+RR calculations for (a,b) \nuclide[76]{Ge} and (c,d) \nuclide[76]{Se}. In the GCM+OC+RR calculation, the $L_{c}$ value is chosen as 0.872 and 0.824 for (a,b) \nuclide[76]{Ge} and (c,d) \nuclide[76]{Se}, respectively.  
}
 \label{fig:D1S_wfs_conv}
\end{figure}

The parameters of our polynomial RR models for norm kernels ${\cal N}$ and the ratios ${\cal H}/{\cal N}$ are optimized as explained in Sec.~\ref{framework:ML}.   Our findings in the training processes are as follows:
\begin{itemize}
    \item The optimal RR model captures the Hermiticity of the kernels, even though it is not strictly enforced in the model's construction at present. 
    
    \item Selecting training data that are uniformly distributed in the $(\beta,\beta')$ plane usually leads to a smaller MSE in (\ref{eq:loss}) than random sampling. We note that this approach does not scale well to larger numbers of generator coordinates, for which  a more robust sampling strategy needs to be explored in future.
    
    \item As shown in Figs.~\ref{fig:D1S_norm} and \ref{fig:D1S_hamiltonian}, the norm kernels with $J=0$ of strongly deformed configurations ($|\beta|>0.4$) are usually small (less than $10^{-3}$). Including these kernels in the training procedure may spoil the description of the model. For kernels with $J\neq0$, we exclude configurations around the spherical shape with $|\beta|<0.06$ to guarantee good performance of the model because they are negligible.
     
\end{itemize}

\begin{figure}[]
		\includegraphics[width=\columnwidth]{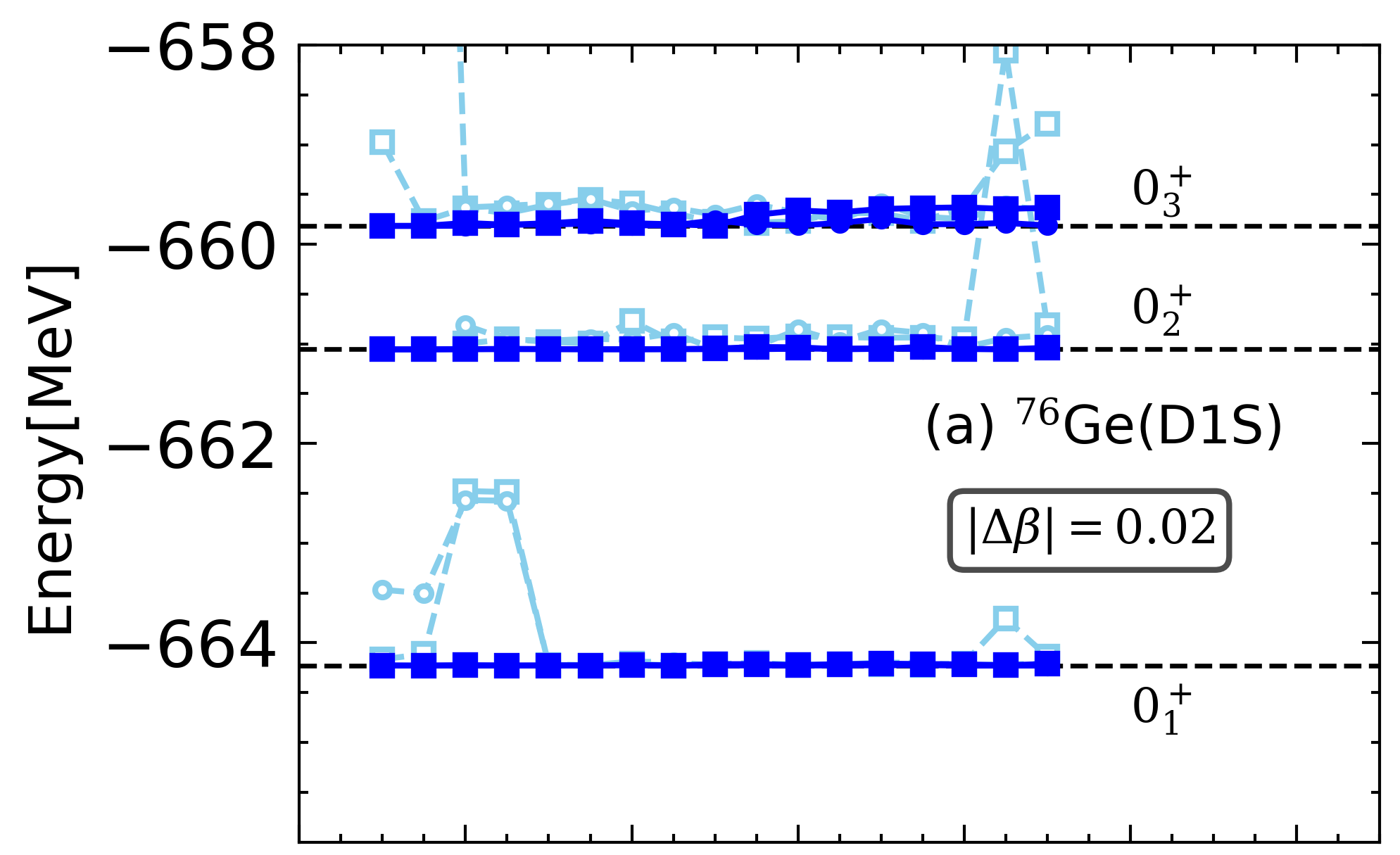}
		\includegraphics[width=\columnwidth]{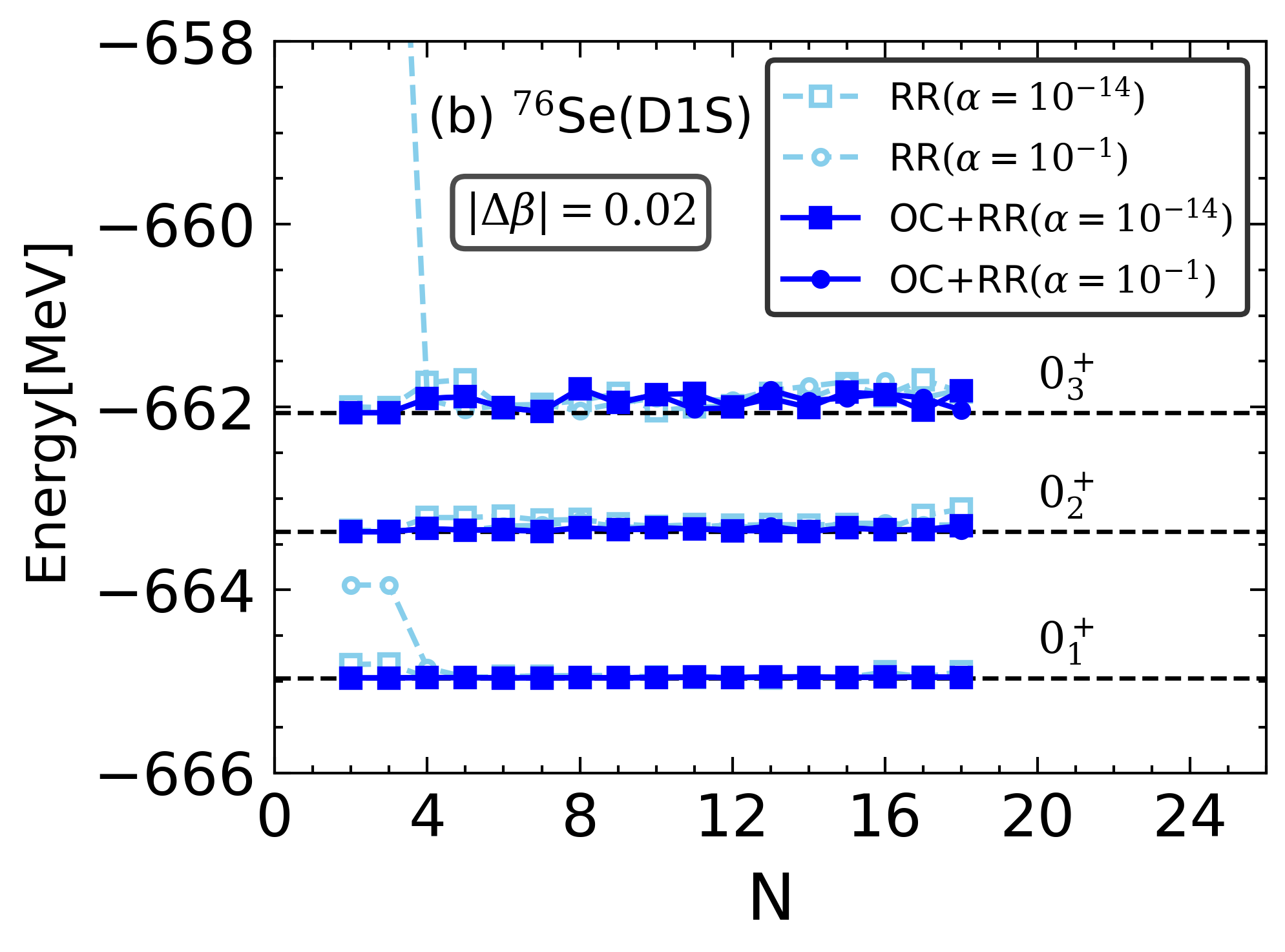}
	\caption{(Color online) The energies of the first three $0^{+}$ states for $^{76}$Ge (a) and $^{76}$Se (b) as a function of the degree parameter $N$ of the polynomials from the GCM+RR and GCM+OC+RR calculations using the D1S force. The black dashed lines indicate the energies from the original GCM calculations using the exactly calculated kernels. }
	\label{fig:D1S_stability_degree}
\end{figure}

\begin{figure}[]
		\includegraphics[width=\columnwidth]{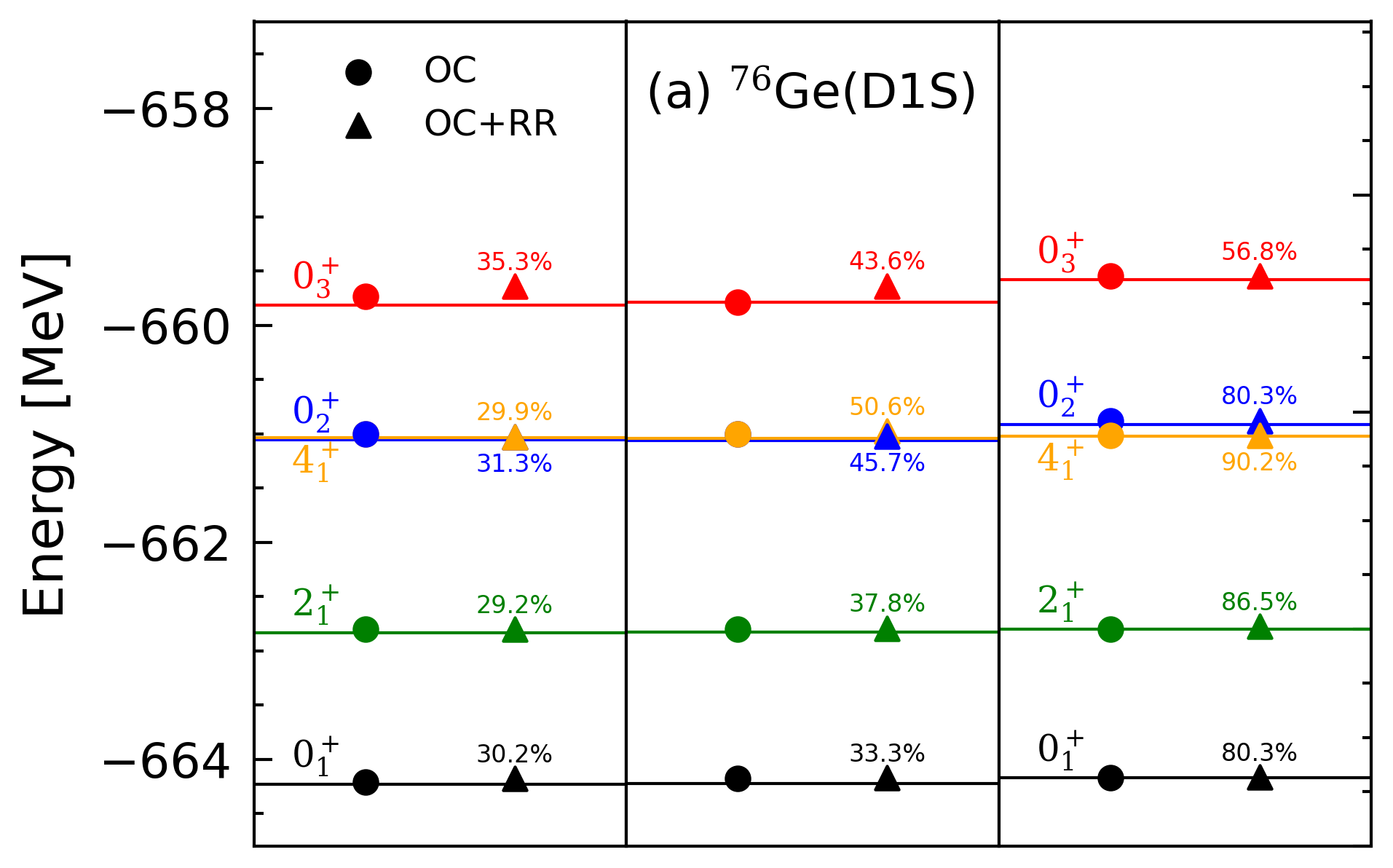}
		\includegraphics[width=\columnwidth]{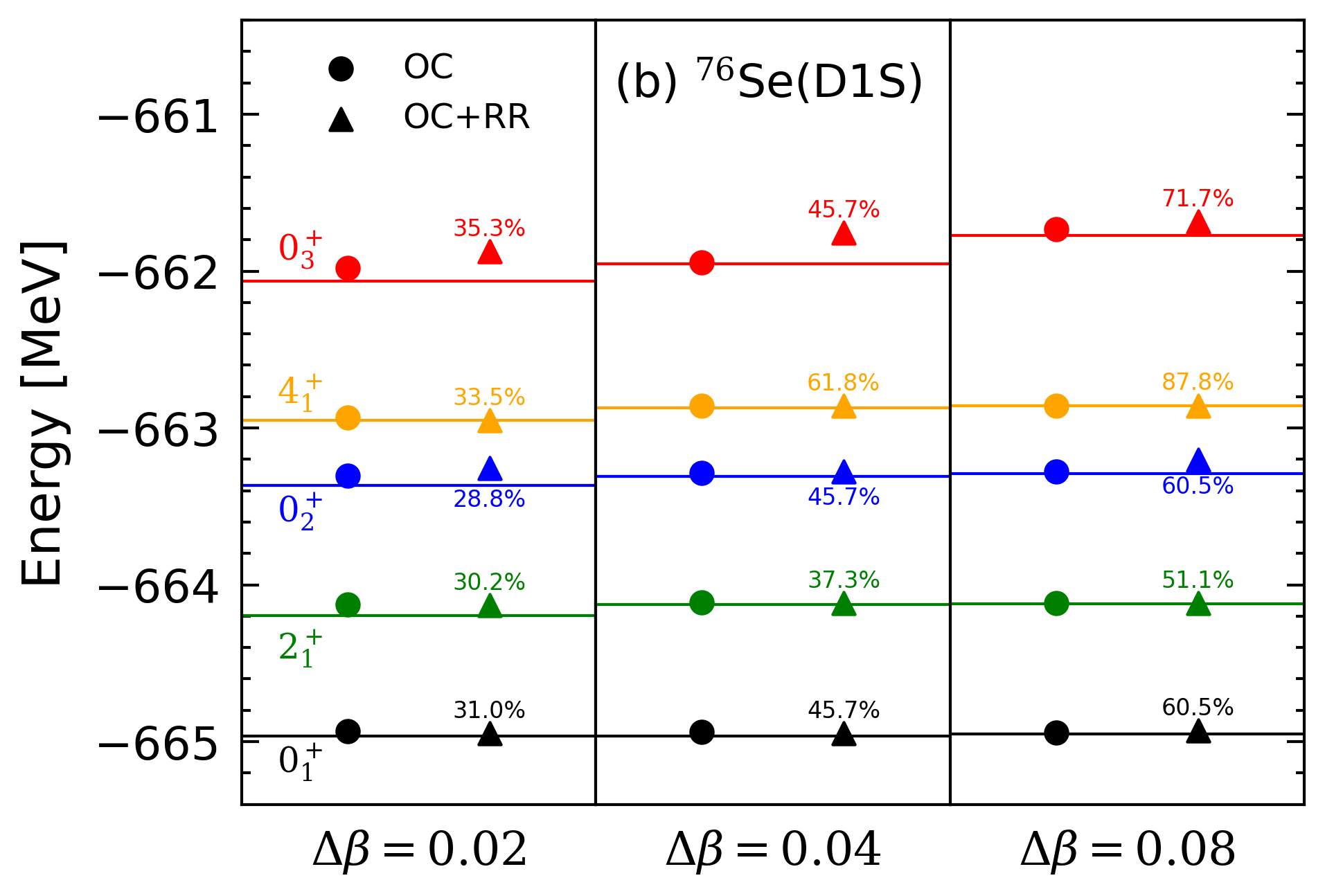}
	\caption{(Color online) The energies of low-lying states in $^{76}$Ge (a) and $^{76}$Se (b) from the GCM (horizontal lines), GCM+OC (circles) and GCM+OC+RR (triangles) calculations,  respectively. From left to right columns shows the results from the calculations with the step size $\Delta\beta=0.02$, $\Delta\beta=0.04$ and $\Delta\beta=0.08$, respectively. The number shown nearby each triangle is determined by the ratio $N_{{\cal S}}/N_{\cal F}$,  where $N_{\cal S}$ is the number of kernels for the configurations within the selected subspace  for a give $L_c$ and those for the training set, while $N_{\cal F}$ is the  number of kernels for the configurations in the full space. }
	\label{fig:D1S_energy_spectrum}
\end{figure}

Figure~\ref{fig:D1S_ML_RMS} shows the  RMSE of the kernels $\ln{\cal N}$ and the ratios ${\cal H}/{\cal N}$ with $J=0$ for both the training set and test set as a function of the degree $N$ of the polynomials in the RR model.  For comparison, the results by the RR models with a small ($10^{-14}$) and large ($10^{-1}$) value of the ridge parameter $\alpha$ are presented. First, we find that the RR model with $\alpha=10^{-14}$ works much better than the RR model with $\alpha=0$, which is the simple linear regression. With the nonzero regularization term in the RR model, the overfitting problem becomes moderate, even though it still appears for $N>12$. Second, with the choice of a larger value of $\alpha(=10^{-1})$, the occurrence of overfitting problem is extended to a larger value of $N(=20)$. Compared to the RR model with $\alpha=10^{-14}$, the RMSE in the model with $\alpha(=10^{-1})$ is systematically larger. Therefore, in this work, the hyperparameters $N=12, \alpha(=10^{-14})$ are employed in the RR model if not mentioned explicitly.  Selecting the value of $N$ giving the best description of the test set, we show the covariance matrix of the RR model for the norm kernels of \nuclide[76]{Ge} with $J=0$ in Fig.~\ref{fig:D1S_Covariance_Matrix}. One can see that the features $(\beta')^{i} (\beta)^{n-i}$ with the even (odd) values of $i$ and $n(\leq N)$ exhibit strong correlation with each other. It indicates that the optimal RR model is still reducible. We will examine the impact of different choices of $N$ on nuclear energy spectra later.

The relative deviations of the norm and Hamiltonian kernels by the optimal RR model are displayed in Fig.~\ref{fig:D1S_deviations}. The deviation for the norm kernels can be up to 10\% for \nuclide[76]{Se}, while that for the ratio ${\cal H}^0/{\cal N}^0$ is less than 0.05\%.  
The noise introduced by the RR model into the norm kernels may spoil the correlation relations among different kernels and thus the orthogonality property of different configurations.  It is shown in Fig.~\ref{fig:D1S_renormalized_energy_platform} that the energy plateau (in particular for the excited states of $\nuclide[76]{Se}$) becomes slightly worse when the RR-model-predicted kernels, instead of the exactly-calculated ones, are used in the GCM calculations. The impact of noise in the kernels of a generalized eigenvalue problem has also been discussed recently in the EC method~\cite{Hicks:2022}, where a trimmed sampling algorithm was proposed to mitigate this issue. We also find that the energy plateau is much worse for the RR model with $\alpha=0$ (not shown here), demonstrating the important role of the regularization term played in the optimization of GCM with the polynomial RR model.

 To mitigate the impact of the noise introduced by the RR model on energy spectra, we  employ the OC method to select the subspace ${\cal S}_{L_c}$ based on the RR-model-predicted kernels, as discussed before.  Fig.~\ref{fig:D1S_renormalized_energy_platform} shows the convergence of the energies of the first three $0^+$ states as a function of the cutoff parameter $L_c$. For comparison, the results from the GCM+OC calculations based on the exactly-calculated norm kernels are also given. One can see that the GCM+OC+RR can reasonably reproduce the convergence behavior of the GCM+OC. The value of $L_c$ is determined based on the convergence behavior, which in principle varies with each state. We note that if the same $L_c$ value is taken for all the states, as in the examples discussed later, the performance of OC is slightly worse for the excited states. This can probably be attributed to the fact that the selection of candidate configurations for the subspace follows their energy ordering. The selected subspace is thus expected to be more complete for the ground state than for the excited states.  Once the subspace is defined, we calculate both norm kernels and Hamiltonian kernels within this subspace and use them to carry out GCM calculations. Interestingly, but perhaps not unexpectedly, we find that the subspaces for different low-lying states differ from each other only by a few configurations. Fig.~\ref{fig:D1S_renormalized_energy_platform} shows that the energy of each state terminates at the number of natural states defined by the dimension of the corresponding subspace.
 
  Figure~\ref{fig:norm_kernels_eigenvalues} shows the distributions of the eigenvalues of the norm kernels with $J=0$ from the exact quantum-number projection calculation and from the RR model prediction. The eigenvalues of the norm kernels within the selected subspace for the ground state are also given for comparison. We see that eigenvalues with values smaller than $10^{-3}$ from the RR model prediction are different from the exactly calculated ones. This explains the degradation of the energy plateaus in Fig.~\ref{fig:D1S_renormalized_energy_platform}. In contrast, the eigenvalues obtained with GCM+OC+RR match the full GCM eigenvalues until the limit of the subspace is reached. Thus, the introduction of noise by the RR model is avoided in the GCM+OC+RR approach.
 
  Figure~\ref{fig:D1S_wfs_conv} displays the measure of the distance in the ground-state wave function $||\Psi^{(n)}-\Psi^{\rm (full)}_{\rm GCM}||$ and in energy, $|E^{(n)}-E^{\rm (full)}_{\rm GCM}|$ with respect to that by the full GCM calculation (with $N_q$ HFB states) as a function of the number of HFB states in both GCM and GCM+OC+RR calculations for  \nuclide[76]{Ge} and  \nuclide[76]{Se}, where the HFB states are ordered in energy, and the measure of the distance in the wave function is defined as~\cite{Broeckhove:1979}
  \beq
  \label{eq:distance_measure}
 ||\Psi^{(n)}-\Psi^{\rm (full)}_{\rm GCM}||=\sqrt{2\left[1-\operatorname{Re}\left(\bra{\Psi^{(n)}}\Psi^{\rm (full)}_{\rm GCM}\rangle\right)\right]}
 \eeq
 with 
 \beq
  \bra{\Psi^{(n)}}\Psi^{\rm (full)}_{\rm GCM}\rangle
  =\sum_{i=1}^{n} \sum_{j=1}^{N_q} f^{(n)^\ast}(\beta_i) {\cal N}^{J=0}\left(\beta_{i}, \beta_{j}\right)f^{\rm (full)}_{\rm GCM}(\beta_j).
 \eeq 
 Here, $f^{(n)}(\beta_i)$ and $f^{\rm (full)}_{\rm GCM}(\beta_j)$ are the weight functions \eqref{eq:wfs} of the ground states in the GCM calculations based on the first $n$ and all $N_q$ HFB states, respectively.  One can see that the ground-state wave function converges faster to the wave function of the full GCM calculation in the GCM+OC+RR than  that in the pure GCM. In the GCM+OC+RR calculation, the residual norm difference between the subspace-projected wave function and the full GCM solution is typically smaller than 0.05, which corresponds to $\bra{\Psi^{(n)}}\Psi^{\rm (full)}_{\rm GCM}\rangle\simeq0.999$.
 
To check how the hyperparameters $(N, \alpha)$ in our RR models affect nuclear energy spectra, we show the energies of the first three $0^+$ states from both the GCM+RR and GCM+OC+RR calculations for both \nuclide[76]{Ge} and \nuclide[76]{Se} as a function of $N$ in Fig.~\ref{fig:D1S_stability_degree}, where two different values of $\alpha$ are employed for comparison. It is shown that the energies  are generally stable under the variations of $N$ for all cases, and consistent with the full GCM results. The observed fluctuations occasionally in the results of GCM+RR calculations can be removed when the OC method is implemented additionally. In other words, the GCM+OC+RR method works well for a large range of values for the hyperparameter $(N, \alpha)$.

\begin{figure}[] 
    \includegraphics[width=\columnwidth]{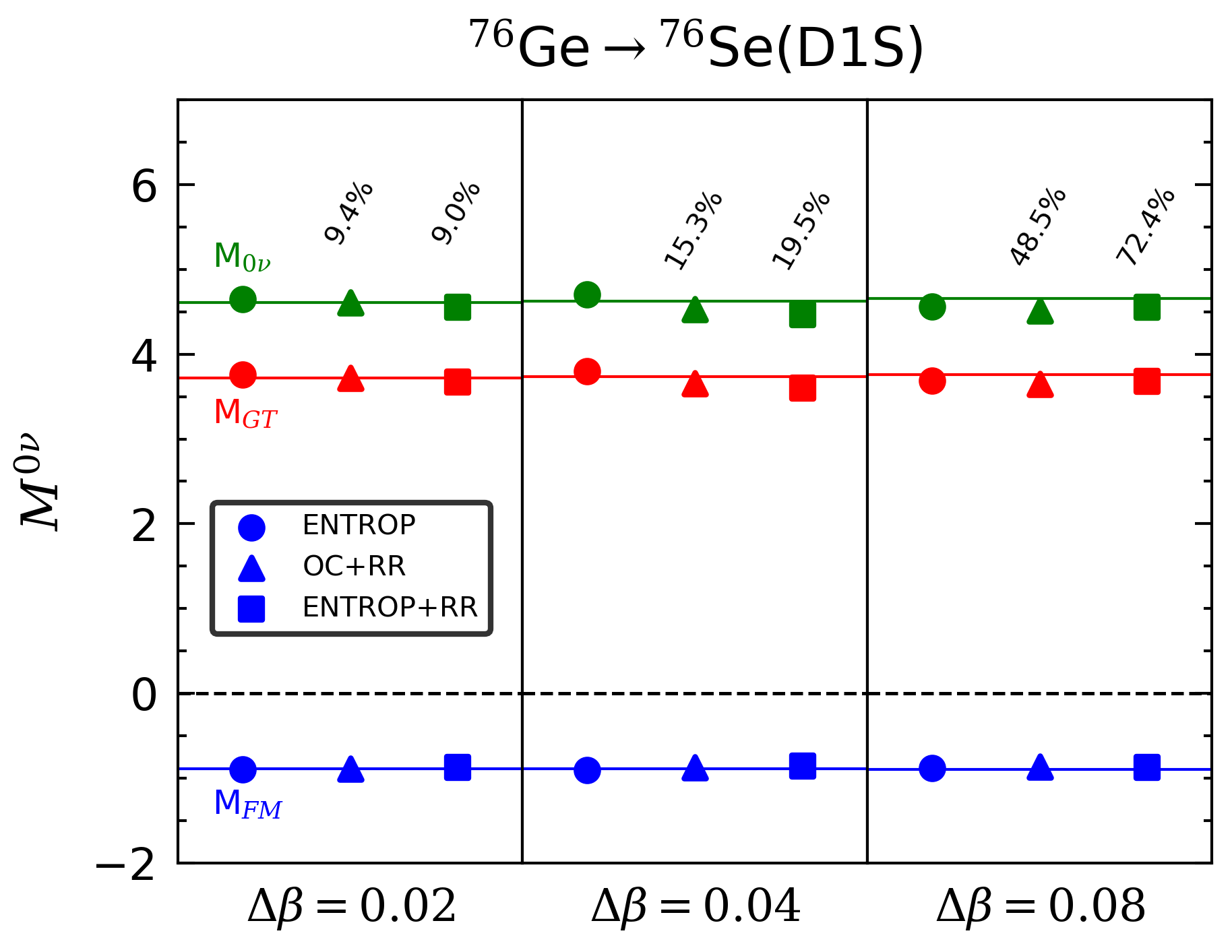}
    \caption{(Color online) The Fermi, GT, and total NMEs of $0\nu\beta\beta$ decay for \nuclide[76]{Ge}. The solid line is obtained from the exact GCM calculation of $^{76}$Ge and $^{76}$Se, while  the circles, triangles, and squares are by GCM+ENTROP, GCM+OC+RR and GCM+ENTROP+RR calculations, respectively. The numbers represent the percentage of cost in computation time compared to the full GCM calculation using all the configurations, and they are obtained from the multiplication of the ratios for \nuclide[76]{Ge} and \nuclide[76]{Se} in Fig.~\ref{fig:D1S_energy_spectrum}. See main text for details.}
    \label{fig:D1S_Ge76_Se76_nme}
\end{figure}
\begin{figure}[] 
		\includegraphics[width=\columnwidth]{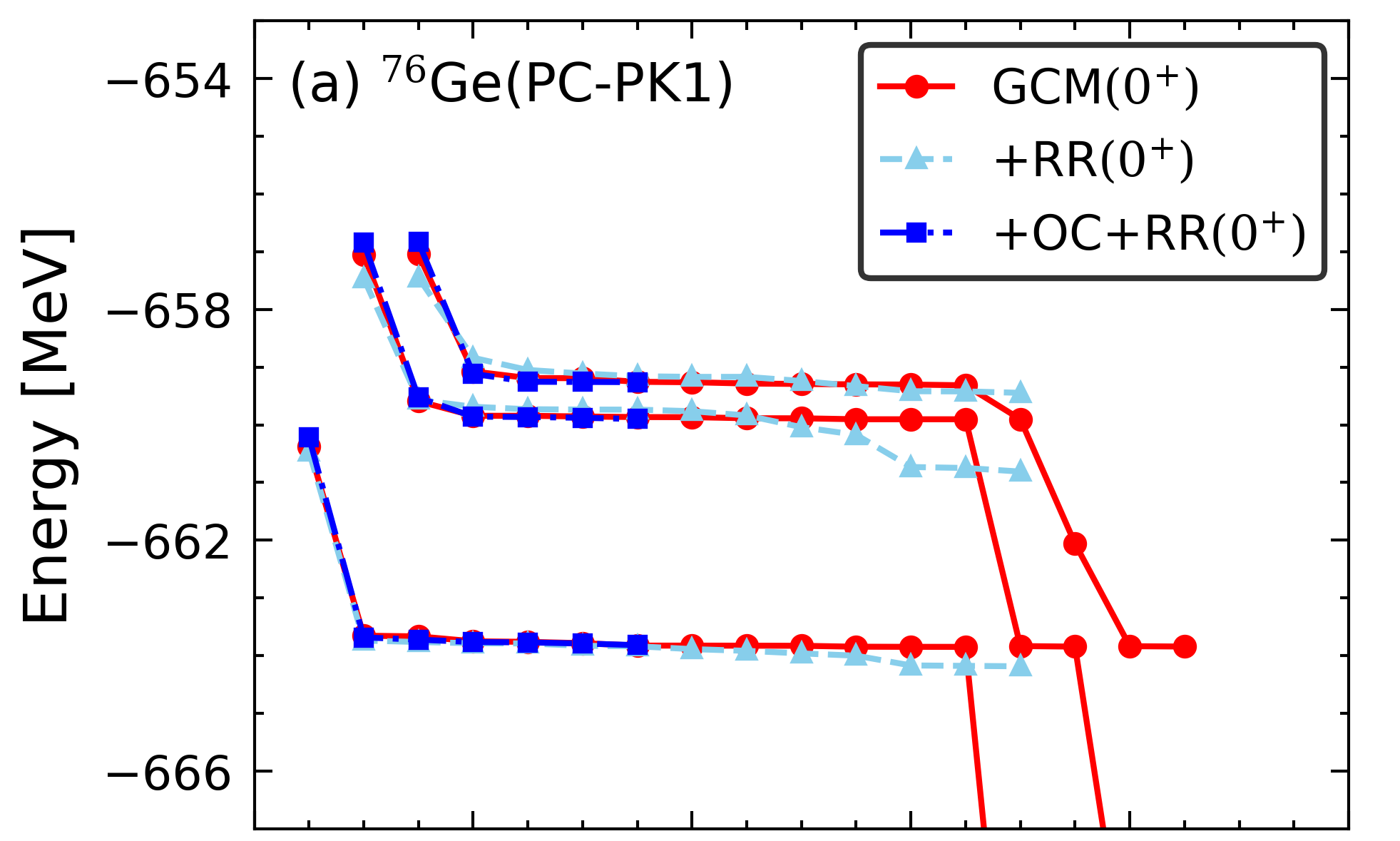}
		\includegraphics[width=\columnwidth]{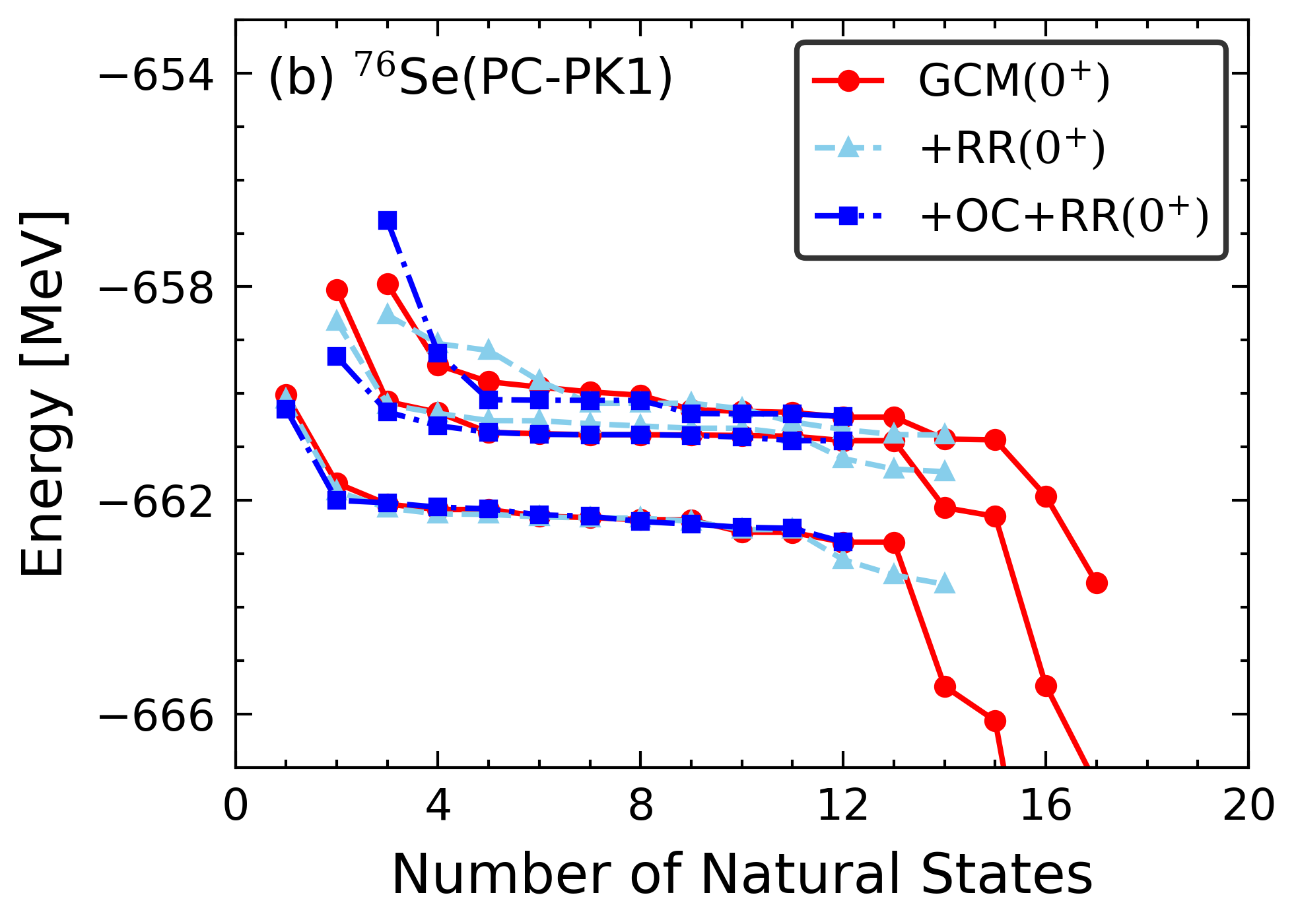}
	\caption{(Color online) Same with Fig.\ref{fig:D1S_renormalized_energy_platform}, but for the relativistic EDF PC-PK1.}
	\label{fig:PC-PK1_energy_platform}
\end{figure}

\begin{figure}[] 
    \includegraphics[width=\columnwidth]{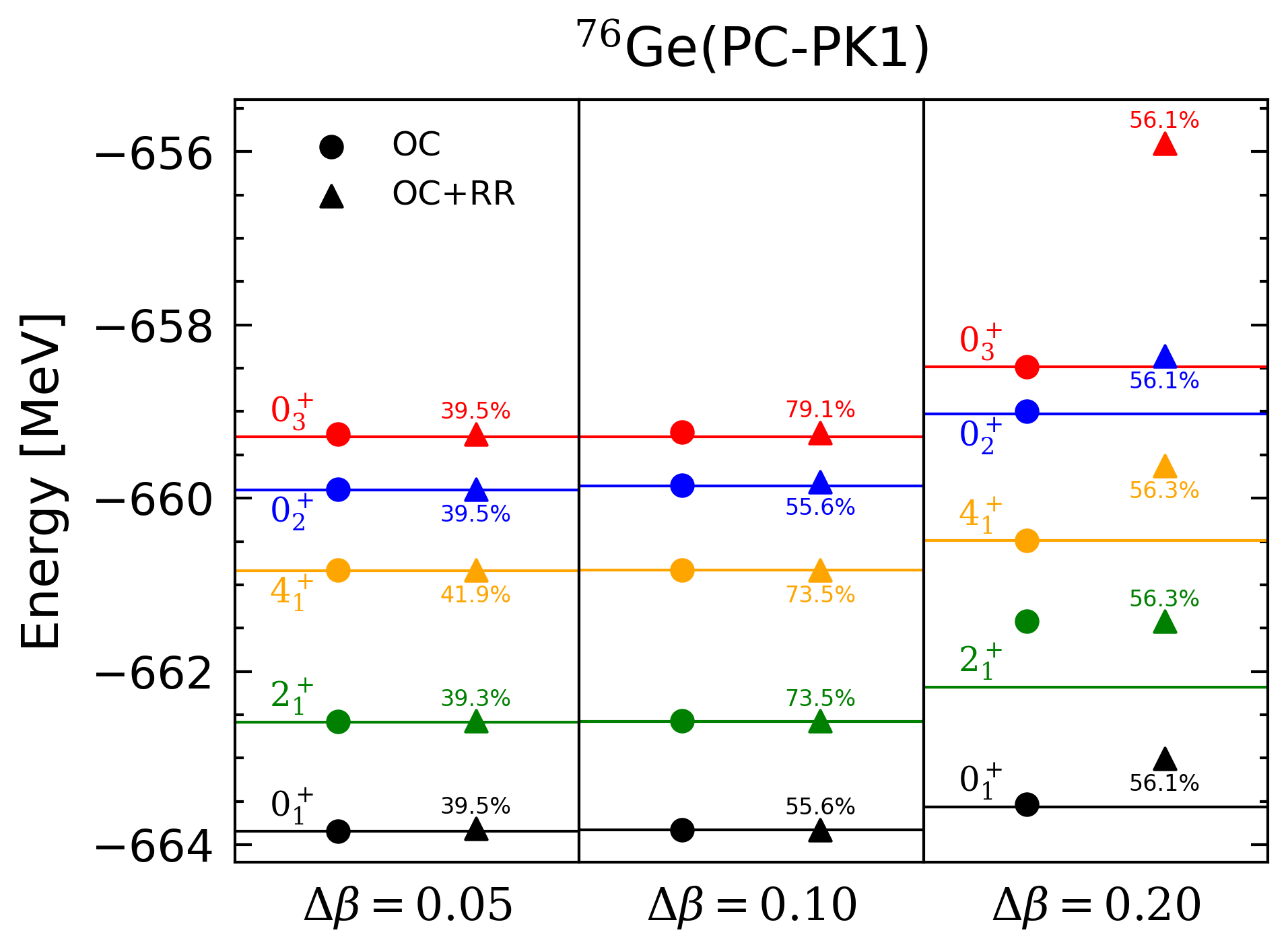}
    \caption{(Color online) Same as Fig.~\ref{fig:D1S_energy_spectrum}, but for the relativistic EDF PC-PK1.}
    \label{fig:PC-PK1_Ge76_energy_spectrum}
\end{figure}

The low-energy spectra from different calculations are shown in Fig.~\ref{fig:D1S_energy_spectrum}. One can see that the decrease of the step size  $\Delta\beta$ in the deformation parameter from 0.08 (total 9 configurations) to 0.02 (total 33 configurations) only weakly affects the spectrum. Quantitatively, the energy difference in the ground state of \nuclide[76]{Ge} introduced by the OC+RR is less than 50 keV for all cases. This error is slightly larger for the $0^+_3$ state, but it is still around 150 keV  for \nuclide[76]{Ge} and around 200 keV for \nuclide[76]{Se}. In both nuclei,  the energy difference between the GCM+OC+RR and GCM+OC is about 20 keV.  In other words, with the application of the OC to GCM calculation, the error introduced by the RR model is negligible. In the current application, the use of the RR model reduces the computational time by a factor of up to three; detailed numbers for each state are included in Fig.~\ref{fig:D1S_energy_spectrum}. As expected, the denser the mesh for discretizing the quadrupole deformation parameter $\beta$ in the original set, the more computational time one can save with the  statistical ML technique.

\begin{figure}[] 
		\includegraphics[width=\columnwidth]{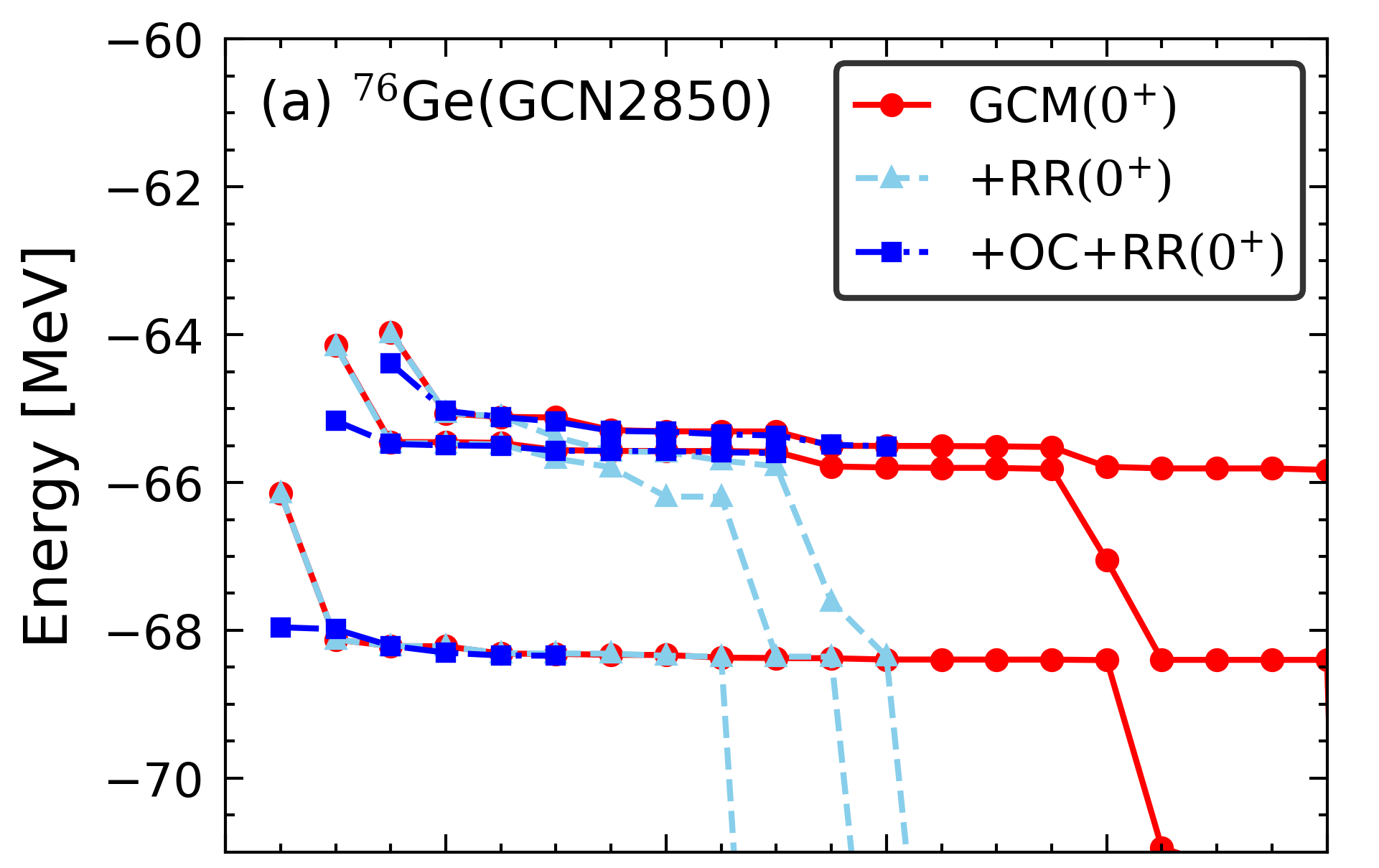}
		\includegraphics[width=\columnwidth]{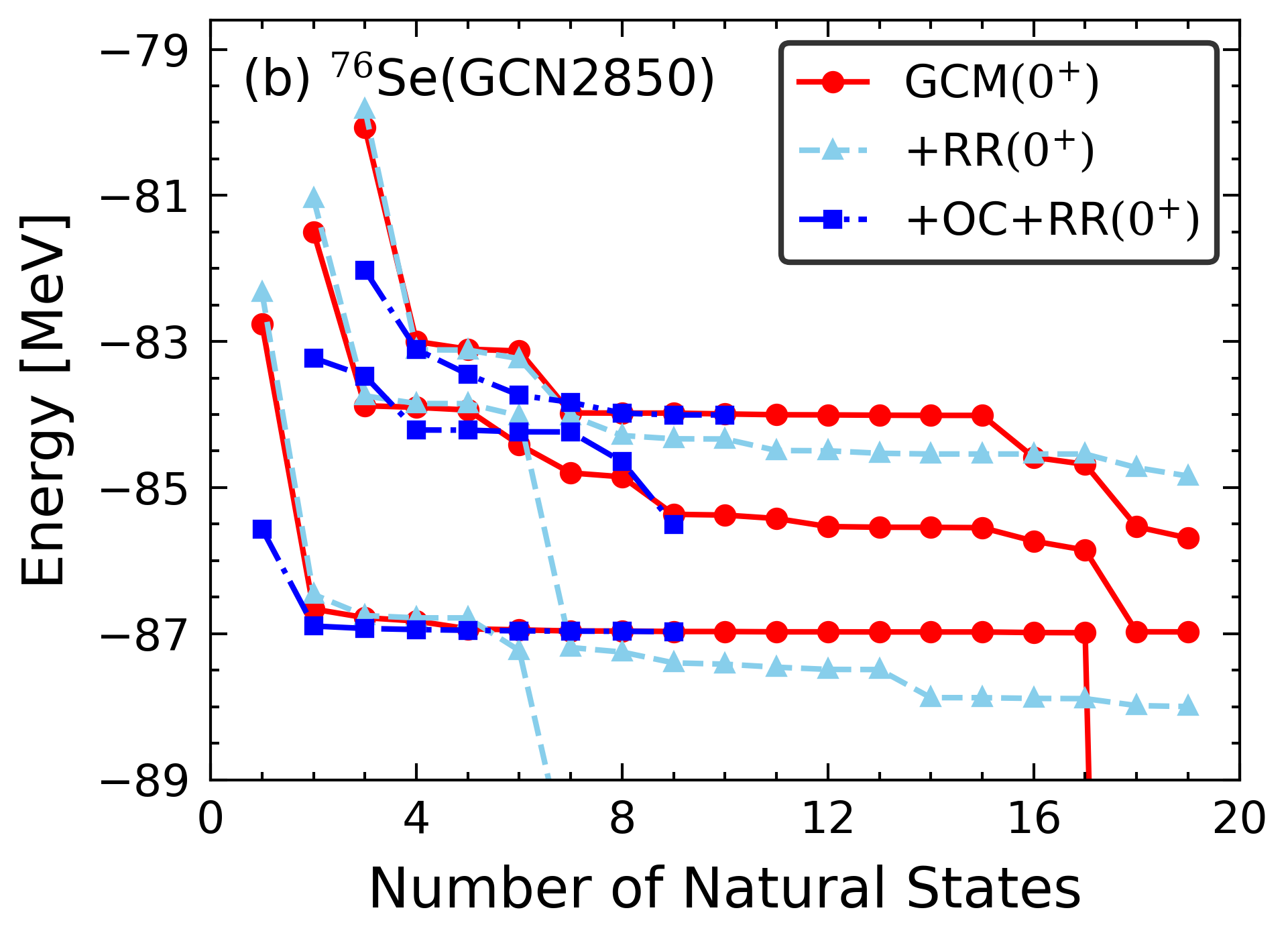}
	\caption{(Color online) Same as Fig.\ref{fig:D1S_renormalized_energy_platform}, but for the shell-model interaction GCN2850.}
	\label{fig:GCN2850_energy_platform}
\end{figure}

\begin{figure}[] 
		\includegraphics[width=\columnwidth]{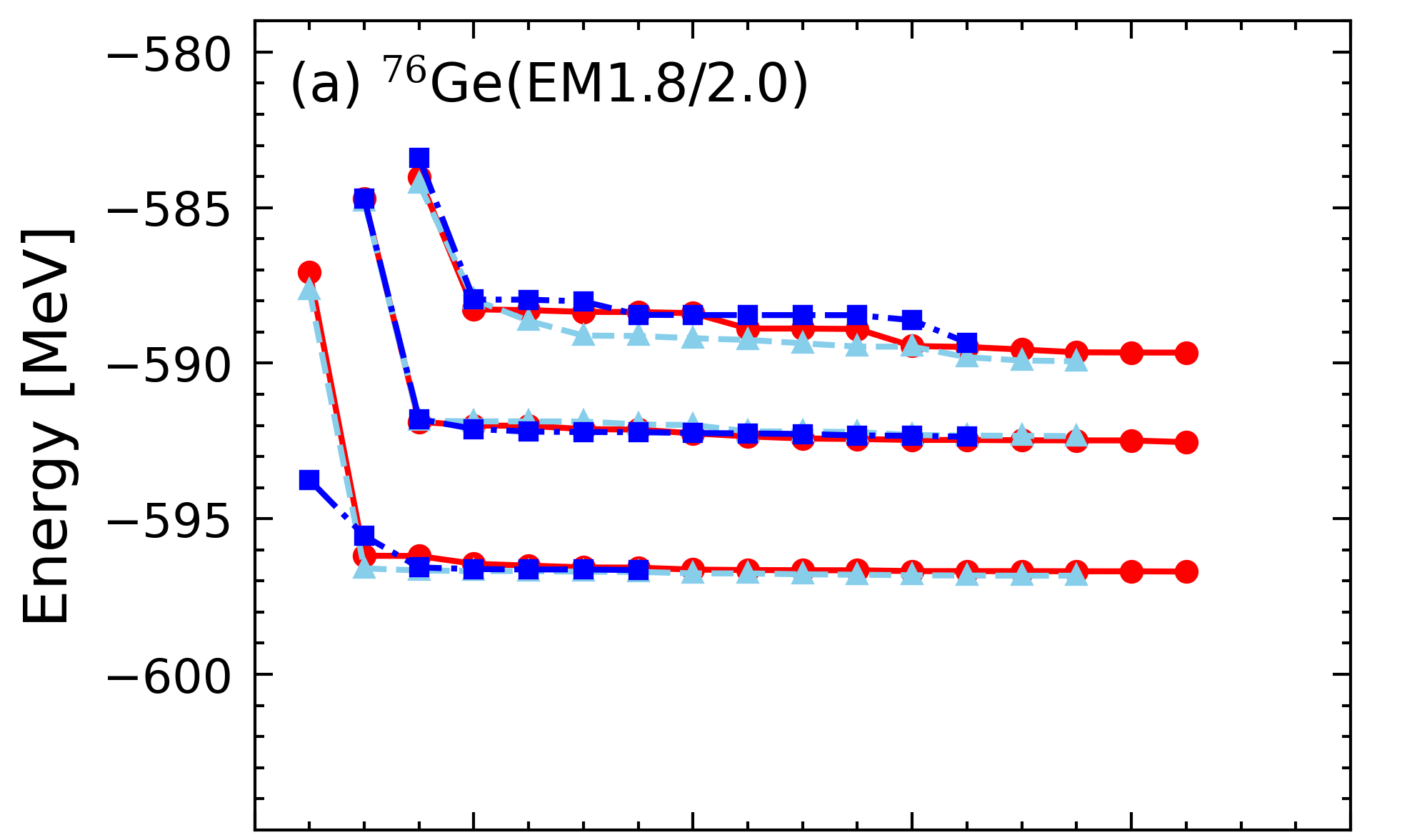}
		\includegraphics[width=\columnwidth]{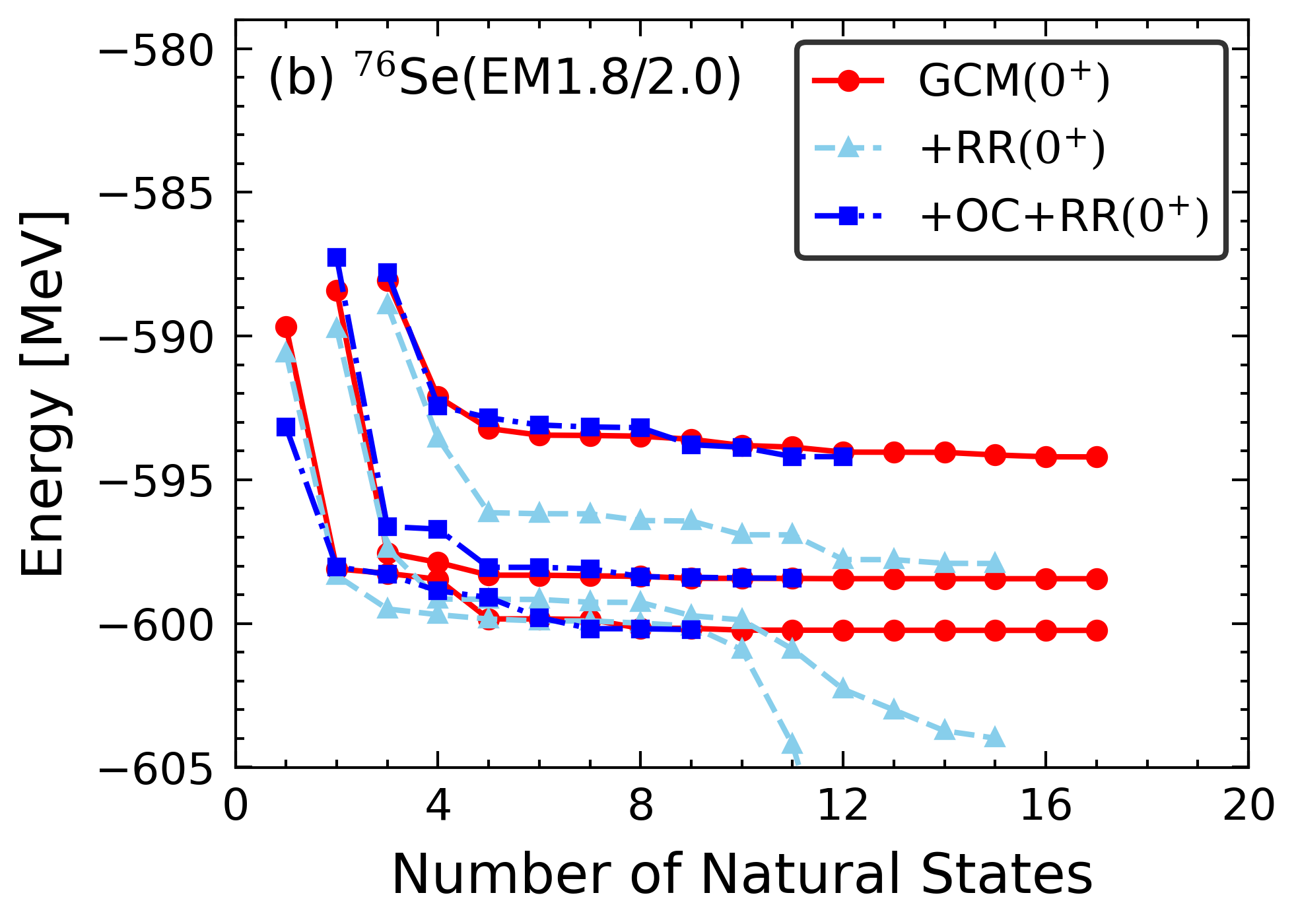}
	\caption{(Color online) Same as Fig.~\ref{fig:D1S_renormalized_energy_platform}, but for the  chiral 2N+3N interaction EM1.8/2.0 with $e_{\rm Max}=6$ and $\hbar\omega=12$ MeV.}
	\label{fig:chiral_energy_platform}
\end{figure}

\begin{figure*}[]
		\includegraphics[width=0.9\columnwidth]{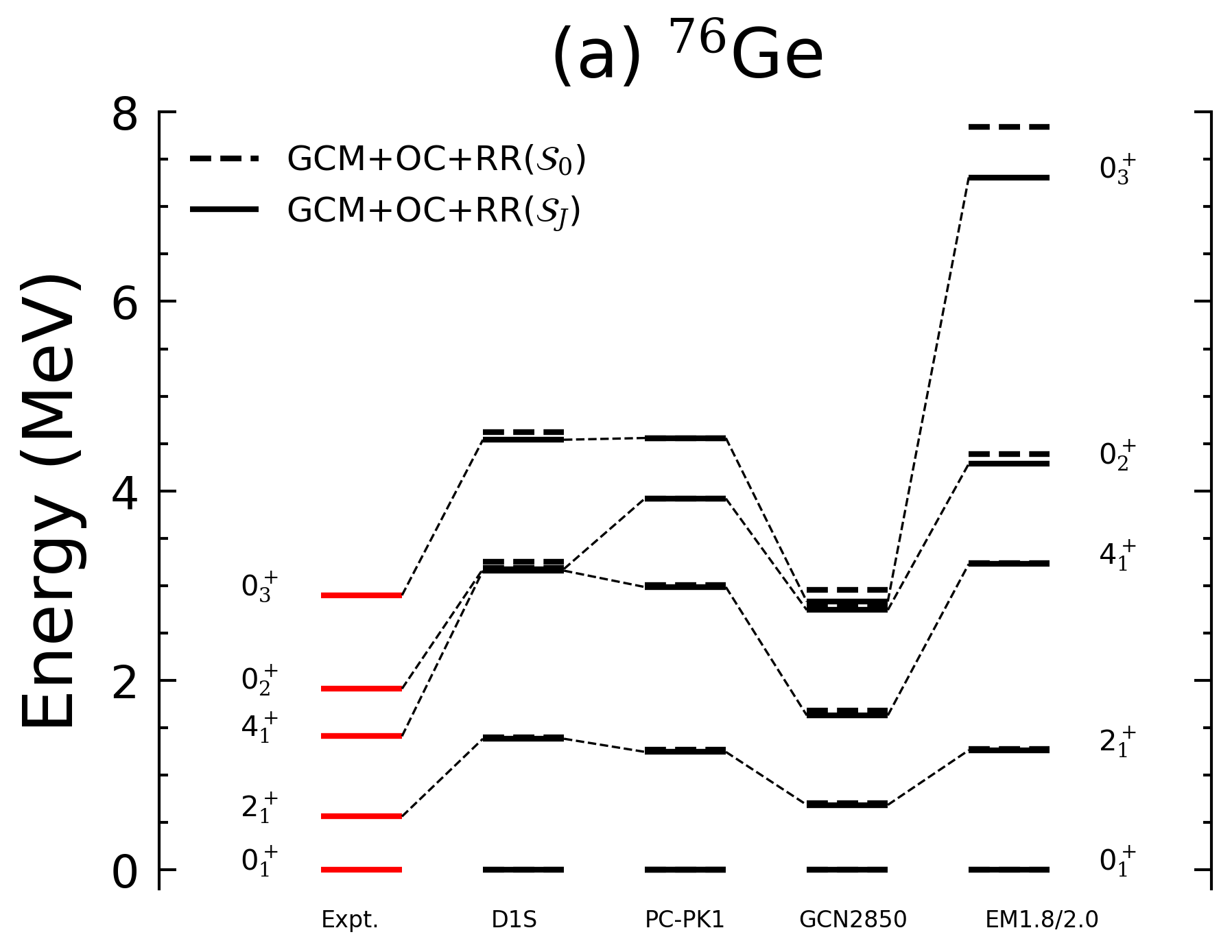}
		\includegraphics[width=0.9\columnwidth]{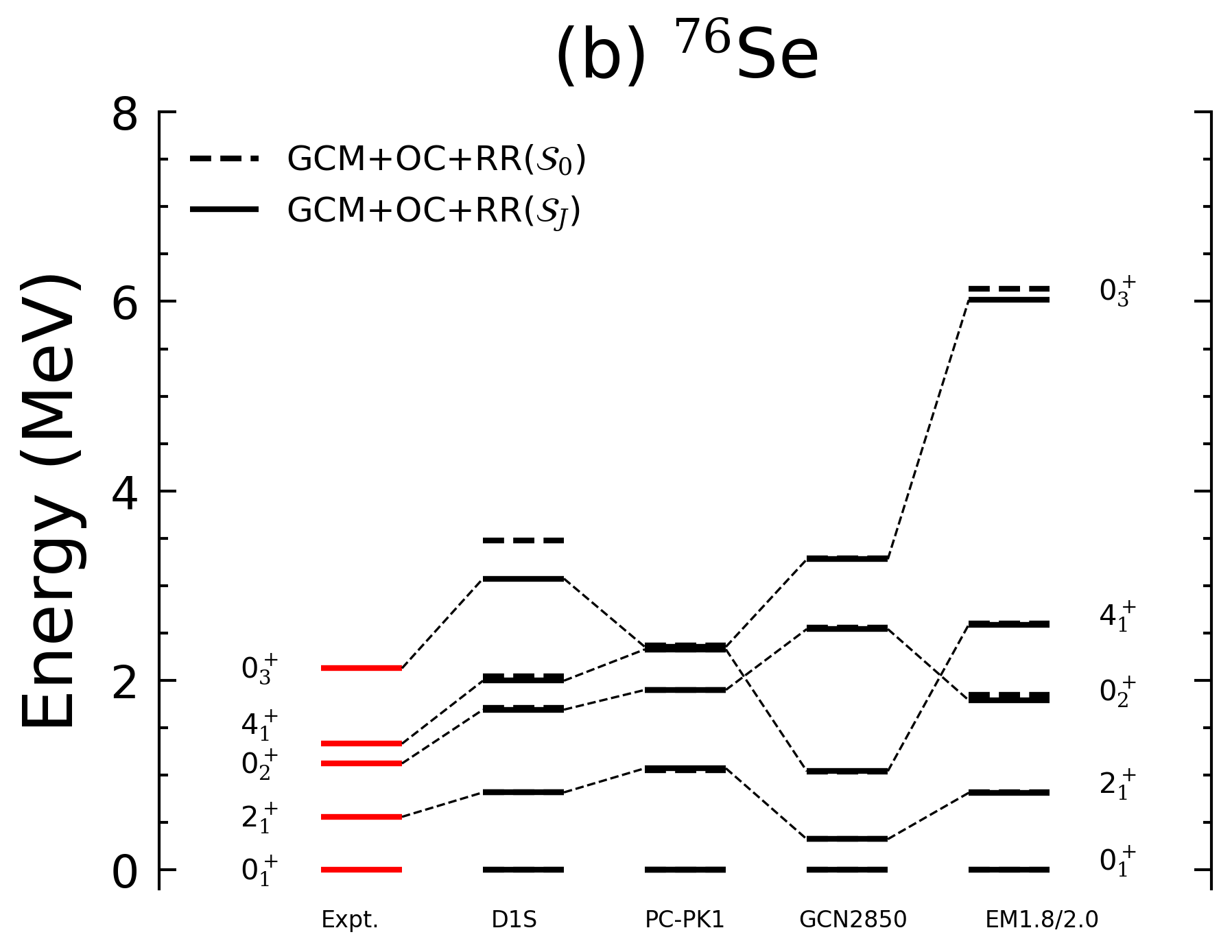}
		\caption{(Color online) The low-lying energy spectra of (a) $^{76}$Ge and (b) $^{76}$Se from GCM+OC+RR calculations based on two different nuclear Hamiltonian and two EDFs. The solid lines are the results of calculations using the configurations within the subspace ($S_{J}$) for each state. The dashed  lines are the results of calculations for all the states using the configurations in the same subspace  ($S_{0}$) as that for the ground state. The results are compared to data from Ref.~\cite{NNDC}.}
		\label{fig:comparison_spectra_Ge6_Se76}
\end{figure*}

 \begin{figure} 
  \includegraphics[width=0.48\columnwidth]{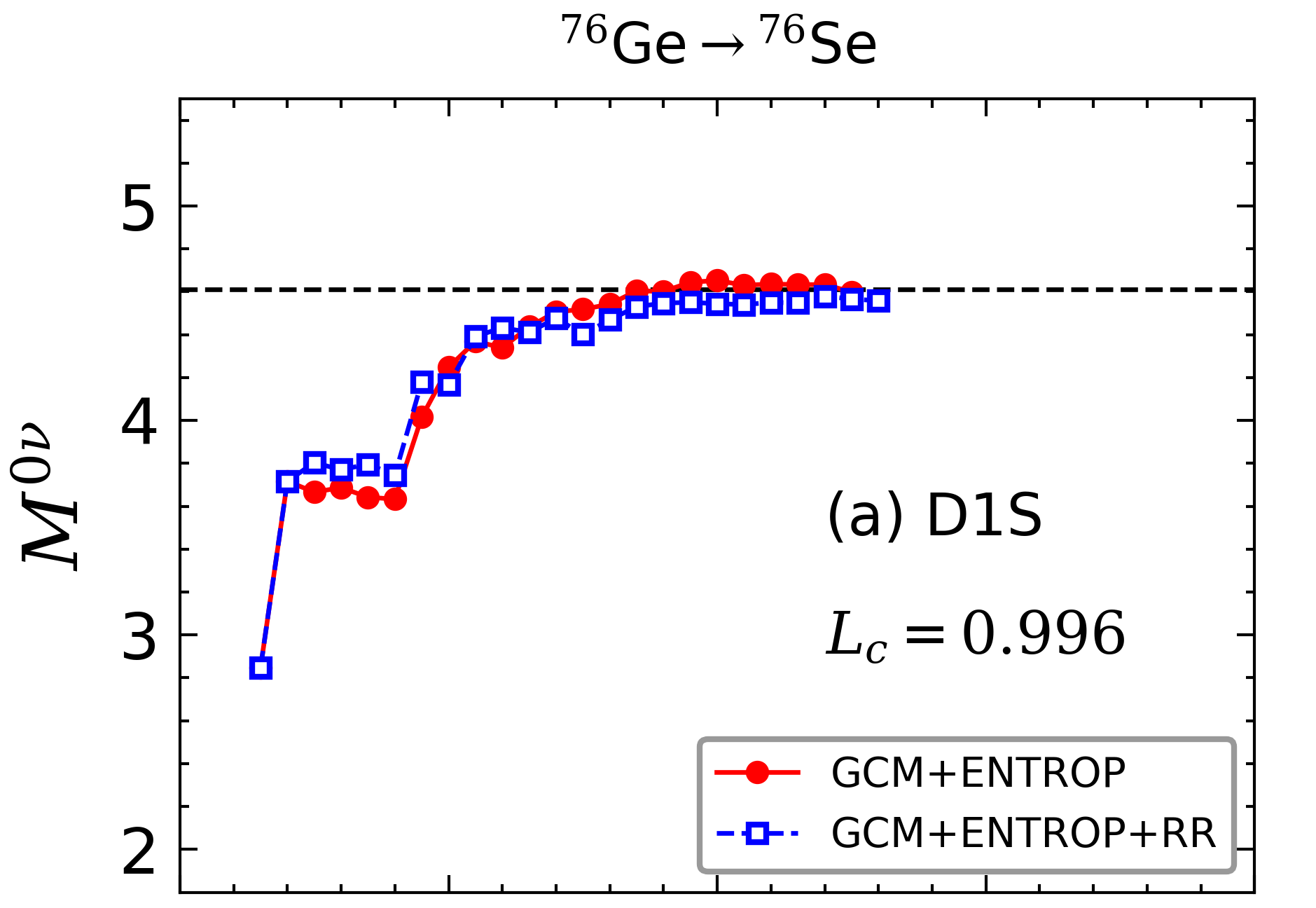} 
 \includegraphics[width=0.48\columnwidth]{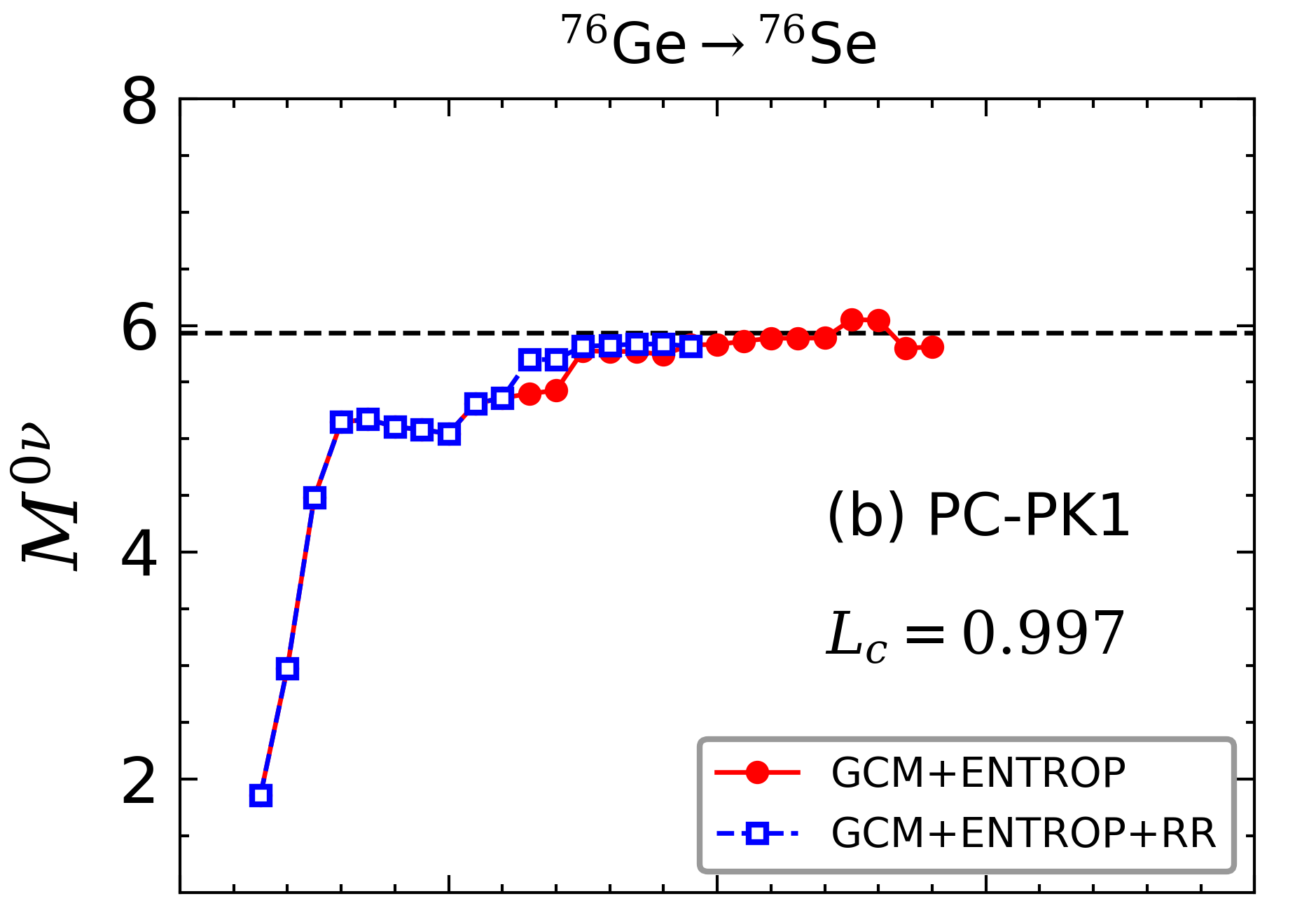} 
 \includegraphics[width=0.48\columnwidth]{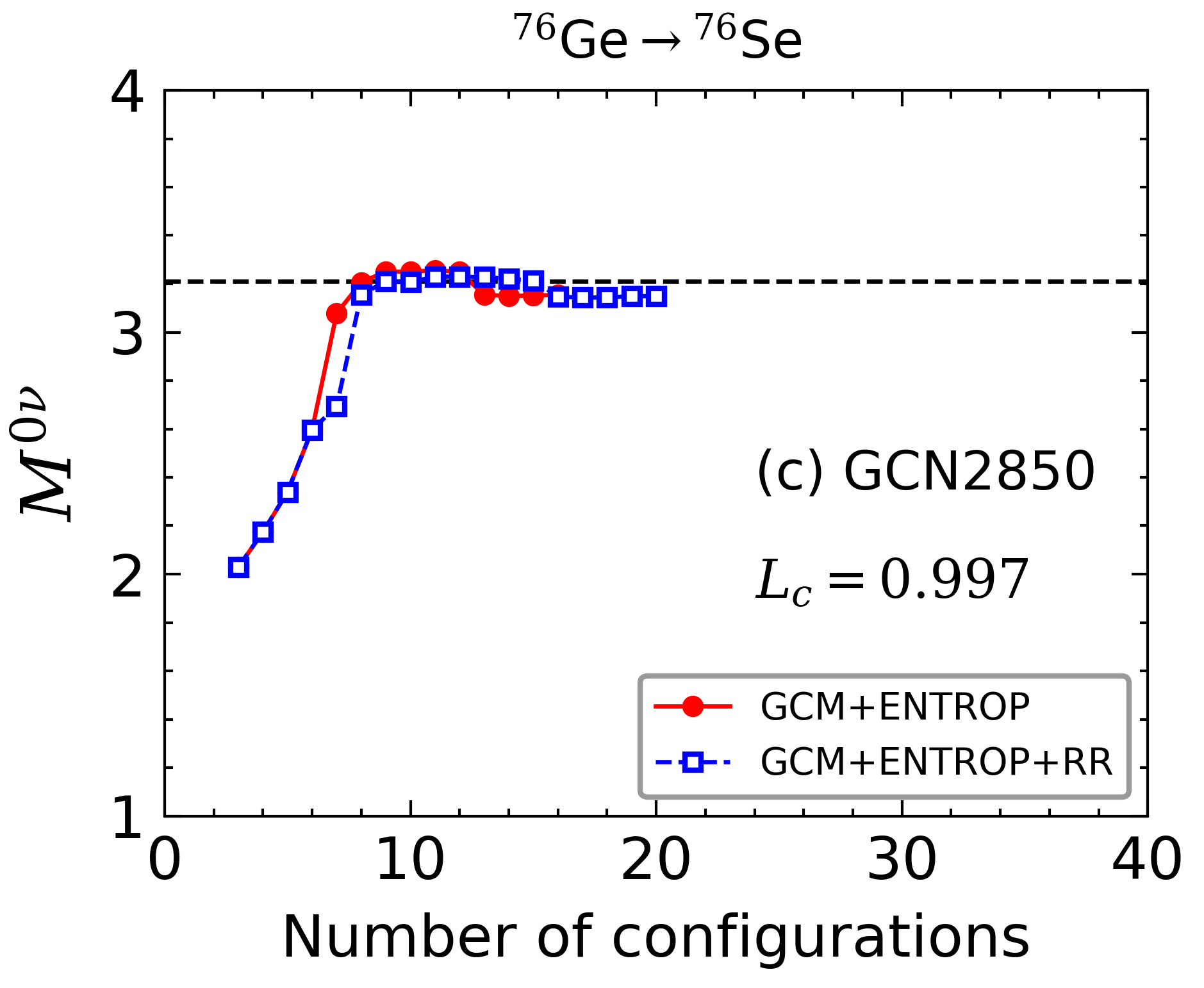} 
 \includegraphics[width=0.48\columnwidth]{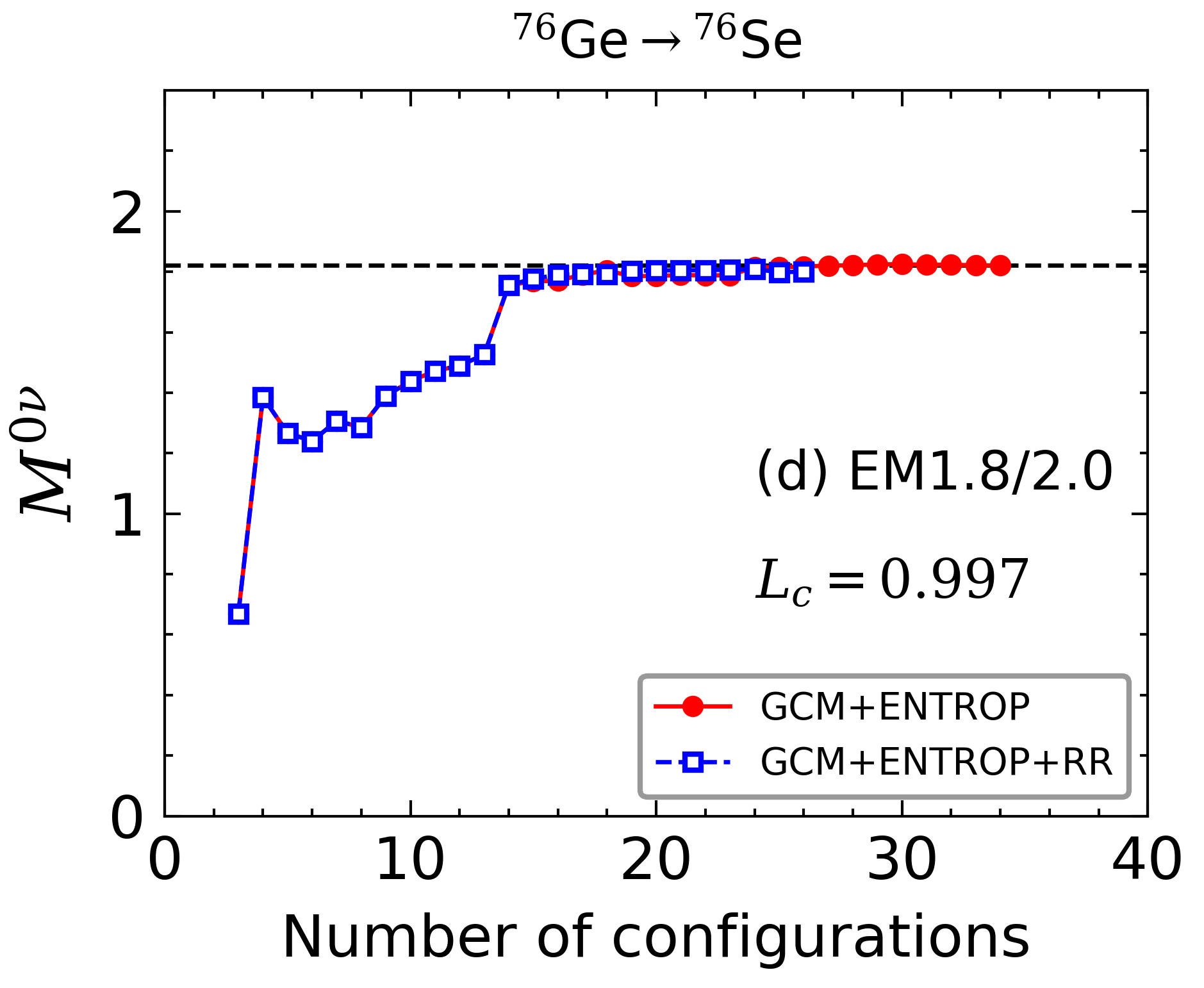} 
\caption{(Color online) The convergence of the NME for the $0\nu\beta\beta$ decay of \nuclide[76]{Ge} against the number of states in \nuclide[76]{Ge} and \nuclide[76]{Se} from the GCM+ENTROP (red line) and GCM+ENTROP+RR (blue dotted line) with (a) D1S, (b) PC-PK1, (c) GCN2850, (d) EM1.8/2.0 ($e_{\rm Max}=6$),  respectively. The results by the original GCM calculation in full space for both nucleus are indicated with dashed lines.}
	\label{fig:convergence_NME}
\end{figure}

\begin{figure}[]
		\includegraphics[width=0.9\columnwidth]{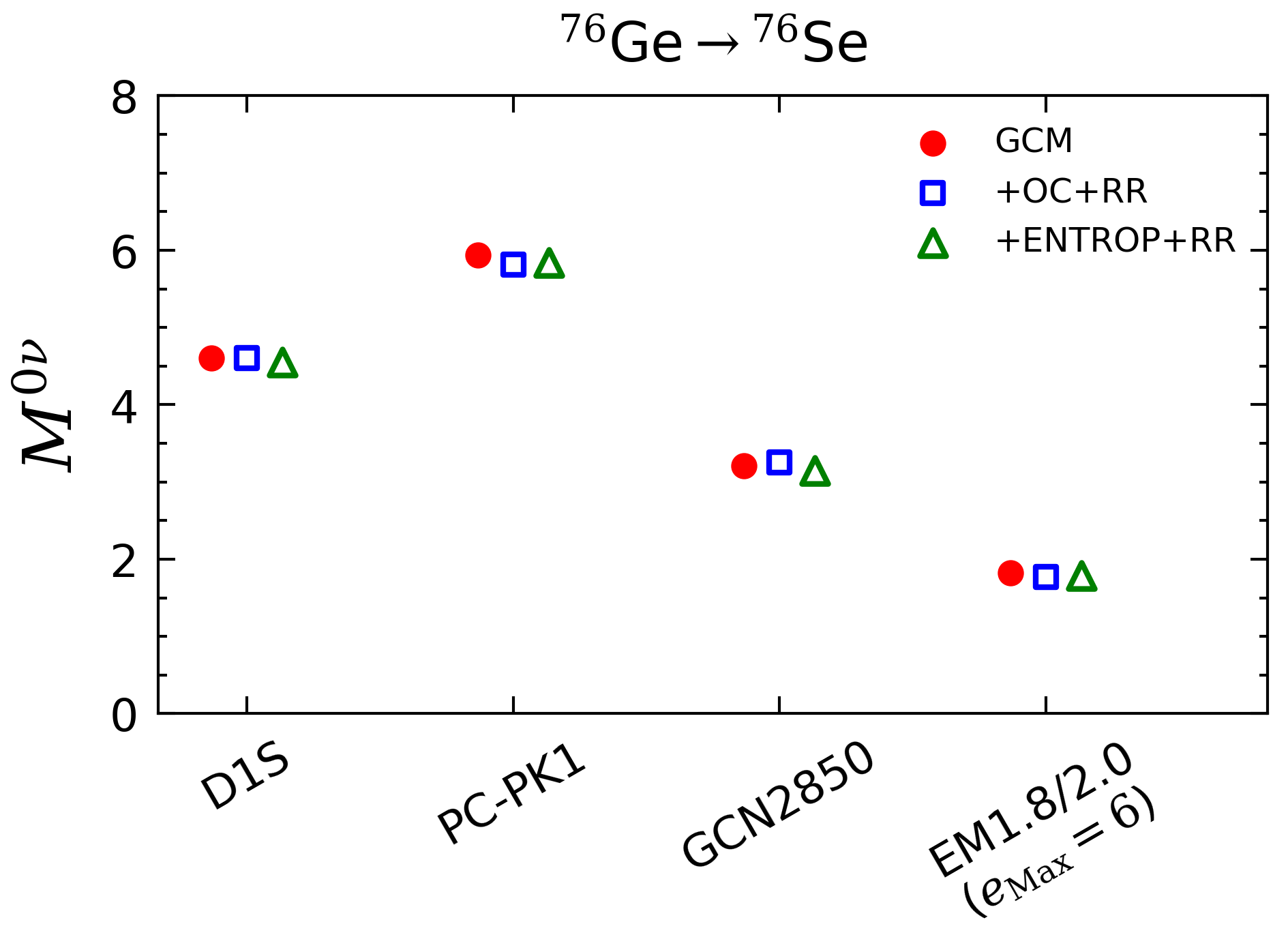} 
		\caption{(Color online) Comparison of NMEs for the $0\nu\beta\beta$ decay of \nuclide[76]{Ge} from the full GCM, GCM+OL+RR and GCM+ENTROP+RR calculations.}
		\label{fig:comparison_NMEs}
\end{figure}

Next, we examine the performance of the method for the NME of $0\nu\beta\beta$ decay. In addition to the use of the OC method, the ENTROP algorithm~\cite{Romero:2021PRC} is also employed for comparison. As the calculation of the NME of $0\nu\beta\beta$ decay requires the ground-state wave functions of two nuclei, it imposes a more stringent test on the RR model than energy spectra.  Fig.~\ref{fig:D1S_Ge76_Se76_nme} shows the  NMEs of $0\nu\beta\beta$ decay for \nuclide[76]{Ge} from GCM+ENTROP, GCM+OC+RR, and GCM+ENTROP+RR calculations.  The cost of computational time compared to the full GCM calculation is indicated in each case.  
One can see that the values by the GCM+OC+RR and GCM+ENTROP+RR are generally close to each other, both are slightly different from those by the GCM+ENTROP and full GCM calculations. Quantitatively, the total NME $M^{0\nu}$ from the GCM+OC+RR calculation  with $\Delta\beta=0.02, 0.04, 0.08$ is 4.61, 4.53, and 4.52, respectively, which should be compared to the values 4.61, 4.62, 4.66 from the full GCM calculation without the use of the statistical ML technique. If the subspaces for \nuclide[76]{Ge}  and \nuclide[76]{Se}  are selected based on the ENTROP+RR, the total NME becomes 4.55, 4.46 and 4.55, respectively. The difference in the three numbers is negligibly small. It implies that the NME in the GCM+ENTROP+RR calculation is less sensitive to the choice of the step size in the deformation parameter $\beta$ than the GCM+OC+RR calculations. In many cases, the predicted NMEs are essentially close to each other and consistent with the value of 4.60 given in Ref.~\cite{Rodriguez:2010}.

\subsubsection{The relativistic EDF: PC-PK1}

In this subsection, we present results for $^{76}$Ge and $^{76}$Se starting from the relativistic EDF PC-PK1~\cite{Zhao:2010PRC} . Again, only axially deformed configurations   are employed. The short-range correlation that has been taken into account in Ref.~\cite{Song:2017} is not included here. More details about the calculations can be found in Ref.~\cite{Yao:2015}.

Figure~\ref{fig:PC-PK1_energy_platform} shows the convergence behavior of the energies of states with $J=0$ in three different calculations with the PC-PK1. Similar to the case of the Gogny D1S, the energy plateaus in the GCM+RR are slightly worse than those by the full GCM with the increase of the number of natural states due to the errors in the kernels introduced by the RR models. With the OC method, the energy plateaus terminate at the number defined by the selected subspace. Fig.~\ref{fig:PC-PK1_Ge76_energy_spectrum} displays the low-energy spectra for \nuclide[76]{Ge} from the GCM+OC and GCM+OC+RR calculations. In the $\Delta\beta=0.2$ case, the energies of states are underestimated evidently  by the GCM+OC+RR compared to the full GCM calculation. We note that in the case with sparsely distributed HFB states, the results are somewhat sensitive to the choice of the degree $N$ in the polynomial regression.  When a set of denser mesh points with  $\Delta\beta=0.1$ or $\Delta\beta=0.05$ is employed, the GCM+OC+RR can reproduce the results of the full GCM calculation, and the results are much more robust against variations of the parameter $N$.  In short, the main findings in the results of the relativistic PC-PK1 EDF are generally similar to those of the Gogny D1S force.

\subsection{Hamiltonian-based GCM calculations}
 In this subsection, we present results for  Hamiltonian-based GCM calculations, where the shell-model interaction GCN2850~\cite{Menendez:2009} and the {\em magic} chiral NN +3N interaction EM1.8/2.0~\cite{Hebeler:2011,Nogga:2004il} are employed. The EM1.8/2.0 interaction provides excellent empirical agreement between ground-state energies and data through at least the $A\sim 60 - 70$ mass region \cite{Stroberg:2021PRL}, which is why it has been frequent used in recent \emph{ab initio} studies, including the in-medium GCM studies of light nuclei~\cite{Yao:2021PRC,Yao:2022DGT} and $^{48}$Ca \cite{Yao:2020PRL} by some of the authors of this work. This interaction, together with the decay operator, is  evolved with the multi-reference in-medium similarity renormalization group~\cite{Hergert:2016,Hergert:2016jk} with a reference ensemble comprising prolate, spherical, and oblate HFB minima in both $^{76}$Ge and $^{76}$Se and with $e_{\rm{max}}=6$. Here, only axially deformed configurations are employed in the GCM calculations. Therefore, the results are somewhat different from those published in Ref.~\cite{Romero:2021PRC}, where triaxially deformed configurations and those with neutron-proton pairing correlations are also included.

 Figure~\ref{fig:GCN2850_energy_platform} displays the energies of states from the GCM calculations based on the shell-model interaction GCN2850 for \nuclide[76]{Ge} and \nuclide[76]{Se} as a function of the number of natural states. The energy plateaus are much worse than those found in the two EDF-based cases. It is even difficult to determine the energies of the states in the GCM+RR calculations. This is especially true for \nuclide[76]{Se}.  The inclusion of the OC method improves the situation, and the energies of both nuclei by the full GCM calculations are reasonably reproduced by the GCM+OC+RR.

 Figure~\ref{fig:chiral_energy_platform} displays the energies of the first three $0^+$ states starting from the chiral nuclear force EM1.8/2.0. We find that the energy plateaus in the full GCM calculation are comparable to those found in the EDF-based calculations, and they are well reproduced in the GCM+OC+RR calculations. The energies of the states in \nuclide[76]{Se} can hardly be reproduced by the GCM+RR only, demonstrating again the necessity and success of implementing the OC method additionally. 

 \subsection{Energy spectra and nuclear matrix elements of $0\nu\beta\beta$ decay }

 Figure~\ref{fig:comparison_spectra_Ge6_Se76} shows the data on the low-energy spectra of both \nuclide[76]{Ge} and \nuclide[76]{Se}, in comparison with the GCM+OC+RR calculations based on two EDFs and two Hamiltonians.
 Results of calculations using the configurations tailored for each state and those of calculations using the same configurations as that for the ground state are shown. One can see that with the mixing of axially deformed configurations, the main feature of the sequences ($0^+_1, 2^+_1, 4^+_1$) in the energy spectra is reasonably reproduced, even though the predicted energy spectra from the calculations with the EDFs (D1S, PC-PK1), and chiral interaction EM1.8/2.0 are more spread out, in contrast to the results of GCN2850. Besides, it is shown that the two different subspace selection strategies yield only slightly different excitation energies for the high-lying states in all the calculations except for GCN2850, where the excited $0^+$ states are shifted much higher when restricted to the configurations of the subspace for the ground state. This is probably due to the limited number of valence single-particle states based on which one cannot sample the configurations with large quadrupole deformation.
 
 Figure~\ref{fig:convergence_NME} displays the convergence of the NME for the $0\nu\beta\beta$ decay of \nuclide[76]{Ge} with respect to the total number of configurations in \nuclide[76]{Ge} and \nuclide[76]{Se} from the calculations with different EDFs and interactions for a given cutoff value $L_c$. The $L_c$ is determined in such a way that the convergent value of the NME is rather stable with the increase of the $L_c$, see Ref.~\cite{Romero:2021PRC}. It is shown in Fig.~\ref{fig:convergence_NME} that the NME in each case converges reasonably well with the increase of the number of states within the subspace, even though the convergence behavior is slightly worse for the PC-PK1 case. Fig.~\ref{fig:comparison_NMEs} summarizes all the NMEs from different calculations.  One can see that the NMEs by the GCM calculations in full space can be excellently reproduced in the GCM+OC/ENTROP+RR, with about one order of magnitude reduced computational cost. Again, it should be emphasized that these NMEs cannot be interpreted as the final NME by each EDF or interaction as only axially deformed configurations are considered here.

\section{Concluding remarks} 
 \label{sec:summary}
 
  The representation of wave functions for the nuclear states of interest in terms of a set of nonorthogonal basis functions is the core idea of GCM. In this approach, the nuclear many-body problem is transformed into a generalized eigenvalue problem, where the dimension of the norm and Hamiltonian kernels grows significantly with the number of the collective coordinates in the GCM.  Therefore, finding an efficient way to sample the basis functions in the multi-dimensional parameter space is important to accelerate or emulate nuclear model calculations without loss of accuracy.  
 
  In this work, we have explored different procedures for implementing statistical ML techniques into GCM calculation to reduce its computational complexity. To mitigate the impact of noise in the predicted kernels by the RR model, we have proposed
  a subspace-reduction algorithm in which optimal ML models are used as a surrogate method for the exact quantum-number-projection calculation of the norm and Hamiltonian kernels. The efficiency and accuracy of each procedure are demonstrated with both non-relativistic and relativistic EDFs, a valence-space shell-model Hamiltonian, and a chiral nuclear interaction in calculations for the low-lying energy spectra of $^{76}$Ge and $^{76}$Se, as well as the $0\nu\beta\beta$-decay NME between their ground states. For the present proof-of-concept study, only axially deformed configurations have been considered.  A polynomial RR model was used to learn the norm and Hamiltonian kernels.  The results have shown that the noise introduced by the optimal RR model may spoil the description of GCM for nuclear spectra, but this issue can be overcome by applying the subspace-reduction algorithms based on the linear dependence/orthogonality conditions for the basis functions. For the NME of ground-state to ground-state $0\nu\beta\beta$ decay, which requires the wave functions of two different nuclei, we have used both the OC and ENTROP methods to select the subspace, which produce similar results. We have found that in the present study the space-reduction algorithm can speed up the GCM calculation by a factor up to about three to nine for the energy spectra and NME, respectively, with negligible loss in accuracy. One can anticipate that this factor will be significantly larger in GCM calculations with multiple generator coordinates. Extensions of our approach in this direction are in progress.

\section*{Acknowledgments}

 We thank B. Bally, J. J. Chen, J. Engel,  Z. M. Niu, L. G. Pang, R. Wirth, X. L. Zhang, and Y. N. Zhang for fruitful discussions.  This work is partly supported by the National Natural Science Foundation of China (Grant Nos. 12141501 and 12275369) and the Fundamental Research Funds for the Central Universities, Sun Yat-sen University. A.M.R. acknowledges the support from NextGenerationEU/PRTR funding. The work of T.R.R. was funded by the Spanish MCIN under contracts PGC2018-094583-B-I00 and PID2021-127890NB-I00. 
 H. H. acknowledges the support of the U.S. Department of Energy, Office of Science, Office of Nuclear Physics under Awards No. DE-SC0017887 and DE-SC0018083 (NUCLEI SciDAC-4 Collaboration).

% \bibliographystyle{apsrev4-1}  
 %\bibliography{ref}  

\begin{thebibliography}{93}%
\makeatletter
\providecommand \@ifxundefined [1]{%
 \@ifx{#1\undefined}
}%
\providecommand \@ifnum [1]{%
 \ifnum #1\expandafter \@firstoftwo
 \else \expandafter \@secondoftwo
 \fi
}%
\providecommand \@ifx [1]{%
 \ifx #1\expandafter \@firstoftwo
 \else \expandafter \@secondoftwo
 \fi
}%
\providecommand \natexlab [1]{#1}%
\providecommand \enquote  [1]{``#1''}%
\providecommand \bibnamefont  [1]{#1}%
\providecommand \bibfnamefont [1]{#1}%
\providecommand \citenamefont [1]{#1}%
\providecommand \href@noop [0]{\@secondoftwo}%
\providecommand \href [0]{\begingroup \@sanitize@url \@href}%
\providecommand \@href[1]{\@@startlink{#1}\@@href}%
\providecommand \@@href[1]{\endgroup#1\@@endlink}%
\providecommand \@sanitize@url [0]{\catcode `\\12\catcode `\$12\catcode
  `\&12\catcode `\#12\catcode `\^12\catcode `\_12\catcode `\%12\relax}%
\providecommand \@@startlink[1]{}%
\providecommand \@@endlink[0]{}%
\providecommand \url  [0]{\begingroup\@sanitize@url \@url }%
\providecommand \@url [1]{\endgroup\@href {#1}{\urlprefix }}%
\providecommand \urlprefix  [0]{URL }%
\providecommand \Eprint [0]{\href }%
\providecommand \doibase [0]{http://dx.doi.org/}%
\providecommand \selectlanguage [0]{\@gobble}%
\providecommand \bibinfo  [0]{\@secondoftwo}%
\providecommand \bibfield  [0]{\@secondoftwo}%
\providecommand \translation [1]{[#1]}%
\providecommand \BibitemOpen [0]{}%
\providecommand \bibitemStop [0]{}%
\providecommand \bibitemNoStop [0]{.\EOS\space}%
\providecommand \EOS [0]{\spacefactor3000\relax}%
\providecommand \BibitemShut  [1]{\csname bibitem#1\endcsname}%
\let\auto@bib@innerbib\@empty
%</preamble>
\bibitem [{\citenamefont {Hill}\ and\ \citenamefont
  {Wheeler}(1953)}]{Hill:1953}%
  \BibitemOpen
  \bibfield  {author} {\bibinfo {author} {\bibfnamefont {D.~L.}\ \bibnamefont
  {Hill}}\ and\ \bibinfo {author} {\bibfnamefont {J.~A.}\ \bibnamefont
  {Wheeler}},\ }\href {\doibase 10.1103/PhysRev.89.1102} {\bibfield  {journal}
  {\bibinfo  {journal} {Phys. Rev.}\ }\textbf {\bibinfo {volume} {89}},\
  \bibinfo {pages} {1102} (\bibinfo {year} {1953})}\BibitemShut {NoStop}%
\bibitem [{\citenamefont {Griffin}\ and\ \citenamefont
  {Wheeler}(1957)}]{Griffin:1957}%
  \BibitemOpen
  \bibfield  {author} {\bibinfo {author} {\bibfnamefont {J.~J.}\ \bibnamefont
  {Griffin}}\ and\ \bibinfo {author} {\bibfnamefont {J.~A.}\ \bibnamefont
  {Wheeler}},\ }\href {\doibase 10.1103/PhysRev.108.311} {\bibfield  {journal}
  {\bibinfo  {journal} {Phys. Rev.}\ }\textbf {\bibinfo {volume} {108}},\
  \bibinfo {pages} {311} (\bibinfo {year} {1957})}\BibitemShut {NoStop}%
\bibitem [{\citenamefont {Ring}\ and\ \citenamefont
  {Schuck}(1980)}]{Ring:1980}%
  \BibitemOpen
  \bibfield  {author} {\bibinfo {author} {\bibfnamefont {P.}~\bibnamefont
  {Ring}}\ and\ \bibinfo {author} {\bibfnamefont {P.}~\bibnamefont {Schuck}},\
  }\href@noop {} {\emph {\bibinfo {title} {The nuclear many-body problem}}}\
  (\bibinfo  {publisher} {Springer-Verlag},\ \bibinfo {address} {New York},\
  \bibinfo {year} {1980})\BibitemShut {NoStop}%
\bibitem [{\citenamefont {{Reinhard}}\ and\ \citenamefont
  {{Goeke}}(1987)}]{Reinhard:1987RPP}%
  \BibitemOpen
  \bibfield  {author} {\bibinfo {author} {\bibfnamefont {P.~G.}\ \bibnamefont
  {{Reinhard}}}\ and\ \bibinfo {author} {\bibfnamefont {K.}~\bibnamefont
  {{Goeke}}},\ }\href {\doibase 10.1088/0034-4885/50/1/001} {\bibfield
  {journal} {\bibinfo  {journal} {Reports on Progress in Physics}\ }\textbf
  {\bibinfo {volume} {50}},\ \bibinfo {pages} {1} (\bibinfo {year}
  {1987})}\BibitemShut {NoStop}%
\bibitem [{\citenamefont {Capelle}(2003)}]{Capelle:2003}%
  \BibitemOpen
  \bibfield  {author} {\bibinfo {author} {\bibfnamefont {K.}~\bibnamefont
  {Capelle}},\ }\href {\doibase 10.1063/1.1593014} {\bibfield  {journal}
  {\bibinfo  {journal} {The Journal of Chemical Physics}\ }\textbf {\bibinfo
  {volume} {119}},\ \bibinfo {pages} {1285} (\bibinfo {year}
  {2003})}\BibitemShut {NoStop}%
\bibitem [{\citenamefont {Alon}\ \emph {et~al.}(2005)\citenamefont {Alon},
  \citenamefont {Streltsov},\ and\ \citenamefont {Cederbaum}}]{Alon:2005}%
  \BibitemOpen
  \bibfield  {author} {\bibinfo {author} {\bibfnamefont {O.~E.}\ \bibnamefont
  {Alon}}, \bibinfo {author} {\bibfnamefont {A.~I.}\ \bibnamefont {Streltsov}},
  \ and\ \bibinfo {author} {\bibfnamefont {L.~S.}\ \bibnamefont {Cederbaum}},\
  }\href {\doibase 10.1103/PhysRevB.71.125113} {\bibfield  {journal} {\bibinfo
  {journal} {Phys. Rev. B}\ }\textbf {\bibinfo {volume} {71}},\ \bibinfo
  {pages} {125113} (\bibinfo {year} {2005})}\BibitemShut {NoStop}%
\bibitem [{\citenamefont {Orestes}\ \emph {et~al.}(2007)\citenamefont
  {Orestes}, \citenamefont {Capelle}, \citenamefont {da~Silva},\ and\
  \citenamefont {Ullrich}}]{Orestes:2007}%
  \BibitemOpen
  \bibfield  {author} {\bibinfo {author} {\bibfnamefont {E.}~\bibnamefont
  {Orestes}}, \bibinfo {author} {\bibfnamefont {K.}~\bibnamefont {Capelle}},
  \bibinfo {author} {\bibfnamefont {A.~B.~F.}\ \bibnamefont {da~Silva}}, \ and\
  \bibinfo {author} {\bibfnamefont {C.~A.}\ \bibnamefont {Ullrich}},\ }\href
  {\doibase 10.1063/1.2768368} {\bibfield  {journal} {\bibinfo  {journal} {The
  Journal of Chemical Physics}\ }\textbf {\bibinfo {volume} {127}},\ \bibinfo
  {pages} {124101} (\bibinfo {year} {2007})}\BibitemShut {NoStop}%
\bibitem [{\citenamefont {Bender}\ \emph {et~al.}(2003)\citenamefont {Bender},
  \citenamefont {Heenen},\ and\ \citenamefont {Reinhard}}]{Bender:2003RMP}%
  \BibitemOpen
  \bibfield  {author} {\bibinfo {author} {\bibfnamefont {M.}~\bibnamefont
  {Bender}}, \bibinfo {author} {\bibfnamefont {P.-H.}\ \bibnamefont {Heenen}},
  \ and\ \bibinfo {author} {\bibfnamefont {P.-G.}\ \bibnamefont {Reinhard}},\
  }\href {\doibase 10.1103/RevModPhys.75.121} {\bibfield  {journal} {\bibinfo
  {journal} {Rev. Mod. Phys.}\ }\textbf {\bibinfo {volume} {75}},\ \bibinfo
  {pages} {121} (\bibinfo {year} {2003})}\BibitemShut {NoStop}%
\bibitem [{\citenamefont {Nikšić}\ \emph {et~al.}(2011)\citenamefont
  {Nikšić}, \citenamefont {Vretenar},\ and\ \citenamefont
  {Ring}}]{Niksic:2011PPNP}%
  \BibitemOpen
  \bibfield  {author} {\bibinfo {author} {\bibfnamefont {T.}~\bibnamefont
  {Nikšić}}, \bibinfo {author} {\bibfnamefont {D.}~\bibnamefont {Vretenar}},
  \ and\ \bibinfo {author} {\bibfnamefont {P.}~\bibnamefont {Ring}},\ }\href
  {\doibase https://doi.org/10.1016/j.ppnp.2011.01.055} {\bibfield  {journal}
  {\bibinfo  {journal} {Prog. Part. Nucl. Phys.}\ }\textbf {\bibinfo {volume}
  {66}},\ \bibinfo {pages} {519} (\bibinfo {year} {2011})}\BibitemShut
  {NoStop}%
\bibitem [{\citenamefont {Egido}(2016)}]{Egido:2016PS}%
  \BibitemOpen
  \bibfield  {author} {\bibinfo {author} {\bibfnamefont {J.~L.}\ \bibnamefont
  {Egido}},\ }\href {\doibase 10.1088/0031-8949/91/7/073003} {\bibfield
  {journal} {\bibinfo  {journal} {Physica Scripta}\ }\textbf {\bibinfo {volume}
  {91}},\ \bibinfo {pages} {073003} (\bibinfo {year} {2016})}\BibitemShut
  {NoStop}%
\bibitem [{\citenamefont {Robledo}\ \emph {et~al.}(2019)\citenamefont
  {Robledo}, \citenamefont {Rodr\'\i{}guez},\ and\ \citenamefont
  {Rodr\'\i{}guez-Guzm\'an}}]{Robledo:2018JPG}%
  \BibitemOpen
  \bibfield  {author} {\bibinfo {author} {\bibfnamefont {L.~M.}\ \bibnamefont
  {Robledo}}, \bibinfo {author} {\bibfnamefont {T.~R.}\ \bibnamefont
  {Rodr\'\i{}guez}}, \ and\ \bibinfo {author} {\bibfnamefont {R.~R.}\
  \bibnamefont {Rodr\'\i{}guez-Guzm\'an}},\ }\href {\doibase
  10.1088/1361-6471/aadebd} {\bibfield  {journal} {\bibinfo  {journal} {J.
  Phys. G}\ }\textbf {\bibinfo {volume} {46}},\ \bibinfo {pages} {013001}
  (\bibinfo {year} {2019})},\ \Eprint {http://arxiv.org/abs/1807.02518}
  {arXiv:1807.02518 [nucl-th]} \BibitemShut {NoStop}%
\bibitem [{\citenamefont {Sheikh}\ \emph {et~al.}(2021)\citenamefont {Sheikh},
  \citenamefont {Dobaczewski}, \citenamefont {Ring}, \citenamefont {Robledo},\
  and\ \citenamefont {Yannouleas}}]{Sheikh:2021qv}%
  \BibitemOpen
  \bibfield  {author} {\bibinfo {author} {\bibfnamefont {J.~A.}\ \bibnamefont
  {Sheikh}}, \bibinfo {author} {\bibfnamefont {J.}~\bibnamefont {Dobaczewski}},
  \bibinfo {author} {\bibfnamefont {P.}~\bibnamefont {Ring}}, \bibinfo {author}
  {\bibfnamefont {L.~M.}\ \bibnamefont {Robledo}}, \ and\ \bibinfo {author}
  {\bibfnamefont {C.}~\bibnamefont {Yannouleas}},\ }\href {\doibase
  10.1088/1361-6471/ac288a} {\bibfield  {journal} {\bibinfo  {journal} {Journal
  of Physics G: Nuclear and Particle Physics}\ }\textbf {\bibinfo {volume}
  {48}},\ \bibinfo {pages} {123001} (\bibinfo {year} {2021})}\BibitemShut
  {NoStop}%
\bibitem [{\citenamefont {Yao}\ \emph {et~al.}(2014)\citenamefont {Yao},
  \citenamefont {Hagino}, \citenamefont {Li}, \citenamefont {Meng},\ and\
  \citenamefont {Ring}}]{Yao:2014}%
  \BibitemOpen
  \bibfield  {author} {\bibinfo {author} {\bibfnamefont {J.~M.}\ \bibnamefont
  {Yao}}, \bibinfo {author} {\bibfnamefont {K.}~\bibnamefont {Hagino}},
  \bibinfo {author} {\bibfnamefont {Z.~P.}\ \bibnamefont {Li}}, \bibinfo
  {author} {\bibfnamefont {J.}~\bibnamefont {Meng}}, \ and\ \bibinfo {author}
  {\bibfnamefont {P.}~\bibnamefont {Ring}},\ }\href {\doibase
  10.1103/PhysRevC.89.054306} {\bibfield  {journal} {\bibinfo  {journal} {Phys.
  Rev. C}\ }\textbf {\bibinfo {volume} {89}},\ \bibinfo {pages} {054306}
  (\bibinfo {year} {2014})},\ \Eprint {http://arxiv.org/abs/1403.4812}
  {arXiv:1403.4812 [nucl-th]} \BibitemShut {NoStop}%
\bibitem [{\citenamefont {Bally}\ \emph {et~al.}(2014)\citenamefont {Bally},
  \citenamefont {Avez}, \citenamefont {Bender},\ and\ \citenamefont
  {Heenen}}]{Bally:2014}%
  \BibitemOpen
  \bibfield  {author} {\bibinfo {author} {\bibfnamefont {B.}~\bibnamefont
  {Bally}}, \bibinfo {author} {\bibfnamefont {B.}~\bibnamefont {Avez}},
  \bibinfo {author} {\bibfnamefont {M.}~\bibnamefont {Bender}}, \ and\ \bibinfo
  {author} {\bibfnamefont {P.~H.}\ \bibnamefont {Heenen}},\ }\href {\doibase
  10.1103/PhysRevLett.113.162501} {\bibfield  {journal} {\bibinfo  {journal}
  {Phys. Rev. Lett.}\ }\textbf {\bibinfo {volume} {113}},\ \bibinfo {pages}
  {162501} (\bibinfo {year} {2014})},\ \Eprint {http://arxiv.org/abs/1406.5984}
  {arXiv:1406.5984 [nucl-th]} \BibitemShut {NoStop}%
\bibitem [{\citenamefont {Rodr\'\i{}guez}(2014)}]{Rodriguez:2014Kr}%
  \BibitemOpen
  \bibfield  {author} {\bibinfo {author} {\bibfnamefont {T.~R.}\ \bibnamefont
  {Rodr\'\i{}guez}},\ }\href {\doibase 10.1103/PhysRevC.90.034306} {\bibfield
  {journal} {\bibinfo  {journal} {Phys. Rev. C}\ }\textbf {\bibinfo {volume}
  {90}},\ \bibinfo {pages} {034306} (\bibinfo {year} {2014})},\ \Eprint
  {http://arxiv.org/abs/1408.5170} {arXiv:1408.5170 [nucl-th]} \BibitemShut
  {NoStop}%
\bibitem [{\citenamefont {Egido}\ \emph {et~al.}(2016)\citenamefont {Egido},
  \citenamefont {Borrajo},\ and\ \citenamefont
  {Rodr\'{\i}guez}}]{Egido:2016PRL}%
  \BibitemOpen
  \bibfield  {author} {\bibinfo {author} {\bibfnamefont {J.~L.}\ \bibnamefont
  {Egido}}, \bibinfo {author} {\bibfnamefont {M.}~\bibnamefont {Borrajo}}, \
  and\ \bibinfo {author} {\bibfnamefont {T.~R.}\ \bibnamefont
  {Rodr\'{\i}guez}},\ }\href {\doibase 10.1103/PhysRevLett.116.052502}
  {\bibfield  {journal} {\bibinfo  {journal} {Phys. Rev. Lett.}\ }\textbf
  {\bibinfo {volume} {116}},\ \bibinfo {pages} {052502} (\bibinfo {year}
  {2016})}\BibitemShut {NoStop}%
\bibitem [{\citenamefont {Yao}\ \emph {et~al.}(2015)\citenamefont {Yao},
  \citenamefont {Song}, \citenamefont {Hagino}, \citenamefont {Ring},\ and\
  \citenamefont {Meng}}]{Yao:2015}%
  \BibitemOpen
  \bibfield  {author} {\bibinfo {author} {\bibfnamefont {J.~M.}\ \bibnamefont
  {Yao}}, \bibinfo {author} {\bibfnamefont {L.~S.}\ \bibnamefont {Song}},
  \bibinfo {author} {\bibfnamefont {K.}~\bibnamefont {Hagino}}, \bibinfo
  {author} {\bibfnamefont {P.}~\bibnamefont {Ring}}, \ and\ \bibinfo {author}
  {\bibfnamefont {J.}~\bibnamefont {Meng}},\ }\href {\doibase
  10.1103/PhysRevC.91.024316} {\bibfield  {journal} {\bibinfo  {journal} {Phys.
  Rev. C}\ }\textbf {\bibinfo {volume} {91}},\ \bibinfo {pages} {024316}
  (\bibinfo {year} {2015})}\BibitemShut {NoStop}%
\bibitem [{\citenamefont {Zhou}\ \emph {et~al.}(2016)\citenamefont {Zhou},
  \citenamefont {Yao}, \citenamefont {Li}, \citenamefont {Meng},\ and\
  \citenamefont {Ring}}]{Zhou:2016}%
  \BibitemOpen
  \bibfield  {author} {\bibinfo {author} {\bibfnamefont {E.~F.}\ \bibnamefont
  {Zhou}}, \bibinfo {author} {\bibfnamefont {J.~M.}\ \bibnamefont {Yao}},
  \bibinfo {author} {\bibfnamefont {Z.~P.}\ \bibnamefont {Li}}, \bibinfo
  {author} {\bibfnamefont {J.}~\bibnamefont {Meng}}, \ and\ \bibinfo {author}
  {\bibfnamefont {P.}~\bibnamefont {Ring}},\ }\href {\doibase
  10.1016/j.physletb.2015.12.028} {\bibfield  {journal} {\bibinfo  {journal}
  {Phys. Lett. B}\ }\textbf {\bibinfo {volume} {753}},\ \bibinfo {pages} {227}
  (\bibinfo {year} {2016})},\ \Eprint {http://arxiv.org/abs/1510.05232}
  {arXiv:1510.05232 [nucl-th]} \BibitemShut {NoStop}%
\bibitem [{\citenamefont {Bernard}\ \emph {et~al.}(2016)\citenamefont
  {Bernard}, \citenamefont {Robledo},\ and\ \citenamefont
  {Rodr\'\i{}guez}}]{Bernard:2016}%
  \BibitemOpen
  \bibfield  {author} {\bibinfo {author} {\bibfnamefont {R.~N.}\ \bibnamefont
  {Bernard}}, \bibinfo {author} {\bibfnamefont {L.~M.}\ \bibnamefont
  {Robledo}}, \ and\ \bibinfo {author} {\bibfnamefont {T.~R.}\ \bibnamefont
  {Rodr\'\i{}guez}},\ }\href {\doibase 10.1103/PhysRevC.93.061302} {\bibfield
  {journal} {\bibinfo  {journal} {Phys. Rev. C}\ }\textbf {\bibinfo {volume}
  {93}},\ \bibinfo {pages} {061302} (\bibinfo {year} {2016})},\ \Eprint
  {http://arxiv.org/abs/1604.06706} {arXiv:1604.06706 [nucl-th]} \BibitemShut
  {NoStop}%
\bibitem [{\citenamefont {Borrajo}\ and\ \citenamefont
  {Egido}(2017)}]{Borrajo:2017PLB}%
  \BibitemOpen
  \bibfield  {author} {\bibinfo {author} {\bibfnamefont {M.}~\bibnamefont
  {Borrajo}}\ and\ \bibinfo {author} {\bibfnamefont {J.~L.}\ \bibnamefont
  {Egido}},\ }\href {\doibase 10.1016/j.physletb.2016.11.037} {\bibfield
  {journal} {\bibinfo  {journal} {Phys. Lett. B}\ }\textbf {\bibinfo {volume}
  {764}},\ \bibinfo {pages} {328} (\bibinfo {year} {2017})},\ \Eprint
  {http://arxiv.org/abs/1611.06982} {arXiv:1611.06982 [nucl-th]} \BibitemShut
  {NoStop}%
\bibitem [{\citenamefont {Rodr\'{\i}guez}\ and\ \citenamefont
  {Mart\'{\i}nez-Pinedo}(2010)}]{Rodriguez:2010}%
  \BibitemOpen
  \bibfield  {author} {\bibinfo {author} {\bibfnamefont {T.~R.}\ \bibnamefont
  {Rodr\'{\i}guez}}\ and\ \bibinfo {author} {\bibfnamefont {G.}~\bibnamefont
  {Mart\'{\i}nez-Pinedo}},\ }\href {\doibase 10.1103/PhysRevLett.105.252503}
  {\bibfield  {journal} {\bibinfo  {journal} {Phys. Rev. Lett.}\ }\textbf
  {\bibinfo {volume} {105}},\ \bibinfo {pages} {252503} (\bibinfo {year}
  {2010})}\BibitemShut {NoStop}%
\bibitem [{\citenamefont {Vaquero}\ \emph {et~al.}(2013)\citenamefont
  {Vaquero}, \citenamefont {Rodr\'{\i}guez},\ and\ \citenamefont
  {Egido}}]{Vaquero:2013}%
  \BibitemOpen
  \bibfield  {author} {\bibinfo {author} {\bibfnamefont {N.~L.}\ \bibnamefont
  {Vaquero}}, \bibinfo {author} {\bibfnamefont {T.~R.}\ \bibnamefont
  {Rodr\'{\i}guez}}, \ and\ \bibinfo {author} {\bibfnamefont {J.~L.}\
  \bibnamefont {Egido}},\ }\href {\doibase 10.1103/PhysRevLett.111.142501}
  {\bibfield  {journal} {\bibinfo  {journal} {Phys. Rev. Lett.}\ }\textbf
  {\bibinfo {volume} {111}},\ \bibinfo {pages} {142501} (\bibinfo {year}
  {2013})}\BibitemShut {NoStop}%
\bibitem [{\citenamefont {Song}\ \emph {et~al.}(2014)\citenamefont {Song},
  \citenamefont {Yao}, \citenamefont {Ring},\ and\ \citenamefont
  {Meng}}]{Song:2014}%
  \BibitemOpen
  \bibfield  {author} {\bibinfo {author} {\bibfnamefont {L.~S.}\ \bibnamefont
  {Song}}, \bibinfo {author} {\bibfnamefont {J.~M.}\ \bibnamefont {Yao}},
  \bibinfo {author} {\bibfnamefont {P.}~\bibnamefont {Ring}}, \ and\ \bibinfo
  {author} {\bibfnamefont {J.}~\bibnamefont {Meng}},\ }\href {\doibase
  https://doi.org/10.1103/PhysRevC.90.054309} {\bibfield  {journal} {\bibinfo
  {journal} {Phys. Rev. C}\ }\textbf {\bibinfo {volume} {90}},\ \bibinfo
  {pages} {054309} (\bibinfo {year} {2014})}\BibitemShut {NoStop}%
\bibitem [{\citenamefont {Yao}\ and\ \citenamefont
  {Engel}(2016)}]{Yao:2016PRC}%
  \BibitemOpen
  \bibfield  {author} {\bibinfo {author} {\bibfnamefont {J.~M.}\ \bibnamefont
  {Yao}}\ and\ \bibinfo {author} {\bibfnamefont {J.}~\bibnamefont {Engel}},\
  }\href {\doibase 10.1103/PhysRevC.94.014306} {\bibfield  {journal} {\bibinfo
  {journal} {Phys. Rev. C}\ }\textbf {\bibinfo {volume} {94}},\ \bibinfo
  {pages} {014306} (\bibinfo {year} {2016})}\BibitemShut {NoStop}%
\bibitem [{\citenamefont {Jiao}\ \emph {et~al.}(2017)\citenamefont {Jiao},
  \citenamefont {Engel},\ and\ \citenamefont {Holt}}]{Jiao:2017}%
  \BibitemOpen
  \bibfield  {author} {\bibinfo {author} {\bibfnamefont {C.~F.}\ \bibnamefont
  {Jiao}}, \bibinfo {author} {\bibfnamefont {J.}~\bibnamefont {Engel}}, \ and\
  \bibinfo {author} {\bibfnamefont {J.~D.}\ \bibnamefont {Holt}},\ }\href
  {\doibase 10.1103/PhysRevC.96.054310} {\bibfield  {journal} {\bibinfo
  {journal} {Phys. Rev. C}\ }\textbf {\bibinfo {volume} {96}},\ \bibinfo
  {pages} {054310} (\bibinfo {year} {2017})}\BibitemShut {NoStop}%
\bibitem [{\citenamefont {Yao}\ \emph {et~al.}(2018)\citenamefont {Yao},
  \citenamefont {Engel}, \citenamefont {Wang}, \citenamefont {Jiao},\ and\
  \citenamefont {Hergert}}]{Yao:2018wq}%
  \BibitemOpen
  \bibfield  {author} {\bibinfo {author} {\bibfnamefont {J.~M.}\ \bibnamefont
  {Yao}}, \bibinfo {author} {\bibfnamefont {J.}~\bibnamefont {Engel}}, \bibinfo
  {author} {\bibfnamefont {L.~J.}\ \bibnamefont {Wang}}, \bibinfo {author}
  {\bibfnamefont {C.~F.}\ \bibnamefont {Jiao}}, \ and\ \bibinfo {author}
  {\bibfnamefont {H.}~\bibnamefont {Hergert}},\ }\href {\doibase
  10.1103/PhysRevC.98.054311} {\bibfield  {journal} {\bibinfo  {journal} {Phys.
  Rev. C}\ }\textbf {\bibinfo {volume} {98}},\ \bibinfo {pages} {054311}
  (\bibinfo {year} {2018})}\BibitemShut {NoStop}%
\bibitem [{\citenamefont {Yao}\ \emph {et~al.}(2020)\citenamefont {Yao},
  \citenamefont {Bally}, \citenamefont {Engel}, \citenamefont {Wirth},
  \citenamefont {Rodr\'{\i}guez},\ and\ \citenamefont {Hergert}}]{Yao:2020PRL}%
  \BibitemOpen
  \bibfield  {author} {\bibinfo {author} {\bibfnamefont {J.~M.}\ \bibnamefont
  {Yao}}, \bibinfo {author} {\bibfnamefont {B.}~\bibnamefont {Bally}}, \bibinfo
  {author} {\bibfnamefont {J.}~\bibnamefont {Engel}}, \bibinfo {author}
  {\bibfnamefont {R.}~\bibnamefont {Wirth}}, \bibinfo {author} {\bibfnamefont
  {T.~R.}\ \bibnamefont {Rodr\'{\i}guez}}, \ and\ \bibinfo {author}
  {\bibfnamefont {H.}~\bibnamefont {Hergert}},\ }\href {\doibase
  10.1103/PhysRevLett.124.232501} {\bibfield  {journal} {\bibinfo  {journal}
  {Phys. Rev. Lett.}\ }\textbf {\bibinfo {volume} {124}},\ \bibinfo {pages}
  {232501} (\bibinfo {year} {2020})}\BibitemShut {NoStop}%
\bibitem [{\citenamefont {Engel}\ and\ \citenamefont
  {Men{\'{e}}ndez}(2017)}]{Engel:2017}%
  \BibitemOpen
  \bibfield  {author} {\bibinfo {author} {\bibfnamefont {J.}~\bibnamefont
  {Engel}}\ and\ \bibinfo {author} {\bibfnamefont {J.}~\bibnamefont
  {Men{\'{e}}ndez}},\ }\href {\doibase 10.1088/1361-6633/aa5bc5} {\bibfield
  {journal} {\bibinfo  {journal} {Rep. Prog. Phys.}\ }\textbf {\bibinfo
  {volume} {80}},\ \bibinfo {pages} {046301} (\bibinfo {year}
  {2017})}\BibitemShut {NoStop}%
\bibitem [{\citenamefont {Yao}\ \emph {et~al.}(2022{\natexlab{a}})\citenamefont
  {Yao}, \citenamefont {Meng}, \citenamefont {Niu},\ and\ \citenamefont
  {Ring}}]{Yao:2022PPNP}%
  \BibitemOpen
  \bibfield  {author} {\bibinfo {author} {\bibfnamefont {J.~M.}\ \bibnamefont
  {Yao}}, \bibinfo {author} {\bibfnamefont {J.}~\bibnamefont {Meng}}, \bibinfo
  {author} {\bibfnamefont {Y.~F.}\ \bibnamefont {Niu}}, \ and\ \bibinfo
  {author} {\bibfnamefont {P.}~\bibnamefont {Ring}},\ }\href {\doibase
  10.1016/j.ppnp.2022.103965} {\bibfield  {journal} {\bibinfo  {journal} {Prog.
  Part. Nucl. Phys.}\ }\textbf {\bibinfo {volume} {126}},\ \bibinfo {pages}
  {103965} (\bibinfo {year} {2022}{\natexlab{a}})},\ \Eprint
  {http://arxiv.org/abs/2111.15543} {arXiv:2111.15543 [nucl-th]} \BibitemShut
  {NoStop}%
\bibitem [{\citenamefont {Agostini}\ \emph {et~al.}(2022)\citenamefont
  {Agostini}, \citenamefont {Benato}, \citenamefont {Detwiler}, \citenamefont
  {Men\'endez},\ and\ \citenamefont {Vissani}}]{Agostini:2022RMP}%
  \BibitemOpen
  \bibfield  {author} {\bibinfo {author} {\bibfnamefont {M.}~\bibnamefont
  {Agostini}}, \bibinfo {author} {\bibfnamefont {G.}~\bibnamefont {Benato}},
  \bibinfo {author} {\bibfnamefont {J.~A.}\ \bibnamefont {Detwiler}}, \bibinfo
  {author} {\bibfnamefont {J.}~\bibnamefont {Men\'endez}}, \ and\ \bibinfo
  {author} {\bibfnamefont {F.}~\bibnamefont {Vissani}},\ }\href@noop {} {\
  (\bibinfo {year} {2022})},\ \Eprint {http://arxiv.org/abs/2202.01787}
  {arXiv:2202.01787 [hep-ex]} \BibitemShut {NoStop}%
\bibitem [{\citenamefont {Romero}\ \emph {et~al.}(2021)\citenamefont {Romero},
  \citenamefont {Yao}, \citenamefont {Bally}, \citenamefont {Rodr\'\i{}guez},\
  and\ \citenamefont {Engel}}]{Romero:2021PRC}%
  \BibitemOpen
  \bibfield  {author} {\bibinfo {author} {\bibfnamefont {A.~M.}\ \bibnamefont
  {Romero}}, \bibinfo {author} {\bibfnamefont {J.~M.}\ \bibnamefont {Yao}},
  \bibinfo {author} {\bibfnamefont {B.}~\bibnamefont {Bally}}, \bibinfo
  {author} {\bibfnamefont {T.~R.}\ \bibnamefont {Rodr\'\i{}guez}}, \ and\
  \bibinfo {author} {\bibfnamefont {J.}~\bibnamefont {Engel}},\ }\href
  {\doibase 10.1103/PhysRevC.104.054317} {\bibfield  {journal} {\bibinfo
  {journal} {Phys. Rev. C}\ }\textbf {\bibinfo {volume} {104}},\ \bibinfo
  {pages} {054317} (\bibinfo {year} {2021})},\ \Eprint
  {http://arxiv.org/abs/2105.03471} {arXiv:2105.03471 [nucl-th]} \BibitemShut
  {NoStop}%
\bibitem [{\citenamefont {Frosini}\ \emph
  {et~al.}(2022{\natexlab{a}})\citenamefont {Frosini}, \citenamefont {Duguet},
  \citenamefont {Ebran}, \citenamefont {Bally}, \citenamefont {Hergert},
  \citenamefont {Rodr\'\i{}guez}, \citenamefont {Roth}, \citenamefont {Yao},\
  and\ \citenamefont {Som\`a}}]{Frosini:2022EPJA}%
  \BibitemOpen
  \bibfield  {author} {\bibinfo {author} {\bibfnamefont {M.}~\bibnamefont
  {Frosini}}, \bibinfo {author} {\bibfnamefont {T.}~\bibnamefont {Duguet}},
  \bibinfo {author} {\bibfnamefont {J.-P.}\ \bibnamefont {Ebran}}, \bibinfo
  {author} {\bibfnamefont {B.}~\bibnamefont {Bally}}, \bibinfo {author}
  {\bibfnamefont {H.}~\bibnamefont {Hergert}}, \bibinfo {author} {\bibfnamefont
  {T.~R.}\ \bibnamefont {Rodr\'\i{}guez}}, \bibinfo {author} {\bibfnamefont
  {R.}~\bibnamefont {Roth}}, \bibinfo {author} {\bibfnamefont {J.}~\bibnamefont
  {Yao}}, \ and\ \bibinfo {author} {\bibfnamefont {V.}~\bibnamefont {Som\`a}},\
  }\href {\doibase 10.1140/epja/s10050-022-00694-x} {\bibfield  {journal}
  {\bibinfo  {journal} {Eur. Phys. J. A}\ }\textbf {\bibinfo {volume} {58}},\
  \bibinfo {pages} {64} (\bibinfo {year} {2022}{\natexlab{a}})},\ \Eprint
  {http://arxiv.org/abs/2111.01461} {arXiv:2111.01461 [nucl-th]} \BibitemShut
  {NoStop}%
\bibitem [{\citenamefont {Frosini}\ \emph
  {et~al.}(2022{\natexlab{b}})\citenamefont {Frosini}, \citenamefont {Duguet},
  \citenamefont {Ebran}, \citenamefont {Bally}, \citenamefont {Mongelli},
  \citenamefont {Rodr{\'\i}guez}, \citenamefont {Roth},\ and\ \citenamefont
  {Som{\`a}}}]{Frosini:2022xg}%
  \BibitemOpen
  \bibfield  {author} {\bibinfo {author} {\bibfnamefont {M.}~\bibnamefont
  {Frosini}}, \bibinfo {author} {\bibfnamefont {T.}~\bibnamefont {Duguet}},
  \bibinfo {author} {\bibfnamefont {J.~P.}\ \bibnamefont {Ebran}}, \bibinfo
  {author} {\bibfnamefont {B.}~\bibnamefont {Bally}}, \bibinfo {author}
  {\bibfnamefont {T.}~\bibnamefont {Mongelli}}, \bibinfo {author}
  {\bibfnamefont {T.~R.}\ \bibnamefont {Rodr{\'\i}guez}}, \bibinfo {author}
  {\bibfnamefont {R.}~\bibnamefont {Roth}}, \ and\ \bibinfo {author}
  {\bibfnamefont {V.}~\bibnamefont {Som{\`a}}},\ }\href
  {https://doi.org/10.1140/epja/s10050-022-00693-y} {\bibfield  {journal}
  {\bibinfo  {journal} {Eur. Phys. J. A}\ }\textbf {\bibinfo {volume} {58}}
  (\bibinfo {year} {2022}{\natexlab{b}})}\BibitemShut {NoStop}%
\bibitem [{\citenamefont {Frosini}\ \emph
  {et~al.}(2022{\natexlab{c}})\citenamefont {Frosini}, \citenamefont {Duguet},
  \citenamefont {Ebran}, \citenamefont {Bally}, \citenamefont {Hergert},
  \citenamefont {Rodr\'\i{}guez}, \citenamefont {Roth}, \citenamefont {Yao},\
  and\ \citenamefont {Som\`a}}]{Frosini:2022EPJA3}%
  \BibitemOpen
  \bibfield  {author} {\bibinfo {author} {\bibfnamefont {M.}~\bibnamefont
  {Frosini}}, \bibinfo {author} {\bibfnamefont {T.}~\bibnamefont {Duguet}},
  \bibinfo {author} {\bibfnamefont {J.-P.}\ \bibnamefont {Ebran}}, \bibinfo
  {author} {\bibfnamefont {B.}~\bibnamefont {Bally}}, \bibinfo {author}
  {\bibfnamefont {H.}~\bibnamefont {Hergert}}, \bibinfo {author} {\bibfnamefont
  {T.~R.}\ \bibnamefont {Rodr\'\i{}guez}}, \bibinfo {author} {\bibfnamefont
  {R.}~\bibnamefont {Roth}}, \bibinfo {author} {\bibfnamefont {J.~M.}\
  \bibnamefont {Yao}}, \ and\ \bibinfo {author} {\bibfnamefont
  {V.}~\bibnamefont {Som\`a}},\ }\href {\doibase
  10.1140/epja/s10050-022-00694-x} {\bibfield  {journal} {\bibinfo  {journal}
  {Eur. Phys. J. A}\ }\textbf {\bibinfo {volume} {58}},\ \bibinfo {pages} {64}
  (\bibinfo {year} {2022}{\natexlab{c}})},\ \Eprint
  {http://arxiv.org/abs/2111.01461} {arXiv:2111.01461 [nucl-th]} \BibitemShut
  {NoStop}%
\bibitem [{\citenamefont {Duguet}\ \emph {et~al.}(2022)\citenamefont {Duguet},
  \citenamefont {Ebran}, \citenamefont {Frosini}, \citenamefont {Hergert},\
  and\ \citenamefont {Som\`a}}]{Duguet:2022}%
  \BibitemOpen
  \bibfield  {author} {\bibinfo {author} {\bibfnamefont {T.}~\bibnamefont
  {Duguet}}, \bibinfo {author} {\bibfnamefont {J.~P.}\ \bibnamefont {Ebran}},
  \bibinfo {author} {\bibfnamefont {M.}~\bibnamefont {Frosini}}, \bibinfo
  {author} {\bibfnamefont {H.}~\bibnamefont {Hergert}}, \ and\ \bibinfo
  {author} {\bibfnamefont {V.}~\bibnamefont {Som\`a}},\ }\href@noop {} {\
  (\bibinfo {year} {2022})},\ \Eprint {http://arxiv.org/abs/2209.03424}
  {arXiv:2209.03424 [nucl-th]} \BibitemShut {NoStop}%
\bibitem [{\citenamefont {Bender}\ \emph {et~al.}(2005)\citenamefont {Bender},
  \citenamefont {Bertsch},\ and\ \citenamefont {Heenen}}]{Bender:2004Global}%
  \BibitemOpen
  \bibfield  {author} {\bibinfo {author} {\bibfnamefont {M.}~\bibnamefont
  {Bender}}, \bibinfo {author} {\bibfnamefont {G.~F.}\ \bibnamefont {Bertsch}},
  \ and\ \bibinfo {author} {\bibfnamefont {P.~H.}\ \bibnamefont {Heenen}},\
  }\href {\doibase 10.1103/PhysRevLett.94.102503} {\bibfield  {journal}
  {\bibinfo  {journal} {Phys. Rev. Lett.}\ }\textbf {\bibinfo {volume} {94}},\
  \bibinfo {pages} {102503} (\bibinfo {year} {2005})},\ \Eprint
  {http://arxiv.org/abs/nucl-th/0410023} {arXiv:nucl-th/0410023} \BibitemShut
  {NoStop}%
\bibitem [{\citenamefont {Rodr\'\i{}guez}\ \emph {et~al.}(2015)\citenamefont
  {Rodr\'\i{}guez}, \citenamefont {Arzhanov},\ and\ \citenamefont
  {Mart\'\i{}nez-Pinedo}}]{Rodriguez:2014Global}%
  \BibitemOpen
  \bibfield  {author} {\bibinfo {author} {\bibfnamefont {T.~R.}\ \bibnamefont
  {Rodr\'\i{}guez}}, \bibinfo {author} {\bibfnamefont {A.}~\bibnamefont
  {Arzhanov}}, \ and\ \bibinfo {author} {\bibfnamefont {G.}~\bibnamefont
  {Mart\'\i{}nez-Pinedo}},\ }\href {\doibase 10.1103/PhysRevC.91.044315}
  {\bibfield  {journal} {\bibinfo  {journal} {Phys. Rev. C}\ }\textbf {\bibinfo
  {volume} {91}},\ \bibinfo {pages} {044315} (\bibinfo {year} {2015})},\
  \Eprint {http://arxiv.org/abs/1407.7699} {arXiv:1407.7699 [nucl-th]}
  \BibitemShut {NoStop}%
\bibitem [{\citenamefont {Broeckhove}\ and\ \citenamefont
  {Deumens}(1979)}]{Broeckhove:1979}%
  \BibitemOpen
  \bibfield  {author} {\bibinfo {author} {\bibfnamefont {J.}~\bibnamefont
  {Broeckhove}}\ and\ \bibinfo {author} {\bibfnamefont {E.}~\bibnamefont
  {Deumens}},\ }\href {\doibase 10.1007/BF01547468} {\bibfield  {journal}
  {\bibinfo  {journal} {Z. Phys. A}\ }\textbf {\bibinfo {volume} {292}},\
  \bibinfo {pages} {243} (\bibinfo {year} {1979})}\BibitemShut {NoStop}%
\bibitem [{\citenamefont {Mart\'\i{}nez-Larraz}\ and\ \citenamefont
  {Rodr\'\i{}guez}(2022)}]{Martinez-Larraz:2022}%
  \BibitemOpen
  \bibfield  {author} {\bibinfo {author} {\bibfnamefont {J.}~\bibnamefont
  {Mart\'\i{}nez-Larraz}}\ and\ \bibinfo {author} {\bibfnamefont {T.~R.}\
  \bibnamefont {Rodr\'\i{}guez}},\ }\href@noop {} {\  (\bibinfo {year}
  {2022})},\ \Eprint {http://arxiv.org/abs/2208.10870} {arXiv:2208.10870
  [nucl-th]} \BibitemShut {NoStop}%
\bibitem [{\citenamefont {Frame}\ \emph {et~al.}(2018)\citenamefont {Frame},
  \citenamefont {He}, \citenamefont {Ipsen}, \citenamefont {Lee}, \citenamefont
  {Lee},\ and\ \citenamefont {Rrapaj}}]{Frame:2018PRL}%
  \BibitemOpen
  \bibfield  {author} {\bibinfo {author} {\bibfnamefont {D.}~\bibnamefont
  {Frame}}, \bibinfo {author} {\bibfnamefont {R.}~\bibnamefont {He}}, \bibinfo
  {author} {\bibfnamefont {I.}~\bibnamefont {Ipsen}}, \bibinfo {author}
  {\bibfnamefont {D.}~\bibnamefont {Lee}}, \bibinfo {author} {\bibfnamefont
  {D.}~\bibnamefont {Lee}}, \ and\ \bibinfo {author} {\bibfnamefont
  {E.}~\bibnamefont {Rrapaj}},\ }\href {\doibase
  10.1103/PhysRevLett.121.032501} {\bibfield  {journal} {\bibinfo  {journal}
  {Phys. Rev. Lett.}\ }\textbf {\bibinfo {volume} {121}},\ \bibinfo {pages}
  {032501} (\bibinfo {year} {2018})}\BibitemShut {NoStop}%
\bibitem [{\citenamefont {Sarkar}\ and\ \citenamefont
  {Lee}(2021)}]{Sarkar:2021}%
  \BibitemOpen
  \bibfield  {author} {\bibinfo {author} {\bibfnamefont {A.}~\bibnamefont
  {Sarkar}}\ and\ \bibinfo {author} {\bibfnamefont {D.}~\bibnamefont {Lee}},\
  }\href {\doibase 10.1103/PhysRevLett.126.032501} {\bibfield  {journal}
  {\bibinfo  {journal} {Phys. Rev. Lett.}\ }\textbf {\bibinfo {volume} {126}},\
  \bibinfo {pages} {032501} (\bibinfo {year} {2021})}\BibitemShut {NoStop}%
\bibitem [{\citenamefont {Ekstr\"om}\ and\ \citenamefont
  {Hagen}(2019)}]{Ekstrom:2019PRL}%
  \BibitemOpen
  \bibfield  {author} {\bibinfo {author} {\bibfnamefont {A.}~\bibnamefont
  {Ekstr\"om}}\ and\ \bibinfo {author} {\bibfnamefont {G.}~\bibnamefont
  {Hagen}},\ }\href {\doibase 10.1103/PhysRevLett.123.252501} {\bibfield
  {journal} {\bibinfo  {journal} {Phys. Rev. Lett.}\ }\textbf {\bibinfo
  {volume} {123}},\ \bibinfo {pages} {252501} (\bibinfo {year} {2019})},\
  \Eprint {http://arxiv.org/abs/1910.02922} {arXiv:1910.02922 [nucl-th]}
  \BibitemShut {NoStop}%
\bibitem [{\citenamefont {K{\"o}nig}\ \emph {et~al.}(2020)\citenamefont
  {K{\"o}nig}, \citenamefont {Ekstr{\"o}m}, \citenamefont {Hebeler},
  \citenamefont {Lee},\ and\ \citenamefont {Schwenk}}]{Konig:2020fn}%
  \BibitemOpen
  \bibfield  {author} {\bibinfo {author} {\bibfnamefont {S.}~\bibnamefont
  {K{\"o}nig}}, \bibinfo {author} {\bibfnamefont {A.}~\bibnamefont
  {Ekstr{\"o}m}}, \bibinfo {author} {\bibfnamefont {K.}~\bibnamefont
  {Hebeler}}, \bibinfo {author} {\bibfnamefont {D.}~\bibnamefont {Lee}}, \ and\
  \bibinfo {author} {\bibfnamefont {A.}~\bibnamefont {Schwenk}},\ }\href
  {\doibase https://doi.org/10.1016/j.physletb.2020.135814} {\bibfield
  {journal} {\bibinfo  {journal} {Physics Letters B}\ }\textbf {\bibinfo
  {volume} {810}},\ \bibinfo {pages} {135814} (\bibinfo {year}
  {2020})}\BibitemShut {NoStop}%
\bibitem [{\citenamefont {Furnstahl}\ \emph {et~al.}(2020)\citenamefont
  {Furnstahl}, \citenamefont {Garcia}, \citenamefont {Millican},\ and\
  \citenamefont {Zhang}}]{Furnstahl:2020PLB}%
  \BibitemOpen
  \bibfield  {author} {\bibinfo {author} {\bibfnamefont {R.~J.}\ \bibnamefont
  {Furnstahl}}, \bibinfo {author} {\bibfnamefont {A.~J.}\ \bibnamefont
  {Garcia}}, \bibinfo {author} {\bibfnamefont {P.~J.}\ \bibnamefont
  {Millican}}, \ and\ \bibinfo {author} {\bibfnamefont {X.}~\bibnamefont
  {Zhang}},\ }\href {\doibase 10.1016/j.physletb.2020.135719} {\bibfield
  {journal} {\bibinfo  {journal} {Phys. Lett. B}\ }\textbf {\bibinfo {volume}
  {809}},\ \bibinfo {pages} {135719} (\bibinfo {year} {2020})},\ \Eprint
  {http://arxiv.org/abs/2007.03635} {arXiv:2007.03635 [nucl-th]} \BibitemShut
  {NoStop}%
\bibitem [{\citenamefont {Drischler}\ \emph {et~al.}(2021)\citenamefont
  {Drischler}, \citenamefont {Quinonez}, \citenamefont {Giuliani},
  \citenamefont {Lovell},\ and\ \citenamefont {Nunes}}]{Drischler:2021PLB}%
  \BibitemOpen
  \bibfield  {author} {\bibinfo {author} {\bibfnamefont {C.}~\bibnamefont
  {Drischler}}, \bibinfo {author} {\bibfnamefont {M.}~\bibnamefont {Quinonez}},
  \bibinfo {author} {\bibfnamefont {P.~G.}\ \bibnamefont {Giuliani}}, \bibinfo
  {author} {\bibfnamefont {A.~E.}\ \bibnamefont {Lovell}}, \ and\ \bibinfo
  {author} {\bibfnamefont {F.~M.}\ \bibnamefont {Nunes}},\ }\href {\doibase
  10.1016/j.physletb.2021.136777} {\bibfield  {journal} {\bibinfo  {journal}
  {Phys. Lett. B}\ }\textbf {\bibinfo {volume} {823}},\ \bibinfo {pages}
  {136777} (\bibinfo {year} {2021})},\ \Eprint
  {http://arxiv.org/abs/2108.08269} {arXiv:2108.08269 [nucl-th]} \BibitemShut
  {NoStop}%
\bibitem [{\citenamefont {Bai}\ and\ \citenamefont {Ren}(2021)}]{Bai:2021PRC}%
  \BibitemOpen
  \bibfield  {author} {\bibinfo {author} {\bibfnamefont {D.}~\bibnamefont
  {Bai}}\ and\ \bibinfo {author} {\bibfnamefont {Z.}~\bibnamefont {Ren}},\
  }\href {\doibase 10.1103/PhysRevC.103.014612} {\bibfield  {journal} {\bibinfo
   {journal} {Phys. Rev. C}\ }\textbf {\bibinfo {volume} {103}},\ \bibinfo
  {pages} {014612} (\bibinfo {year} {2021})},\ \Eprint
  {http://arxiv.org/abs/2101.06336} {arXiv:2101.06336 [nucl-th]} \BibitemShut
  {NoStop}%
\bibitem [{\citenamefont {Kanada-En'yo}(1998)}]{Kanada-Enyo:1998PRL}%
  \BibitemOpen
  \bibfield  {author} {\bibinfo {author} {\bibfnamefont {Y.}~\bibnamefont
  {Kanada-En'yo}},\ }\href {\doibase 10.1103/PhysRevLett.81.5291} {\bibfield
  {journal} {\bibinfo  {journal} {Phys. Rev. Lett.}\ }\textbf {\bibinfo
  {volume} {81}},\ \bibinfo {pages} {5291} (\bibinfo {year}
  {1998})}\BibitemShut {NoStop}%
\bibitem [{\citenamefont {Ohta}\ \emph {et~al.}(2004)\citenamefont {Ohta},
  \citenamefont {Yabana},\ and\ \citenamefont {Nakatsukasa}}]{Ohta:2004PRC}%
  \BibitemOpen
  \bibfield  {author} {\bibinfo {author} {\bibfnamefont {H.}~\bibnamefont
  {Ohta}}, \bibinfo {author} {\bibfnamefont {K.}~\bibnamefont {Yabana}}, \ and\
  \bibinfo {author} {\bibfnamefont {T.}~\bibnamefont {Nakatsukasa}},\ }\href
  {\doibase 10.1103/PhysRevC.70.014301} {\bibfield  {journal} {\bibinfo
  {journal} {Phys. Rev. C}\ }\textbf {\bibinfo {volume} {70}},\ \bibinfo
  {pages} {014301} (\bibinfo {year} {2004})}\BibitemShut {NoStop}%
\bibitem [{\citenamefont {Gao}(2022)}]{Gao:2022PLB}%
  \BibitemOpen
  \bibfield  {author} {\bibinfo {author} {\bibfnamefont {Z.-C.}\ \bibnamefont
  {Gao}},\ }\href {\doibase 10.1016/j.physletb.2021.136795} {\bibfield
  {journal} {\bibinfo  {journal} {Phys. Lett. B}\ }\textbf {\bibinfo {volume}
  {824}},\ \bibinfo {pages} {136795} (\bibinfo {year} {2022})},\ \Eprint
  {http://arxiv.org/abs/2102.04044} {arXiv:2102.04044 [nucl-th]} \BibitemShut
  {NoStop}%
\bibitem [{\citenamefont {Otsuka}\ \emph {et~al.}(2001)\citenamefont {Otsuka},
  \citenamefont {Honma}, \citenamefont {Mizusaki}, \citenamefont {Shimizu},\
  and\ \citenamefont {Utsuno}}]{Otsuka:2001PPNP}%
  \BibitemOpen
  \bibfield  {author} {\bibinfo {author} {\bibfnamefont {T.}~\bibnamefont
  {Otsuka}}, \bibinfo {author} {\bibfnamefont {M.}~\bibnamefont {Honma}},
  \bibinfo {author} {\bibfnamefont {T.}~\bibnamefont {Mizusaki}}, \bibinfo
  {author} {\bibfnamefont {N.}~\bibnamefont {Shimizu}}, \ and\ \bibinfo
  {author} {\bibfnamefont {Y.}~\bibnamefont {Utsuno}},\ }\href {\doibase
  https://doi.org/10.1016/S0146-6410(01)00157-0} {\bibfield  {journal}
  {\bibinfo  {journal} {Prog. Part. Nucl. Phys.}\ }\textbf {\bibinfo {volume}
  {47}},\ \bibinfo {pages} {319} (\bibinfo {year} {2001})}\BibitemShut
  {NoStop}%
\bibitem [{\citenamefont {Shinohara}\ \emph {et~al.}(2006)\citenamefont
  {Shinohara}, \citenamefont {Ohta}, \citenamefont {Nakatsukasa},\ and\
  \citenamefont {Yabana}}]{Shinohara:2006PRC}%
  \BibitemOpen
  \bibfield  {author} {\bibinfo {author} {\bibfnamefont {S.}~\bibnamefont
  {Shinohara}}, \bibinfo {author} {\bibfnamefont {H.}~\bibnamefont {Ohta}},
  \bibinfo {author} {\bibfnamefont {T.}~\bibnamefont {Nakatsukasa}}, \ and\
  \bibinfo {author} {\bibfnamefont {K.}~\bibnamefont {Yabana}},\ }\href
  {\doibase 10.1103/PhysRevC.74.054315} {\bibfield  {journal} {\bibinfo
  {journal} {Phys. Rev. C}\ }\textbf {\bibinfo {volume} {74}},\ \bibinfo
  {pages} {054315} (\bibinfo {year} {2006})},\ \Eprint
  {http://arxiv.org/abs/nucl-th/0607004} {arXiv:nucl-th/0607004} \BibitemShut
  {NoStop}%
\bibitem [{\citenamefont {Ichikawa}\ and\ \citenamefont
  {Itagaki}(2022)}]{Ichikawa:2021}%
  \BibitemOpen
  \bibfield  {author} {\bibinfo {author} {\bibfnamefont {T.}~\bibnamefont
  {Ichikawa}}\ and\ \bibinfo {author} {\bibfnamefont {N.}~\bibnamefont
  {Itagaki}},\ }\href {\doibase 10.1103/PhysRevC.105.024314} {\bibfield
  {journal} {\bibinfo  {journal} {Phys. Rev. C}\ }\textbf {\bibinfo {volume}
  {105}},\ \bibinfo {pages} {024314} (\bibinfo {year} {2022})},\ \Eprint
  {http://arxiv.org/abs/2110.12869} {arXiv:2110.12869 [nucl-th]} \BibitemShut
  {NoStop}%
\bibitem [{\citenamefont {Jiao}\ and\ \citenamefont
  {Johnson}(2019)}]{Jiao:2019PRC}%
  \BibitemOpen
  \bibfield  {author} {\bibinfo {author} {\bibfnamefont {C.}~\bibnamefont
  {Jiao}}\ and\ \bibinfo {author} {\bibfnamefont {C.~W.}\ \bibnamefont
  {Johnson}},\ }\href {\doibase 10.1103/PhysRevC.100.031303} {\bibfield
  {journal} {\bibinfo  {journal} {Phys. Rev. C}\ }\textbf {\bibinfo {volume}
  {100}},\ \bibinfo {pages} {031303} (\bibinfo {year} {2019})}\BibitemShut
  {NoStop}%
\bibitem [{\citenamefont {Dao}\ and\ \citenamefont {Nowacki}(2022)}]{Dao:2022}%
  \BibitemOpen
  \bibfield  {author} {\bibinfo {author} {\bibfnamefont {D.~D.}\ \bibnamefont
  {Dao}}\ and\ \bibinfo {author} {\bibfnamefont {F.}~\bibnamefont {Nowacki}},\
  }\href {\doibase 10.1103/PhysRevC.105.054314} {\bibfield  {journal} {\bibinfo
   {journal} {Phys. Rev. C}\ }\textbf {\bibinfo {volume} {105}},\ \bibinfo
  {pages} {054314} (\bibinfo {year} {2022})}\BibitemShut {NoStop}%
\bibitem [{\citenamefont {Boehnlein}\ \emph {et~al.}(2022)\citenamefont
  {Boehnlein}, \citenamefont {Diefenthaler}, \citenamefont {Sato},
  \citenamefont {Schram}, \citenamefont {Ziegler}, \citenamefont {Fanelli},
  \citenamefont {Hjorth-Jensen}, \citenamefont {Horn}, \citenamefont {Kuchera},
  \citenamefont {Lee}, \citenamefont {Nazarewicz}, \citenamefont {Ostroumov},
  \citenamefont {Orginos}, \citenamefont {Poon}, \citenamefont {Wang},
  \citenamefont {Scheinker}, \citenamefont {Smith},\ and\ \citenamefont
  {Pang}}]{Boehnlein:2022RMP}%
  \BibitemOpen
  \bibfield  {author} {\bibinfo {author} {\bibfnamefont {A.}~\bibnamefont
  {Boehnlein}}, \bibinfo {author} {\bibfnamefont {M.}~\bibnamefont
  {Diefenthaler}}, \bibinfo {author} {\bibfnamefont {N.}~\bibnamefont {Sato}},
  \bibinfo {author} {\bibfnamefont {M.}~\bibnamefont {Schram}}, \bibinfo
  {author} {\bibfnamefont {V.}~\bibnamefont {Ziegler}}, \bibinfo {author}
  {\bibfnamefont {C.}~\bibnamefont {Fanelli}}, \bibinfo {author} {\bibfnamefont
  {M.}~\bibnamefont {Hjorth-Jensen}}, \bibinfo {author} {\bibfnamefont
  {T.}~\bibnamefont {Horn}}, \bibinfo {author} {\bibfnamefont {M.~P.}\
  \bibnamefont {Kuchera}}, \bibinfo {author} {\bibfnamefont {D.}~\bibnamefont
  {Lee}}, \bibinfo {author} {\bibfnamefont {W.}~\bibnamefont {Nazarewicz}},
  \bibinfo {author} {\bibfnamefont {P.}~\bibnamefont {Ostroumov}}, \bibinfo
  {author} {\bibfnamefont {K.}~\bibnamefont {Orginos}}, \bibinfo {author}
  {\bibfnamefont {A.}~\bibnamefont {Poon}}, \bibinfo {author} {\bibfnamefont
  {X.-N.}\ \bibnamefont {Wang}}, \bibinfo {author} {\bibfnamefont
  {A.}~\bibnamefont {Scheinker}}, \bibinfo {author} {\bibfnamefont {M.~S.}\
  \bibnamefont {Smith}}, \ and\ \bibinfo {author} {\bibfnamefont {L.-G.}\
  \bibnamefont {Pang}},\ }\href {\doibase 10.1103/RevModPhys.94.031003}
  {\bibfield  {journal} {\bibinfo  {journal} {Rev. Mod. Phys.}\ }\textbf
  {\bibinfo {volume} {94}},\ \bibinfo {pages} {031003} (\bibinfo {year}
  {2022})}\BibitemShut {NoStop}%
\bibitem [{\citenamefont {Utama}\ \emph
  {et~al.}(2016{\natexlab{a}})\citenamefont {Utama}, \citenamefont
  {Piekarewicz},\ and\ \citenamefont {Prosper}}]{Utama:2016PRC}%
  \BibitemOpen
  \bibfield  {author} {\bibinfo {author} {\bibfnamefont {R.}~\bibnamefont
  {Utama}}, \bibinfo {author} {\bibfnamefont {J.}~\bibnamefont {Piekarewicz}},
  \ and\ \bibinfo {author} {\bibfnamefont {H.~B.}\ \bibnamefont {Prosper}},\
  }\href {\doibase 10.1103/PhysRevC.93.014311} {\bibfield  {journal} {\bibinfo
  {journal} {Phys. Rev. C}\ }\textbf {\bibinfo {volume} {93}},\ \bibinfo
  {pages} {014311} (\bibinfo {year} {2016}{\natexlab{a}})},\ \Eprint
  {http://arxiv.org/abs/1508.06263} {arXiv:1508.06263 [nucl-th]} \BibitemShut
  {NoStop}%
\bibitem [{\citenamefont {Niu}\ and\ \citenamefont
  {Liang}(2018)}]{Niu:2018PLB}%
  \BibitemOpen
  \bibfield  {author} {\bibinfo {author} {\bibfnamefont {Z.~M.}\ \bibnamefont
  {Niu}}\ and\ \bibinfo {author} {\bibfnamefont {H.~Z.}\ \bibnamefont
  {Liang}},\ }\href {\doibase 10.1016/j.physletb.2018.01.002} {\bibfield
  {journal} {\bibinfo  {journal} {Phys. Lett. B}\ }\textbf {\bibinfo {volume}
  {778}},\ \bibinfo {pages} {48} (\bibinfo {year} {2018})},\ \Eprint
  {http://arxiv.org/abs/1801.04411} {arXiv:1801.04411 [nucl-th]} \BibitemShut
  {NoStop}%
\bibitem [{\citenamefont {Neufcourt}\ \emph {et~al.}(2018)\citenamefont
  {Neufcourt}, \citenamefont {Cao}, \citenamefont {Nazarewicz},\ and\
  \citenamefont {Viens}}]{Neufcourt:2018PLB}%
  \BibitemOpen
  \bibfield  {author} {\bibinfo {author} {\bibfnamefont {L.}~\bibnamefont
  {Neufcourt}}, \bibinfo {author} {\bibfnamefont {Y.}~\bibnamefont {Cao}},
  \bibinfo {author} {\bibfnamefont {W.}~\bibnamefont {Nazarewicz}}, \ and\
  \bibinfo {author} {\bibfnamefont {F.}~\bibnamefont {Viens}},\ }\href
  {\doibase 10.1103/PhysRevC.98.034318} {\bibfield  {journal} {\bibinfo
  {journal} {Phys. Rev. C}\ }\textbf {\bibinfo {volume} {98}},\ \bibinfo
  {pages} {034318} (\bibinfo {year} {2018})},\ \Eprint
  {http://arxiv.org/abs/1806.00552} {arXiv:1806.00552 [nucl-th]} \BibitemShut
  {NoStop}%
\bibitem [{\citenamefont {Niu}\ and\ \citenamefont
  {Liang}(2022)}]{Niu:2022PRC}%
  \BibitemOpen
  \bibfield  {author} {\bibinfo {author} {\bibfnamefont {Z.~M.}\ \bibnamefont
  {Niu}}\ and\ \bibinfo {author} {\bibfnamefont {H.~Z.}\ \bibnamefont
  {Liang}},\ }\href {\doibase 10.1103/PhysRevC.106.L021303} {\bibfield
  {journal} {\bibinfo  {journal} {Phys. Rev. C}\ }\textbf {\bibinfo {volume}
  {106}},\ \bibinfo {pages} {L021303} (\bibinfo {year} {2022})},\ \Eprint
  {http://arxiv.org/abs/2208.04783} {arXiv:2208.04783 [nucl-th]} \BibitemShut
  {NoStop}%
\bibitem [{\citenamefont {Wu}\ \emph {et~al.}(2022{\natexlab{a}})\citenamefont
  {Wu}, \citenamefont {Lu},\ and\ \citenamefont {Zhao}}]{Wu:2022PLB}%
  \BibitemOpen
  \bibfield  {author} {\bibinfo {author} {\bibfnamefont {X.~H.}\ \bibnamefont
  {Wu}}, \bibinfo {author} {\bibfnamefont {Y.~Y.}\ \bibnamefont {Lu}}, \ and\
  \bibinfo {author} {\bibfnamefont {P.~W.}\ \bibnamefont {Zhao}},\ }\href
  {\doibase 10.1016/j.physletb.2022.137394} {\bibfield  {journal} {\bibinfo
  {journal} {Phys. Lett. B}\ }\textbf {\bibinfo {volume} {834}},\ \bibinfo
  {pages} {137394} (\bibinfo {year} {2022}{\natexlab{a}})},\ \Eprint
  {http://arxiv.org/abs/2208.13966} {arXiv:2208.13966 [nucl-th]} \BibitemShut
  {NoStop}%
\bibitem [{\citenamefont {Utama}\ \emph
  {et~al.}(2016{\natexlab{b}})\citenamefont {Utama}, \citenamefont {Chen},\
  and\ \citenamefont {Piekarewicz}}]{Utama:2016}%
  \BibitemOpen
  \bibfield  {author} {\bibinfo {author} {\bibfnamefont {R.}~\bibnamefont
  {Utama}}, \bibinfo {author} {\bibfnamefont {W.-C.}\ \bibnamefont {Chen}}, \
  and\ \bibinfo {author} {\bibfnamefont {J.}~\bibnamefont {Piekarewicz}},\
  }\href {\doibase 10.1088/0954-3899/43/11/114002} {\bibfield  {journal}
  {\bibinfo  {journal} {J. Phys. G}\ }\textbf {\bibinfo {volume} {43}},\
  \bibinfo {pages} {114002} (\bibinfo {year} {2016}{\natexlab{b}})},\ \Eprint
  {http://arxiv.org/abs/1608.03020} {arXiv:1608.03020 [nucl-th]} \BibitemShut
  {NoStop}%
\bibitem [{\citenamefont {Wu}\ \emph {et~al.}(2020)\citenamefont {Wu},
  \citenamefont {Bai}, \citenamefont {Sagawa},\ and\ \citenamefont
  {Zhang}}]{Wu:2020}%
  \BibitemOpen
  \bibfield  {author} {\bibinfo {author} {\bibfnamefont {D.}~\bibnamefont
  {Wu}}, \bibinfo {author} {\bibfnamefont {C.~L.}\ \bibnamefont {Bai}},
  \bibinfo {author} {\bibfnamefont {H.}~\bibnamefont {Sagawa}}, \ and\ \bibinfo
  {author} {\bibfnamefont {H.~Q.}\ \bibnamefont {Zhang}},\ }\href {\doibase
  10.1103/PhysRevC.102.054323} {\bibfield  {journal} {\bibinfo  {journal}
  {Phys. Rev. C}\ }\textbf {\bibinfo {volume} {102}},\ \bibinfo {pages}
  {054323} (\bibinfo {year} {2020})},\ \Eprint
  {http://arxiv.org/abs/2006.09677} {arXiv:2006.09677 [nucl-th]} \BibitemShut
  {NoStop}%
\bibitem [{\citenamefont {Dong}\ \emph {et~al.}(2022)\citenamefont {Dong},
  \citenamefont {An}, \citenamefont {Lu},\ and\ \citenamefont
  {Geng}}]{Dong:2022PRC}%
  \BibitemOpen
  \bibfield  {author} {\bibinfo {author} {\bibfnamefont {X.-X.}\ \bibnamefont
  {Dong}}, \bibinfo {author} {\bibfnamefont {R.}~\bibnamefont {An}}, \bibinfo
  {author} {\bibfnamefont {J.-X.}\ \bibnamefont {Lu}}, \ and\ \bibinfo {author}
  {\bibfnamefont {L.-S.}\ \bibnamefont {Geng}},\ }\href {\doibase
  10.1103/PhysRevC.105.014308} {\bibfield  {journal} {\bibinfo  {journal}
  {Phys. Rev. C}\ }\textbf {\bibinfo {volume} {105}},\ \bibinfo {pages}
  {014308} (\bibinfo {year} {2022})},\ \Eprint
  {http://arxiv.org/abs/2109.09626} {arXiv:2109.09626 [nucl-th]} \BibitemShut
  {NoStop}%
\bibitem [{\citenamefont {Costiris}\ \emph {et~al.}(2009)\citenamefont
  {Costiris}, \citenamefont {Mavrommatis}, \citenamefont {Gernoth},\ and\
  \citenamefont {Clark}}]{Costiris:2009PRC}%
  \BibitemOpen
  \bibfield  {author} {\bibinfo {author} {\bibfnamefont {N.~J.}\ \bibnamefont
  {Costiris}}, \bibinfo {author} {\bibfnamefont {E.}~\bibnamefont
  {Mavrommatis}}, \bibinfo {author} {\bibfnamefont {K.~A.}\ \bibnamefont
  {Gernoth}}, \ and\ \bibinfo {author} {\bibfnamefont {J.~W.}\ \bibnamefont
  {Clark}},\ }\href {\doibase 10.1103/PhysRevC.80.044332} {\bibfield  {journal}
  {\bibinfo  {journal} {Phys. Rev. C}\ }\textbf {\bibinfo {volume} {80}},\
  \bibinfo {pages} {044332} (\bibinfo {year} {2009})}\BibitemShut {NoStop}%
\bibitem [{\citenamefont {Niu}\ \emph {et~al.}(2019)\citenamefont {Niu},
  \citenamefont {Liang}, \citenamefont {Sun}, \citenamefont {Long},\ and\
  \citenamefont {Niu}}]{Niu:2018PRC_BetaDecay}%
  \BibitemOpen
  \bibfield  {author} {\bibinfo {author} {\bibfnamefont {Z.~M.}\ \bibnamefont
  {Niu}}, \bibinfo {author} {\bibfnamefont {H.~Z.}\ \bibnamefont {Liang}},
  \bibinfo {author} {\bibfnamefont {B.~H.}\ \bibnamefont {Sun}}, \bibinfo
  {author} {\bibfnamefont {W.~H.}\ \bibnamefont {Long}}, \ and\ \bibinfo
  {author} {\bibfnamefont {Y.~F.}\ \bibnamefont {Niu}},\ }\href {\doibase
  10.1103/PhysRevC.99.064307} {\bibfield  {journal} {\bibinfo  {journal} {Phys.
  Rev. C}\ }\textbf {\bibinfo {volume} {99}},\ \bibinfo {pages} {064307}
  (\bibinfo {year} {2019})},\ \Eprint {http://arxiv.org/abs/1810.03156}
  {arXiv:1810.03156 [nucl-th]} \BibitemShut {NoStop}%
\bibitem [{\citenamefont {wang}\ \emph {et~al.}(2019)\citenamefont {wang},
  \citenamefont {Pei}, \citenamefont {Liu},\ and\ \citenamefont
  {Qiang}}]{Wang:2019PRL}%
  \BibitemOpen
  \bibfield  {author} {\bibinfo {author} {\bibfnamefont {Z.-A.}\ \bibnamefont
  {wang}}, \bibinfo {author} {\bibfnamefont {J.}~\bibnamefont {Pei}}, \bibinfo
  {author} {\bibfnamefont {Y.}~\bibnamefont {Liu}}, \ and\ \bibinfo {author}
  {\bibfnamefont {Y.}~\bibnamefont {Qiang}},\ }\href {\doibase
  10.1103/PhysRevLett.123.122501} {\bibfield  {journal} {\bibinfo  {journal}
  {Phys. Rev. Lett.}\ }\textbf {\bibinfo {volume} {123}},\ \bibinfo {pages}
  {122501} (\bibinfo {year} {2019})},\ \Eprint
  {http://arxiv.org/abs/1906.04485} {arXiv:1906.04485 [nucl-th]} \BibitemShut
  {NoStop}%
\bibitem [{\citenamefont {Lasseri}\ \emph {et~al.}(2020)\citenamefont
  {Lasseri}, \citenamefont {Regnier}, \citenamefont {Ebran},\ and\
  \citenamefont {Penon}}]{Lasseri:2020PRL}%
  \BibitemOpen
  \bibfield  {author} {\bibinfo {author} {\bibfnamefont {R.-D.}\ \bibnamefont
  {Lasseri}}, \bibinfo {author} {\bibfnamefont {D.}~\bibnamefont {Regnier}},
  \bibinfo {author} {\bibfnamefont {J.-P.}\ \bibnamefont {Ebran}}, \ and\
  \bibinfo {author} {\bibfnamefont {A.}~\bibnamefont {Penon}},\ }\href
  {\doibase 10.1103/PhysRevLett.124.162502} {\bibfield  {journal} {\bibinfo
  {journal} {Phys. Rev. Lett.}\ }\textbf {\bibinfo {volume} {124}},\ \bibinfo
  {pages} {162502} (\bibinfo {year} {2020})},\ \Eprint
  {http://arxiv.org/abs/1910.04132} {arXiv:1910.04132 [nucl-th]} \BibitemShut
  {NoStop}%
\bibitem [{\citenamefont {Yang}\ \emph {et~al.}(2022)\citenamefont {Yang},
  \citenamefont {Fan}, \citenamefont {Naito}, \citenamefont {Niu},
  \citenamefont {Li},\ and\ \citenamefont {Liang}}]{Yang:2022}%
  \BibitemOpen
  \bibfield  {author} {\bibinfo {author} {\bibfnamefont {Z.-X.}\ \bibnamefont
  {Yang}}, \bibinfo {author} {\bibfnamefont {X.-H.}\ \bibnamefont {Fan}},
  \bibinfo {author} {\bibfnamefont {T.}~\bibnamefont {Naito}}, \bibinfo
  {author} {\bibfnamefont {Z.-M.}\ \bibnamefont {Niu}}, \bibinfo {author}
  {\bibfnamefont {Z.-P.}\ \bibnamefont {Li}}, \ and\ \bibinfo {author}
  {\bibfnamefont {H.}~\bibnamefont {Liang}},\ }\href@noop {} {\  (\bibinfo
  {year} {2022})},\ \Eprint {http://arxiv.org/abs/2205.15649} {arXiv:2205.15649
  [nucl-th]} \BibitemShut {NoStop}%
\bibitem [{\citenamefont {Wu}\ \emph {et~al.}(2022{\natexlab{b}})\citenamefont
  {Wu}, \citenamefont {Ren},\ and\ \citenamefont {Zhao}}]{Wu:2022PRC}%
  \BibitemOpen
  \bibfield  {author} {\bibinfo {author} {\bibfnamefont {X.~H.}\ \bibnamefont
  {Wu}}, \bibinfo {author} {\bibfnamefont {Z.~X.}\ \bibnamefont {Ren}}, \ and\
  \bibinfo {author} {\bibfnamefont {P.~W.}\ \bibnamefont {Zhao}},\ }\href
  {\doibase 10.1103/PhysRevC.105.L031303} {\bibfield  {journal} {\bibinfo
  {journal} {Phys. Rev. C}\ }\textbf {\bibinfo {volume} {105}},\ \bibinfo
  {pages} {L031303} (\bibinfo {year} {2022}{\natexlab{b}})},\ \Eprint
  {http://arxiv.org/abs/2105.07696} {arXiv:2105.07696 [nucl-th]} \BibitemShut
  {NoStop}%
\bibitem [{\citenamefont {Islam}\ \emph {et~al.}(1979)\citenamefont {Islam},
  \citenamefont {Mang},\ and\ \citenamefont {Ring}}]{Islam:1979NPA}%
  \BibitemOpen
  \bibfield  {author} {\bibinfo {author} {\bibfnamefont {S.}~\bibnamefont
  {Islam}}, \bibinfo {author} {\bibfnamefont {H.}~\bibnamefont {Mang}}, \ and\
  \bibinfo {author} {\bibfnamefont {P.}~\bibnamefont {Ring}},\ }\href {\doibase
  https://doi.org/10.1016/0375-9474(79)90373-7} {\bibfield  {journal} {\bibinfo
   {journal} {Nucl. Phys. A}\ }\textbf {\bibinfo {volume} {326}},\ \bibinfo
  {pages} {161} (\bibinfo {year} {1979})}\BibitemShut {NoStop}%
\bibitem [{\citenamefont {Yao}\ \emph {et~al.}(2009)\citenamefont {Yao},
  \citenamefont {Meng}, \citenamefont {Ring},\ and\ \citenamefont
  {Arteaga}}]{Yao:2009PRC}%
  \BibitemOpen
  \bibfield  {author} {\bibinfo {author} {\bibfnamefont {J.~M.}\ \bibnamefont
  {Yao}}, \bibinfo {author} {\bibfnamefont {J.}~\bibnamefont {Meng}}, \bibinfo
  {author} {\bibfnamefont {P.}~\bibnamefont {Ring}}, \ and\ \bibinfo {author}
  {\bibfnamefont {D.~P.}\ \bibnamefont {Arteaga}},\ }\href {\doibase
  10.1103/PhysRevC.79.044312} {\bibfield  {journal} {\bibinfo  {journal} {Phys.
  Rev. C}\ }\textbf {\bibinfo {volume} {79}},\ \bibinfo {pages} {044312}
  (\bibinfo {year} {2009})}\BibitemShut {NoStop}%
\bibitem [{\citenamefont {Hagino}\ \emph {et~al.}(2003)\citenamefont {Hagino},
  \citenamefont {Bertsch},\ and\ \citenamefont {Reinhard}}]{Hagino:2003}%
  \BibitemOpen
  \bibfield  {author} {\bibinfo {author} {\bibfnamefont {K.}~\bibnamefont
  {Hagino}}, \bibinfo {author} {\bibfnamefont {G.~F.}\ \bibnamefont {Bertsch}},
  \ and\ \bibinfo {author} {\bibfnamefont {P.-G.}\ \bibnamefont {Reinhard}},\
  }\href {\doibase 10.1103/PhysRevC.68.024306} {\bibfield  {journal} {\bibinfo
  {journal} {Phys. Rev. C}\ }\textbf {\bibinfo {volume} {68}},\ \bibinfo
  {pages} {024306} (\bibinfo {year} {2003})}\BibitemShut {NoStop}%
\bibitem [{\citenamefont {Sabbey}\ \emph {et~al.}(2007)\citenamefont {Sabbey},
  \citenamefont {Bender}, \citenamefont {Bertsch},\ and\ \citenamefont
  {Heenen}}]{Sabbey:2007}%
  \BibitemOpen
  \bibfield  {author} {\bibinfo {author} {\bibfnamefont {B.}~\bibnamefont
  {Sabbey}}, \bibinfo {author} {\bibfnamefont {M.}~\bibnamefont {Bender}},
  \bibinfo {author} {\bibfnamefont {G.~F.}\ \bibnamefont {Bertsch}}, \ and\
  \bibinfo {author} {\bibfnamefont {P.-H.}\ \bibnamefont {Heenen}},\ }\href
  {\doibase 10.1103/PhysRevC.75.044305} {\bibfield  {journal} {\bibinfo
  {journal} {Phys. Rev. C}\ }\textbf {\bibinfo {volume} {75}},\ \bibinfo
  {pages} {044305} (\bibinfo {year} {2007})}\BibitemShut {NoStop}%
\bibitem [{\citenamefont {{Brink}}\ and\ \citenamefont
  {{Weiguny}}(1968)}]{Brink:1968NPA}%
  \BibitemOpen
  \bibfield  {author} {\bibinfo {author} {\bibfnamefont {D.~M.}\ \bibnamefont
  {{Brink}}}\ and\ \bibinfo {author} {\bibfnamefont {A.}~\bibnamefont
  {{Weiguny}}},\ }\href {\doibase 10.1016/0375-9474(68)90059-6} {\bibfield
  {journal} {\bibinfo  {journal} {Nucl. Phys. A}\ }\textbf {\bibinfo {volume}
  {120}},\ \bibinfo {pages} {59} (\bibinfo {year} {1968})}\BibitemShut
  {NoStop}%
\bibitem [{\citenamefont {Shalev-Shwartz}\ and\ \citenamefont
  {Ben-David}(2014)}]{Shwartz:2014Book}%
  \BibitemOpen
  \bibfield  {author} {\bibinfo {author} {\bibfnamefont {S.}~\bibnamefont
  {Shalev-Shwartz}}\ and\ \bibinfo {author} {\bibfnamefont {S.}~\bibnamefont
  {Ben-David}},\ }\href@noop {} {\emph {\bibinfo {title} {Understanding Machine
  Learning - From Theory to Algorithms.}}}\ (\bibinfo  {publisher} {Cambridge
  University Press},\ \bibinfo {year} {2014})\ pp.\ \bibinfo {pages} {I--XVI,
  1--397}\BibitemShut {NoStop}%
\bibitem [{\citenamefont {Geron}(2017)}]{Aurelien:2017Book}%
  \BibitemOpen
  \bibfield  {author} {\bibinfo {author} {\bibfnamefont {A.}~\bibnamefont
  {Geron}},\ }\href@noop {} {\emph {\bibinfo {title} {Hands-on machine learning
  with Scikit-Learn and TensorFlow : concepts, tools, and techniques to build
  intelligent systems}}}\ (\bibinfo  {publisher} {O'Reilly Media},\ \bibinfo
  {address} {Sebastopol, CA},\ \bibinfo {year} {2017})\BibitemShut {NoStop}%
\bibitem [{\citenamefont {Barlow}(1989)}]{Barlow:1989}%
  \BibitemOpen
  \bibfield  {author} {\bibinfo {author} {\bibfnamefont {R.~J.}\ \bibnamefont
  {Barlow}},\ }\href@noop {} {\emph {\bibinfo {title} {{Statistics. A guide to
  the use of statistical methods in the physical sciences}}}}\ (\bibinfo
  {publisher} {WileyBlackwell},\ \bibinfo {year} {1989})\BibitemShut {NoStop}%
\bibitem [{\citenamefont {Mehta}\ \emph {et~al.}(2019)\citenamefont {Mehta},
  \citenamefont {Bukov}, \citenamefont {Wang}, \citenamefont {Day},
  \citenamefont {Richardson}, \citenamefont {Fisher},\ and\ \citenamefont
  {Schwab}}]{Mehta2019PR}%
  \BibitemOpen
  \bibfield  {author} {\bibinfo {author} {\bibfnamefont {P.}~\bibnamefont
  {Mehta}}, \bibinfo {author} {\bibfnamefont {M.}~\bibnamefont {Bukov}},
  \bibinfo {author} {\bibfnamefont {C.-H.}\ \bibnamefont {Wang}}, \bibinfo
  {author} {\bibfnamefont {A.~G.}\ \bibnamefont {Day}}, \bibinfo {author}
  {\bibfnamefont {C.}~\bibnamefont {Richardson}}, \bibinfo {author}
  {\bibfnamefont {C.~K.}\ \bibnamefont {Fisher}}, \ and\ \bibinfo {author}
  {\bibfnamefont {D.~J.}\ \bibnamefont {Schwab}},\ }\href {\doibase
  https://doi.org/10.1016/j.physrep.2019.03.001} {\bibfield  {journal}
  {\bibinfo  {journal} {Physics Reports}\ }\textbf {\bibinfo {volume} {810}},\
  \bibinfo {pages} {1} (\bibinfo {year} {2019})}\BibitemShut {NoStop}%
\bibitem [{\citenamefont {Decharge}\ and\ \citenamefont
  {Gogny}(1980)}]{Decharge:1980zm}%
  \BibitemOpen
  \bibfield  {author} {\bibinfo {author} {\bibfnamefont {J.}~\bibnamefont
  {Decharge}}\ and\ \bibinfo {author} {\bibfnamefont {D.}~\bibnamefont
  {Gogny}},\ }\href {\doibase 10.1103/PhysRevC.21.1568} {\bibfield  {journal}
  {\bibinfo  {journal} {Phys. Rev. C}\ }\textbf {\bibinfo {volume} {21}},\
  \bibinfo {pages} {1568} (\bibinfo {year} {1980})}\BibitemShut {NoStop}%
\bibitem [{\citenamefont {Berger}\ \emph {et~al.}(1991)\citenamefont {Berger},
  \citenamefont {Girod},\ and\ \citenamefont {Gogny}}]{Berger:1991by}%
  \BibitemOpen
  \bibfield  {author} {\bibinfo {author} {\bibfnamefont {J.}~\bibnamefont
  {Berger}}, \bibinfo {author} {\bibfnamefont {M.}~\bibnamefont {Girod}}, \
  and\ \bibinfo {author} {\bibfnamefont {D.}~\bibnamefont {Gogny}},\ }\href
  {\doibase 10.1016/0010-4655(91)90263-K} {\bibfield  {journal} {\bibinfo
  {journal} {Comp. Phys. Comm.}\ }\textbf {\bibinfo {volume} {63}},\ \bibinfo
  {pages} {365} (\bibinfo {year} {1991})}\BibitemShut {NoStop}%
\bibitem [{\citenamefont {Yao}(2022)}]{Yao:2022HBNP}%
  \BibitemOpen
  \bibfield  {author} {\bibinfo {author} {\bibfnamefont {J.~M.}\ \bibnamefont
  {Yao}},\ }\href@noop {} {\  (\bibinfo {year} {2022})},\ \Eprint
  {http://arxiv.org/abs/2204.12126} {arXiv:2204.12126 [nucl-th]} \BibitemShut
  {NoStop}%
\bibitem [{\citenamefont {Hicks}\ and\ \citenamefont {Lee}(2022)}]{Hicks:2022}%
  \BibitemOpen
  \bibfield  {author} {\bibinfo {author} {\bibfnamefont {C.}~\bibnamefont
  {Hicks}}\ and\ \bibinfo {author} {\bibfnamefont {D.}~\bibnamefont {Lee}},\
  }\href@noop {} {\  (\bibinfo {year} {2022})},\ \Eprint
  {http://arxiv.org/abs/2209.02083} {arXiv:2209.02083 [nucl-th]} \BibitemShut
  {NoStop}%
\bibitem [{\citenamefont {{National Nuclear Data Center}}(2020)}]{NNDC}%
  \BibitemOpen
  \bibfield  {author} {\bibinfo {author} {\bibnamefont {{National Nuclear Data
  Center}}},\ }\href {https://www.nndc.bnl.gov/nudat2} {\enquote {\bibinfo
  {title} {{NuDat 2 Database}},}\ } (\bibinfo {year} {2020}),\ \bibinfo {note}
  {\url{https://www.nndc.bnl.gov/nudat2}}\BibitemShut {NoStop}%
\bibitem [{\citenamefont {Zhao}\ \emph {et~al.}(2010)\citenamefont {Zhao},
  \citenamefont {Li}, \citenamefont {Yao},\ and\ \citenamefont
  {Meng}}]{Zhao:2010PRC}%
  \BibitemOpen
  \bibfield  {author} {\bibinfo {author} {\bibfnamefont {P.~W.}\ \bibnamefont
  {Zhao}}, \bibinfo {author} {\bibfnamefont {Z.~P.}\ \bibnamefont {Li}},
  \bibinfo {author} {\bibfnamefont {J.~M.}\ \bibnamefont {Yao}}, \ and\
  \bibinfo {author} {\bibfnamefont {J.}~\bibnamefont {Meng}},\ }\href {\doibase
  10.1103/PhysRevC.82.054319} {\bibfield  {journal} {\bibinfo  {journal} {Phys.
  Rev. C}\ }\textbf {\bibinfo {volume} {82}},\ \bibinfo {pages} {054319}
  (\bibinfo {year} {2010})}\BibitemShut {NoStop}%
\bibitem [{\citenamefont {Song}\ \emph {et~al.}(2017)\citenamefont {Song},
  \citenamefont {Yao}, \citenamefont {Ring},\ and\ \citenamefont
  {Meng}}]{Song:2017}%
  \BibitemOpen
  \bibfield  {author} {\bibinfo {author} {\bibfnamefont {L.~S.}\ \bibnamefont
  {Song}}, \bibinfo {author} {\bibfnamefont {J.~M.}\ \bibnamefont {Yao}},
  \bibinfo {author} {\bibfnamefont {P.}~\bibnamefont {Ring}}, \ and\ \bibinfo
  {author} {\bibfnamefont {J.}~\bibnamefont {Meng}},\ }\href {\doibase
  10.1103/PhysRevC.95.024305} {\bibfield  {journal} {\bibinfo  {journal} {Phys.
  Rev. C}\ }\textbf {\bibinfo {volume} {95}},\ \bibinfo {pages} {024305}
  (\bibinfo {year} {2017})}\BibitemShut {NoStop}%
\bibitem [{\citenamefont {Men\'endez}\ \emph {et~al.}(2009)\citenamefont
  {Men\'endez}, \citenamefont {Poves}, \citenamefont {Caurier},\ and\
  \citenamefont {Nowacki}}]{Menendez:2009}%
  \BibitemOpen
  \bibfield  {author} {\bibinfo {author} {\bibfnamefont {J.}~\bibnamefont
  {Men\'endez}}, \bibinfo {author} {\bibfnamefont {A.}~\bibnamefont {Poves}},
  \bibinfo {author} {\bibfnamefont {E.}~\bibnamefont {Caurier}}, \ and\
  \bibinfo {author} {\bibfnamefont {F.}~\bibnamefont {Nowacki}},\ }\href
  {\doibase 10.1016/j.nuclphysa.2008.12.005} {\bibfield  {journal} {\bibinfo
  {journal} {Nuclear Physics A}\ }\textbf {\bibinfo {volume} {818}},\ \bibinfo
  {pages} {139 } (\bibinfo {year} {2009})}\BibitemShut {NoStop}%
\bibitem [{\citenamefont {Hebeler}\ \emph {et~al.}(2011)\citenamefont
  {Hebeler}, \citenamefont {Bogner}, \citenamefont {Furnstahl}, \citenamefont
  {Nogga},\ and\ \citenamefont {Schwenk}}]{Hebeler:2011}%
  \BibitemOpen
  \bibfield  {author} {\bibinfo {author} {\bibfnamefont {K.}~\bibnamefont
  {Hebeler}}, \bibinfo {author} {\bibfnamefont {S.~K.}\ \bibnamefont {Bogner}},
  \bibinfo {author} {\bibfnamefont {R.~J.}\ \bibnamefont {Furnstahl}}, \bibinfo
  {author} {\bibfnamefont {A.}~\bibnamefont {Nogga}}, \ and\ \bibinfo {author}
  {\bibfnamefont {A.}~\bibnamefont {Schwenk}},\ }\href {\doibase
  10.1103/PhysRevC.83.031301} {\bibfield  {journal} {\bibinfo  {journal} {Phys.
  Rev. C}\ }\textbf {\bibinfo {volume} {83}},\ \bibinfo {pages} {031301}
  (\bibinfo {year} {2011})}\BibitemShut {NoStop}%
\bibitem [{\citenamefont {Nogga}\ \emph {et~al.}(2004)\citenamefont {Nogga},
  \citenamefont {Bogner},\ and\ \citenamefont {Schwenk}}]{Nogga:2004il}%
  \BibitemOpen
  \bibfield  {author} {\bibinfo {author} {\bibfnamefont {A.}~\bibnamefont
  {Nogga}}, \bibinfo {author} {\bibfnamefont {S.~K.}\ \bibnamefont {Bogner}}, \
  and\ \bibinfo {author} {\bibfnamefont {A.}~\bibnamefont {Schwenk}},\ }\href
  {\doibase 10.1103/PhysRevC.70.061002} {\bibfield  {journal} {\bibinfo
  {journal} {Phys. Rev. C}\ }\textbf {\bibinfo {volume} {70}},\ \bibinfo
  {pages} {061002} (\bibinfo {year} {2004})}\BibitemShut {NoStop}%
\bibitem [{\citenamefont {Stroberg}\ \emph {et~al.}(2021)\citenamefont
  {Stroberg}, \citenamefont {Holt}, \citenamefont {Schwenk},\ and\
  \citenamefont {Simonis}}]{Stroberg:2021PRL}%
  \BibitemOpen
  \bibfield  {author} {\bibinfo {author} {\bibfnamefont {S.~R.}\ \bibnamefont
  {Stroberg}}, \bibinfo {author} {\bibfnamefont {J.~D.}\ \bibnamefont {Holt}},
  \bibinfo {author} {\bibfnamefont {A.}~\bibnamefont {Schwenk}}, \ and\
  \bibinfo {author} {\bibfnamefont {J.}~\bibnamefont {Simonis}},\ }\href
  {\doibase 10.1103/PhysRevLett.126.022501} {\bibfield  {journal} {\bibinfo
  {journal} {Phys. Rev. Lett.}\ }\textbf {\bibinfo {volume} {126}},\ \bibinfo
  {pages} {022501} (\bibinfo {year} {2021})}\BibitemShut {NoStop}%
\bibitem [{\citenamefont {Yao}\ \emph {et~al.}(2021)\citenamefont {Yao},
  \citenamefont {Belley}, \citenamefont {Wirth}, \citenamefont {Miyagi},
  \citenamefont {Payne}, \citenamefont {Stroberg}, \citenamefont {Hergert},\
  and\ \citenamefont {Holt}}]{Yao:2021PRC}%
  \BibitemOpen
  \bibfield  {author} {\bibinfo {author} {\bibfnamefont {J.~M.}\ \bibnamefont
  {Yao}}, \bibinfo {author} {\bibfnamefont {A.}~\bibnamefont {Belley}},
  \bibinfo {author} {\bibfnamefont {R.}~\bibnamefont {Wirth}}, \bibinfo
  {author} {\bibfnamefont {T.}~\bibnamefont {Miyagi}}, \bibinfo {author}
  {\bibfnamefont {C.~G.}\ \bibnamefont {Payne}}, \bibinfo {author}
  {\bibfnamefont {S.~R.}\ \bibnamefont {Stroberg}}, \bibinfo {author}
  {\bibfnamefont {H.}~\bibnamefont {Hergert}}, \ and\ \bibinfo {author}
  {\bibfnamefont {J.~D.}\ \bibnamefont {Holt}},\ }\href {\doibase
  10.1103/PhysRevC.103.014315} {\bibfield  {journal} {\bibinfo  {journal}
  {Phys. Rev. C}\ }\textbf {\bibinfo {volume} {103}},\ \bibinfo {pages}
  {014315} (\bibinfo {year} {2021})}\BibitemShut {NoStop}%
\bibitem [{\citenamefont {Yao}\ \emph {et~al.}(2022{\natexlab{b}})\citenamefont
  {Yao}, \citenamefont {Ginnett}, \citenamefont {Belley}, \citenamefont
  {Miyagi}, \citenamefont {Wirth}, \citenamefont {Bogner}, \citenamefont
  {Engel}, \citenamefont {Hergert}, \citenamefont {Holt},\ and\ \citenamefont
  {Stroberg}}]{Yao:2022DGT}%
  \BibitemOpen
  \bibfield  {author} {\bibinfo {author} {\bibfnamefont {J.~M.}\ \bibnamefont
  {Yao}}, \bibinfo {author} {\bibfnamefont {I.}~\bibnamefont {Ginnett}},
  \bibinfo {author} {\bibfnamefont {A.}~\bibnamefont {Belley}}, \bibinfo
  {author} {\bibfnamefont {T.}~\bibnamefont {Miyagi}}, \bibinfo {author}
  {\bibfnamefont {R.}~\bibnamefont {Wirth}}, \bibinfo {author} {\bibfnamefont
  {S.}~\bibnamefont {Bogner}}, \bibinfo {author} {\bibfnamefont
  {J.}~\bibnamefont {Engel}}, \bibinfo {author} {\bibfnamefont
  {H.}~\bibnamefont {Hergert}}, \bibinfo {author} {\bibfnamefont {J.~D.}\
  \bibnamefont {Holt}}, \ and\ \bibinfo {author} {\bibfnamefont {S.~R.}\
  \bibnamefont {Stroberg}},\ }\href {\doibase 10.1103/PhysRevC.106.014315}
  {\bibfield  {journal} {\bibinfo  {journal} {Phys. Rev. C}\ }\textbf {\bibinfo
  {volume} {106}},\ \bibinfo {pages} {014315} (\bibinfo {year}
  {2022}{\natexlab{b}})},\ \Eprint {http://arxiv.org/abs/2204.12971}
  {arXiv:2204.12971 [nucl-th]} \BibitemShut {NoStop}%
\bibitem [{\citenamefont {Hergert}(2016)}]{Hergert:2016}%
  \BibitemOpen
  \bibfield  {author} {\bibinfo {author} {\bibfnamefont {H.}~\bibnamefont
  {Hergert}},\ }\href {\doibase 10.1088/1402-4896/92/2/023002} {\bibfield
  {journal} {\bibinfo  {journal} {Physica Scripta}\ }\textbf {\bibinfo {volume}
  {92}},\ \bibinfo {pages} {023002} (\bibinfo {year} {2016})}\BibitemShut
  {NoStop}%
\bibitem [{\citenamefont {Hergert}\ \emph {et~al.}(2016)\citenamefont
  {Hergert}, \citenamefont {Bogner}, \citenamefont {Morris}, \citenamefont
  {Schwenk},\ and\ \citenamefont {Tsukiyama}}]{Hergert:2016jk}%
  \BibitemOpen
  \bibfield  {author} {\bibinfo {author} {\bibfnamefont {H.}~\bibnamefont
  {Hergert}}, \bibinfo {author} {\bibfnamefont {S.~K.}\ \bibnamefont {Bogner}},
  \bibinfo {author} {\bibfnamefont {T.~D.}\ \bibnamefont {Morris}}, \bibinfo
  {author} {\bibfnamefont {A.}~\bibnamefont {Schwenk}}, \ and\ \bibinfo
  {author} {\bibfnamefont {K.}~\bibnamefont {Tsukiyama}},\ }\href {\doibase
  http://dx.doi.org/10.1016/j.physrep.2015.12.007} {\bibfield  {journal}
  {\bibinfo  {journal} {Physics Reports}\ }\textbf {\bibinfo {volume} {621}},\
  \bibinfo {pages} {165} (\bibinfo {year} {2016})}\BibitemShut {NoStop}%
\end{thebibliography}
  
  %merlin.mbs apsrev4-1.bst 2010-07-25 4.21a (PWD, AO, DPC) hacked
%Control: key (0)
%Control: author (72) initials jnrlst
%Control: editor formatted (1) identically to author
%Control: production of article title (-1) disabled
%Control: page (0) single
%Control: year (1) truncated
%Control: production of eprint (0) enabled
%

\end{document}